\documentclass[a4paper,fleqn,usenatbib]{mnras}

\usepackage{newtxtext,newtxmath}
\usepackage[T1]{fontenc}
\usepackage{ae,aecompl}
\usepackage{graphicx}
\usepackage{amsmath}

\usepackage{color}

\definecolor{drkgrn}{rgb}{0,0.5,0}
\definecolor{myred}{rgb}{0.7,0.1,0.1}

\definecolor{myblue}{rgb}{0.,0.4,0.7}

\definecolor{purple}{rgb}{0.4,0,0.4}

\def\figdir{./Figures}

\title[Isolated UDGs]{The Formation of Isolated Ultra-Diffuse Galaxies in {\sc Romulus25}}
\author[A. C. Wright et al.]{
Anna C. Wright$^{1}$\thanks{E-mail:acwright@jhu.edu}, Michael Tremmel$^{2}$, Alyson M. Brooks$^{1}$, Ferah Munshi$^{3}$, 
\newauthor{ Daisuke Nagai$^{2,4,5}$, Ray S. Sharma$^{1}$, and Thomas R. Quinn$^{6}$}
\\
$^{1}$ Department of Physics \& Astronomy, Rutgers, The State University of New Jersey, 136 Frelinghuysen Road, Piscataway, NJ 08854, USA \\
$^{2}$ Yale Center for Astronomy \& Astrophysics, Physics Department, P.O. Box 208120, New Haven, CT 06520, USA
\\
$^{3}$ Department of Physics \& Astronomy, University of Oklahoma 440 W. Brooks St., Norman, OK 73019\\
$^{4}$ Department of Physics, Yale University, New Haven, CT 06520, USA\\
$^{5}$ Department of Astronomy, Yale University, New Haven, CT 06511, USA\\
$^{6}$ Astronomy Department, University of Washington, Box 351580, Seattle, WA, 98195-1580}
\date{Accepted XXX. Received YYY; in original form ZZZ}

\pubyear{2020}

\begin{document}
\label{firstpage}
\pagerange{\pageref{firstpage}--\pageref{lastpage}}
\maketitle
\begin{abstract}
{We use the \textsc{Romulus25} cosmological simulation volume to identify the largest-ever simulated sample of {\it field} ultra-diffuse galaxies (UDGs). At $z=0$, we find that isolated UDGs have average star formation rates, colors, and virial masses for their stellar masses and environment. UDGs have moderately elevated HI masses, being 70\% (300\%) more HI-rich than typical isolated dwarf galaxies at luminosities brighter (fainter) than M$_\mathrm{B}$=-14. However, UDGs are consistent with the general isolated dwarf galaxy population and make up $\sim$20\% of all field galaxies with 10$^7$<M$_\star$/M$_\odot$<10$^{9}$. The HI masses, effective radii, and overall appearances of our UDGs are consistent with existing observations of field UDGs, but we predict that many isolated UDGs have been missed by current surveys. Despite their isolation at $z=0$, the UDGs in our sample are the products of major mergers. Mergers are no more common in UDG than non-UDG progenitors, but mergers that create UDGs tend to happen earlier -- almost never occurring after $z=1$, produce a temporary boost in spin, and cause star formation to be redistributed to the outskirts of galaxies, resulting in lower central star formation rates. The centers of the galaxies fade as their central stellar populations age, but their global star formation rates are maintained through bursts of star formation at larger radii, producing steeper negative g-r color gradients. This formation channel is unique relative to other proposals for UDG formation in isolated galaxies, demonstrating that UDGs can potentially be formed through multiple mechanisms.\\ \\}
\end{abstract}


\maketitle
\section{Introduction}
\label{intro}
\indent The hypothesis that a large number of galaxies might lurk beneath the limiting brightness of the night sky dates back at least 60 years \cite[e.g.,][]{zwicky1957,disney1976}. In the intervening decades, advances in imaging and image processing techniques have confirmed the existence of a rich low surface brightness (LSB) universe \citep[e.g.,][]{kormendy1974,malin1978,binggeli1985,schwartzenberg1995,abraham2014,mihos2017,danieli2018,borlaff2019}, the inhabitants of which are not only numerous, but incredibly diverse. Of particular interest in recent years are the subset dubbed "ultra-diffuse galaxies" (UDGs), which have stellar masses typical of dwarf galaxies, but physical sizes more akin to Milky-Way-mass galaxies (although see \citeauthor{chamba2020} \citeyear{chamba2020} and \citeauthor{trujillo2020} \citeyear{trujillo2020} for in-depth discussions of galaxy size). Although such objects had been observed in the past \citep[e.g.,][]{sandage1984,caldwell1987,impey1988,conselice2003}, their abundance -- particularly within clusters -- is a recent revelation. The diffuse nature of extreme LSB galaxies has long been viewed as evidence that they are incapable of surviving in high density environments \citep[e.g.,][]{moore1996,gnedin2003}. However, they have now been found in large numbers in several local clusters \citep[e.g.,][]{vandokkum2015forty,koda2015,mihos2015,vanderburg2016,lee2017,wittmann2017}, as well as in a number of nearby galaxy groups \citep[e.g.,][]{merritt2016,roman2017,greco2018illuminating}.\\
\indent Their seeming ability to survive the harsh tides of the cluster environment has sparked debate about the underlying properties of the dark matter halos inhabited by UDGs. \citet{vandokkum2015forty} proposed that UDGs occupy Milky-Way-mass halos, but failed to form enough stars due to early gas loss. This scenario is supported by kinematic studies of individual UDGs in the Virgo and Coma clusters that have measured halo masses of 10$^{11}$-10$^{12}$ M$_\odot$ \citep[e.g.,][]{vandokkum2016,martinnavarro2019,vandokkum2019}, as well as the higher than average specific frequency and luminosity of the globular clusters observed around some UDGs \citep[e.g.,][]{mihos2015,vandokkum2017}. \\
\indent However, a number of authors have also found evidence that UDGs are true dwarfs, inhabiting halos with M$_\mathrm{vir}\leq$10$^{11}$ M$_\odot$ \citep[e.g.,][]{trujillo2017,kovacs2019}. Many UDGs appear to have globular cluster populations consistent with those of more typical dwarf galaxies \citep[e.g.,][]{beasley2016,peng2016,beasleytrujillo2016}, while others, though globular-cluster-rich, have stellar or gas kinematics that suggest that they may lack dark matter halos entirely \citep[e.g.,][]{vandokkum2018,vandokkum2019DF4,danieli2019,vandokkum2019,mancerapina2019btfr}. Increasingly, the evidence seems to point to a picture of UDGs as a diverse population that likely formed through a variety of mechanisms \citep{zaritsky2017,papastergis2017,lee2017,amorisco2018,toloba2018,lim2018,forbes2020}.\\
\indent Many of the proposed mechanisms to form this diverse sample rely on the cluster/group environment. Idealized simulations from \citet{yozin2015} and \citet{safarzadeh2017} and cosmological simulations from \citet{chan2018}, \citet{jiang2019}, and \citet{tremmel2020} suggest that the primary role of the cluster/group environment is to quench and puff up the infalling UDG progenitor through ram pressure stripping and/or strangulation, causing the galaxy to dim as its stellar population passively evolves. Other authors \citep[e.g.,][]{liao2019,sales2020} cite the tidal stripping and heating endemic to dense environments as the primary factor in UDG formation. Semi-analytic models from \citet{carleton2019} show that this process is particularly effective in cored halos, where tidal stripping results in increased expansion of the stellar component of the infalling galaxy. \citet{ogiya2018} used N-body simulations to show that this same process might be responsible for the severely dark-matter-deficient UDGs discovered in the NGC 1052 group \citep[e.g.,][]{vandokkum2018}.\\
\indent However, some authors have also proposed internal formation mechanisms. Analytic models have long suggested that LSB galaxies are the natural inhabitants of dark matter halos that form with higher-than-average angular momentum \citep[e.g.,][]{dalcanton1997}. This work was extended to UDGs by \citet{amorisco2016}, who used semi-analytic models to show that a population of dwarf galaxies with high spin could reproduce many of the properties of observed UDGs \cite[see also][]{rong2017,liao2019}. Alternatively, cosmological simulations from \citet{dicintio2017}, \citet{chan2018}, and \citet{martin2019} have shown that the same repeated bursts of supernova feedback that lead to the creation of dark matter cores in dwarf galaxies \citep[e.g.,][]{governato2010,pontzen2012,dicintio2014,chan2015} can also cause their stellar components to expand, leading to the formation of UDGs \citep[see also follow-up work in][]{jiang2019,freundlich2020,cardonabarrero2020,jackson2020}. Because these mechanisms do not rely on the presence of a dense environment, the authors predict that UDGs ought to be present in isolation, and likely with different properties than those in clusters and groups.\\
\indent Yet, while thousands of UDGs have been found in clusters and groups, fewer than 150 have thus far been discovered in the field (although see \citeauthor{prole2019} \citeyear{prole2019} and \citeauthor{barbosa2020} \citeyear{barbosa2020} for a number of unconfirmed candidates). This is largely due to the fact that the positive identification of a UDG requires not only a central surface brightness, but also a physical effective radius, and therefore a distance. Distance measurements for cluster and group UDGs typically come from their association with brighter (and therefore better studied) galaxies. Isolated UDGs, however, require either new distance measurements or detection by a previous survey -- a tall order, given their inherent faintness \citep[but see][]{greco2020}. Although a few individual field UDGs have been discovered \citep[e.g.,][]{kadowaki2017,greco2018study}, by far the largest sample comes from \citet{leisman2017}, who identified 115 HI-rich UDGs in ALFALFA data \citep[see also follow-up work in][]{jones2018,he2019,mancerapina2019btfr,janowiecki2019,mancerapina2020}. Like most HI-rich galaxies, these UDGs are overwhelmingly star-forming and blue. They also appear to inhabit dwarf-mass halos with slightly elevated spin. However, much remains unknown about the population of UDGs that inhabits low density environments - particularly those galaxies that might lie below the detection threshold of surveys like ALFALFA.\\
\indent In this paper, we use the cosmological simulation \textsc{Romulus25} to study the formation and evolution of isolated UDGs. \textsc{Romulus25} is a 25-Mpc-per-side volume and is run with the same state-of-the-art subgrid physics and resolution as \textsc{RomulusC}, a zoom-in simulation of a 10$^{14}$ M$_\odot$ galaxy cluster that we have previously used to explore the properties and origins of cluster UDGs \citep{tremmel2020}. To date, it is one of the highest resolution volumes ever run (cf., L025N0752 \citep{schaye2014}; TNG50 \citep{pillepich2019}), approaching the resolution of much smaller zoom-in simulations. This unique combination of resolution and volume allows us to study large numbers of dwarf galaxies in depth. We describe the properties of the simulation in greater detail in Section \ref{sims}. In Section \ref{res}, we compare the evolution and $z=0$ properties of our sample of isolated UDGs to a carefully-selected sample of non-UDGs and explore the origin of isolated UDGs. We compare our sample to existing observations of isolated UDGs in Section \ref{disc} and summarize our findings in Section \ref{wrapup}.
\section{The \textsc{Romulus25} Simulation}
\label{sims}
\indent All of the galaxies analyzed in this paper are selected from \textsc{Romulus25} \citep{tremmel2017}, a high resolution cosmological simulation run using the N-body + smoothed particle hydrodynamics (SPH) code \textsc{ChaNGa} \citep{menon2015}. \textsc{ChaNGa} uses many of the same physics modules as its precursor, \textsc{Gasoline} \citep{wadsley2004}, as well as an updated SPH implementation that reduces artificial surface tension, allowing for better capture of fluid instabilities \citep{wadsley2017}. As in \textsc{Gasoline}, unresolved physics (e.g., star formation, supermassive black hole (SMBH) growth, and stellar and SMBH feedback) is governed by subgrid prescriptions. These contain free parameters that have been optimized over a broad parameter space to create galaxies in the halo mass range 10$^{10.5-12}$ M$_\odot$ that match $z=0$ scaling relations, including the stellar mass - halo mass \citep{Moster2013} and stellar mass - SMBH mass relations \citep{schramm2013}. However, \textsc{Romulus25} has also been shown to produce realistic galaxies across its entire resolved range (M$_\mathrm{vir}$=3$\times$10$^9$ - 2$\times$10$^{13}$ M$_\odot$) and to reproduce observations of high redshift star formation and SMBH growth \citep{tremmel2017}. \\
\indent {\sc Romulus25} is a uniform resolution simulation consisting of a co-moving volume measuring 25 Mpc on each side. The simulation is run with a $\Lambda$CDM cosmology following \cite{planck2014} ($\Omega_0$ = 0.3086, $\Lambda$ = 0.6914, $h$ = 0.6777, $\sigma_8$ = 0.8288) and evolved until $z=0$. Gravitational interactions between particles are resolved with a spline gravitational force softening length ($\epsilon_g$) of 350 pc (Plummer equivalent 250 pc), which converges to a Newtonian force at 2$\epsilon_g$. Dark matter particles within \textsc{Romulus25} are oversampled relative to gas particles, such that the simulation initially contains 3.375 times more of the former than the latter. This allows us to use dark matter and gas particles that are more similar in mass than in many comparable simulations. Each dark matter particle has a mass of 3.39$\times$10$^5$ M$_\odot$ while each gas particle has a mass of 2.12$\times$10$^5$ M$_\odot$. In addition to enabling the simulation to track the dynamics of SMBHs \citep{tremmel2015}, this decreases numerical effects due to two-body scattering. This is particularly relevant for the study of UDGs, as energy equipartition has recently been shown to lead to spurious growth in galaxy sizes over time \citep{ludlow2019}. \\
\indent In order to approximate the effects of reionization, {\sc Romulus25} includes a UV background following a revised version of \cite{haardt2012} with self-shielding from \cite{pontzen2008}. The simulation uses primordial cooling for neutral and ionized H and He, which is calculated from collisional ionization rates \citep{abel1997}, radiative recombination \citep{black1981,verner1996}, photoionization, bremsstrahlung, and H and He line cooling \citep{cen1992}. The simulation also includes low-temperature metal-line cooling following \citet{bromm2001}. However, although high resolution, \textsc{Romulus25} does not resolve the multiphase interstellar medium (ISM), and, in particular, cannot track the creation and destruction of molecular hydrogen. Because metal-line cooling in the absence of molecular hydrogen physics has been shown to lead to overcooling in spiral galaxies \citep{christensen2014}, we do not include high-temperature metal-line cooling.\\
\indent Star formation within \textsc{Romulus25} is a stochastic process. Any gas particle that is sufficiently cold (T<10$^4$K) and dense (n>0.2 cm$^{-3}$) has a probability $p$ of forming a star particle:
\begin{equation}
\label{eq:1}
p = \frac{m_{\mathrm{gas}}}{m_{\mathrm{star}}}(1-e^{c^*\Delta t/t_{\mathrm{form}}})
\end{equation}
where m$_{\mathrm{gas}}$ is the mass of the gas particle, m$_{\mathrm{star}}$ is the mass of the resulting star particle, $c^*$ is the star-forming efficiency factor (here set to 0.15), $\Delta t$ is the star formation timescale (10$^6$ yr in this simulation), and $t_{\mathrm{form}}$ is the dynamical time. Any resultant star particle forms with a mass 30\% of the initial gas particle mass ($M_{\star} = $6$\times$10$^4$ M$_\odot$) and represents a simple stellar population, with masses and corresponding lifetimes drawn from a \citet{kroupa2001} initial mass function. \\
\indent While stars with masses greater than 40 M$_\odot$ are assumed to collapse directly to black holes, those with 8$\leq$M/M$_\odot\leq$40 explode as Type II supernovae (SNe). This is implemented via `blastwave' feedback following \citet{stinson2006}. Each SN generates 0.75$\times$10$^{51}$ ergs that is thermally deposited into the surrounding gas particles, where cooling is temporarily disabled to mimic the adiabatic expansion phase of the SN. Lower mass stars also contribute to feedback via Type Ia SNe, in which cooling is not disabled, and stellar winds \citep{kennicutt1994}, which account for 99\% of all mass lost by a given star particle over its lifetime. Mass and metals are returned to the ISM following \citet{shen2010} and \citet{governato2015}. \\
\indent \textsc{Romulus25} also includes a novel implementation of black hole physics \citep{tremmel2015,tremmel2017,tremmel2018dancing,tremmel2018wandering,tremmel2019introducing}. Seed SMBHs (M=$10^6 M_\odot$) form in pristine (Z<3$\times$10$^{-4}Z_{\odot}$), dense (n>3 cm$^{-3}$) gas that has not yet cooled to the temperature required for star formation. This ensures that SMBHs form in regions that are collapsing more quickly than either the cooling or star formation timescales and that, in the vast majority of cases, they form within the first Gyr of the simulation. SMBH orbits are then traced via a dynamical friction sub-grid model following \cite{tremmel2015}, which allows for better tracking of dynamical evolution, including mergers. SMBHs are permitted to grow via modified Bondi-Hoyle accretion. In order to approximate the effects of active galactic nucleus (AGN) feedback, any actively accreting SMBH converts a fraction of the accreted mass to thermal energy, which it injects into the surrounding gas particles, naturally driving a collimated outflow. As with SN feedback, cooling is temporarily disabled in the affected gas particles to prevent them from radiating the energy away too quickly as a result of limited resolution. \\ 
\indent Halos are identified with Amiga's Halo Finder \citep[AHF;][]{knebe2001,gill2004} and tracked across timesteps with {\sc tangos} \citep{pontzen2018}. Halo properties are calculated based on all particles within a halo's virial radius (R$_{\mathrm{vir}}$), which is calculated by AHF via a spherical top-hat collapse technique that varies with redshift following \citet{bryan1998}. However, throughout the paper we also refer to M$_{200}$, which is the mass contained within the radius at which the mean enclosed density of particles bound to the halo drops below 200 times the critical density of the universe at the relevant redshift. In order to facilitate better comparison with observations, we calculate stellar masses based on photometric colors following \citet{munshi2013}.
\section{Results}
\label{res}
\subsection{Classification of UDGs}
\label{class}
\begin{figure*}
\includegraphics[trim=3mm 10mm 28mm 22mm, clip,width=0.97\textwidth]{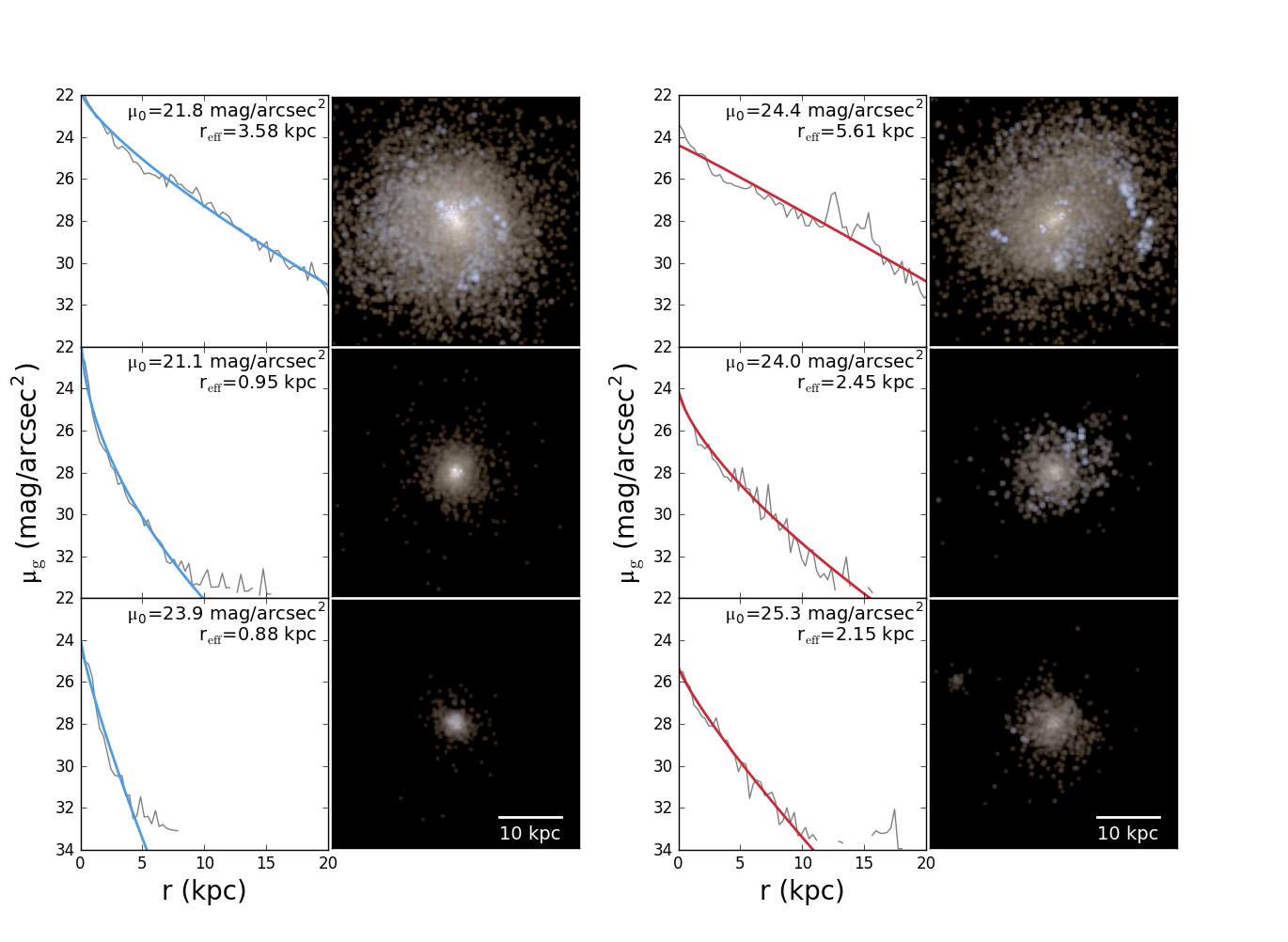}
\caption{g-band surface brightness profiles with accompanying S\'ersic fits and face-on UVI images of representative isolated dwarfs (left) and UDGs (right) from Romulus25. Fit parameters for each galaxy are shown in the upper right hand corner of each surface brightness profile plot. The top panels depict high mass dwarfs (10$^8$<M$_\star$/M$_\odot$<10$^9$), the middle panels intermediate mass dwarfs (10$^{7.5}$<M$_\star$/M$_\odot$<10$^8$), and the bottom panels low mass dwarfs (10$^7$<M$_\star$/M$_\odot$<10$^{7.5}$). Each UVI image is 40 kpc across and reaches a surface brightness of 32 mag/arcsec$^2$. Note that we do not attempt to fit any features fainter than this, as this is roughly the limit of the most sensitive observations. Surface brightness calculations and UVI images were generated using results from stellar population synthesis models \citep[http://stev.oapd.inaf.it/cgi-bin/cmd][]{Marigo2008,Girardi2010}.}
\label{fig:rep}
\end{figure*}

In order to ensure that we are analysing only well-resolved galaxies, we begin by limiting our sample to only those galaxies with M$_{\mathrm{vir}}$>3$\times$10$^9$ M$_\odot$ ($\sim$10,000 dark matter particles) and M$_{\star}$>10$^7$ M$_\odot$ ($\sim$150 star particles) at $z=0$. Within \textsc{Romulus25}, we identify 1799 such galaxies, from which we select isolated galaxies following \citet{geha2012}. In addition to eliminating satellites of any other halos, this definition requires that isolated galaxies be at least 1.5 Mpc away from any galaxy with M$_{\star}$>2.5$\times$10$^{10}$ M$_\odot$. A number of surveys have found that dwarf galaxies with M$_\star$<10$^9$M$_\odot$ are almost exclusively star-forming at this distance from a massive galaxy \citep[e.g.,][]{geha2012,rasmussen2012,penny2016}. As low-mass galaxies are thought to quench only through environmental processes, this finding indicates that galaxies more than 1.5 Mpc from a massive galaxy have likely never interacted with it. Adopting this definition therefore allows us to mitigate the possibility of including galaxies that have been transformed into UDGs through ram pressure stripping or tidal heating in our sample. We find 890 isolated galaxies in \textsc{Romulus25}. \\
\indent UDGs are identified from this isolated sample via the process described in \citet{tremmel2020}. So as to best mimic observational methods, we fit a S\'ersic profile to the $z=0$ g-band surface brightness profile of each galaxy, sampling at the spatial resolution of the simulation (300 pc). We do not attempt to fit any features fainter than 32 mag/arcsec$^2$, as this is roughly the depth of the most sensitive observations \citep[e.g.,][]{trujillo2016,borlaff2019}. The equation for a S\'ersic profile is given by 
\begin{equation}
    \mu(r) = \mu_{\mathrm{eff}}+2.5c_n\Big(\Big(\frac{r}{r_{\mathrm{eff}}}\Big)^{1/n}-1\Big)
\end{equation}
\citep{sersic1963}, where $\mu$ is the surface brightness at radius $r$ and $\mu_{\mathrm{eff}}$ and r$_{\mathrm{eff}}$ are the effective surface brightness and the effective radius, permitted to range between 10-40 mag/arcsec$^2$ and 0-100 kpc, respectively. Following \citet{capaccioli1989}, c$_n$ = 0.868$n$-0.142, where $n$ is the S\'ersic index, which we allow to vary between 0.5 and 16.5. Sample isolated galaxies and their accompanying surface brightness profiles and S\'ersic fits are shown in Figure \ref{fig:rep}.\\
\indent For our well-resolved isolated sample, our fitting procedure fails for only 3 galaxies, although we eliminate another 4 whose fitting parameters' proximity to the fitting bounds indicate a poor fit. Each of the remaining fits is then inspected by eye in order to ensure that a reasonable set of parameters has been found. This results in the removal of a further 6 galaxies, the majority of which have ongoing major mergers. The final sample is, therefore, composed of 877 galaxies. \\ 
\indent Following \cite{vandokkum2015forty}, we identify as a UDG any galaxy with r$_{\mathrm{eff}}\geq$1.5 kpc and $\mu_{\mathrm{0,g}}\geq$24 mag/arcsec$^2$. The former measurement is taken directly from the S\'ersic fit, while the latter is the value of the S\'ersic profile evaluated at $r=0$. Because our classification is based on the S\'ersic fit, rather than the actual surface brightness profile, $\sim$12\% of those galaxies identified as UDGs have actual $\mu_\mathrm{0,g}$ - here defined as the average g-band surface brightness within the inner 300 kpc of the galaxy - brighter than 24 mag/arcsec$^2$. However, the majority of these are relatively high mass (M$_\star$>10$^8$ M$_\odot$) UDGs that have actual $\mu_{\mathrm{0,g}}$>23.6 mag/arcsec$^2$ (see, for example, the top right panel of Figure \ref{fig:rep}). As these galaxies still, therefore, represent the low surface brightness tail of the galaxy distribution and this method is consistent with that used by observers \citep[e.g.,][]{martinezdelgado2016}, we do not feel that this introduces significant contamination into our sample.\\
\indent We find a total of 134 isolated UDGs in the \textsc{Romulus25} simulation. We discuss how this number compares to current estimates from observations in Section \ref{disc}.
\subsection{Properties of UDGs at z=0}
\label{iniprops}
\begin{figure}
\includegraphics[trim=17mm 10mm 17mm 27mm, clip, width=0.47\textwidth]{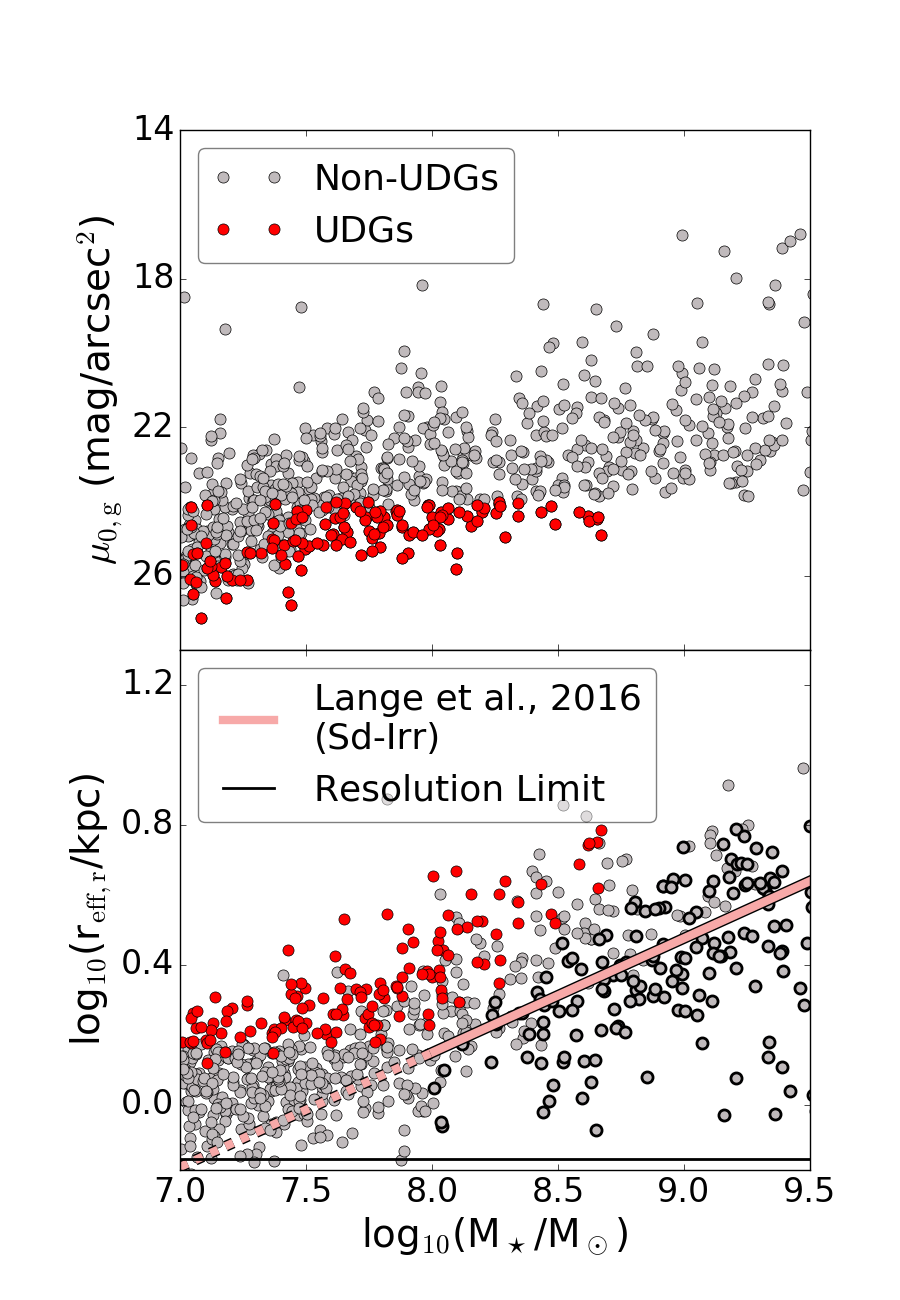}
\caption{Central surface brightness (top) and effective radius (bottom) plotted against stellar mass for all of the isolated galaxies in \textsc{Romulus25} with M$_{\star}$<10$^{9.5}$ M$_\odot$. There is no obvious separation between UDGs (red) and non-UDGs (gray) on these axes. UDGs are not a separate population, but the high effective radius - low central surface brightness tail of the galaxy population. Note that, while we classify our UDGs using g-band data and show g-band central surface brightnesses here, the effective radii shown in this figure are based on fits to r-band surface brightness profiles to provide a more accurate comparison to the Sd-Irr M$_{\star}$-r$_{\mathrm{eff}}$ relation from \protect\citet{lange2016}, which is based on r-band data from the GAMA survey and is extrapolated below M$_{\star} = 10^8$ M$_\odot$ (indicated by the dashed line). This is why some UDGs appear to have r$_{\mathrm{eff}}$<1.5 kpc. In the lower panel, points with thick outlines indicate galaxies that have $\mu_\mathrm{eff,r}$<24.5 mag/arcsec$^2$ and M$_\star$>10$^8$ M$_\odot$ and would therefore likely be observable by GAMA. As we would expect, they adhere more closely to the published relation.}
\label{fig:smreffmu0}
\end{figure}

\begin{figure}
\includegraphics[trim= 4mm 5mm 5mm 5mm,clip, width=0.47\textwidth]{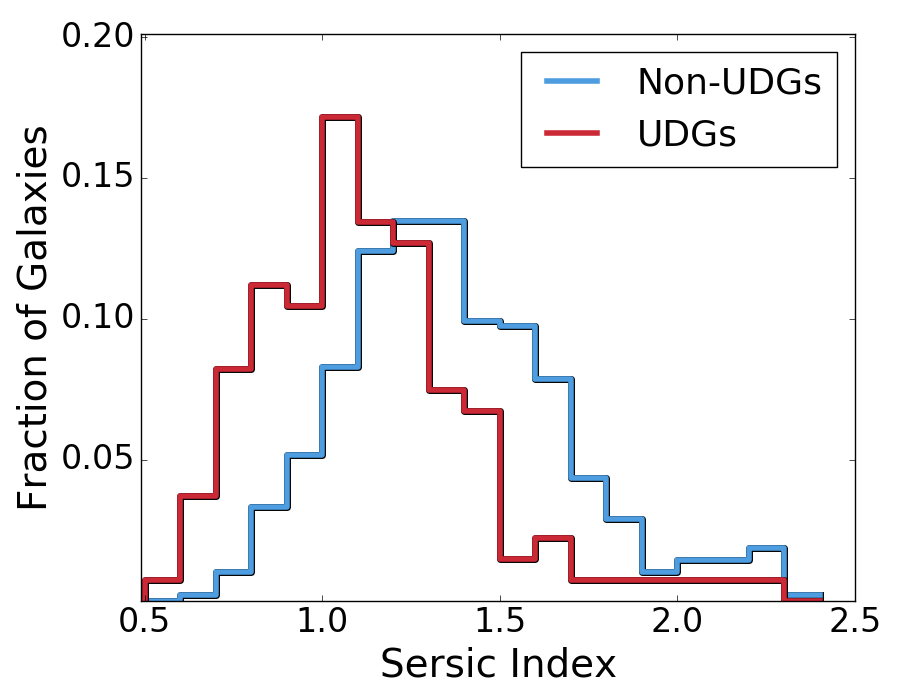}
\caption{Distribution of S\'ersic indices for isolated UDGs, shown in red, and isolated non-UDGs, shown in blue, in \textsc{Romulus25}. We limit this figure to galaxies with M$_{\star}$<10$^{8.7}$ M$_\odot$, corresponding to the most massive UDG. UDGs tend to have slightly lower S\'ersic indices than non-UDGs, peaking at $n=1$.}
\label{fig:SI}
\end{figure}

\begin{figure}
\includegraphics[trim=4mm 5mm 4mm 4mm,clip,  width=0.47\textwidth]{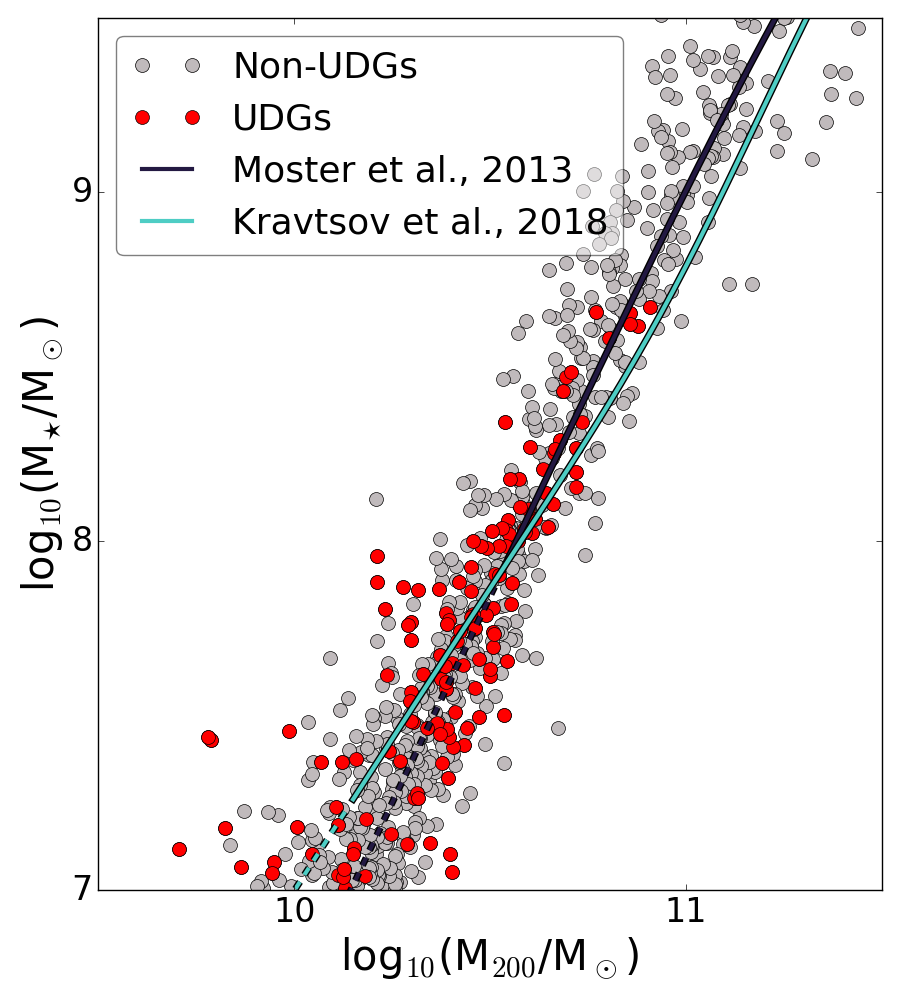}
\caption{Stellar mass - halo mass relation for all isolated galaxies in \textsc{Romulus25} with M$_{\star}$<10$^{9.5}$ M$_\odot$ compared to abundance matching data from \protect\citet{Moster2013} and \protect\citet{kravtsov2018}. Stellar and halo masses are shown following \protect\citet{munshi2013}. UDGs are shown in red, while non-UDGs are shown in gray. All of the UDGs in our sample have M$_{\star}$<10$^{8.7}$ M$_\odot$ and M$_{\mathrm{200}}$<10$^{11}$ M$_\odot$. They are, therefore, true dwarf galaxies, falling on the stellar mass - halo mass relation.}
\label{fig:smhm}
\end{figure}

\begin{figure}
\includegraphics[trim= 5mm 5mm 5mm 4mm, clip, width=0.47\textwidth]{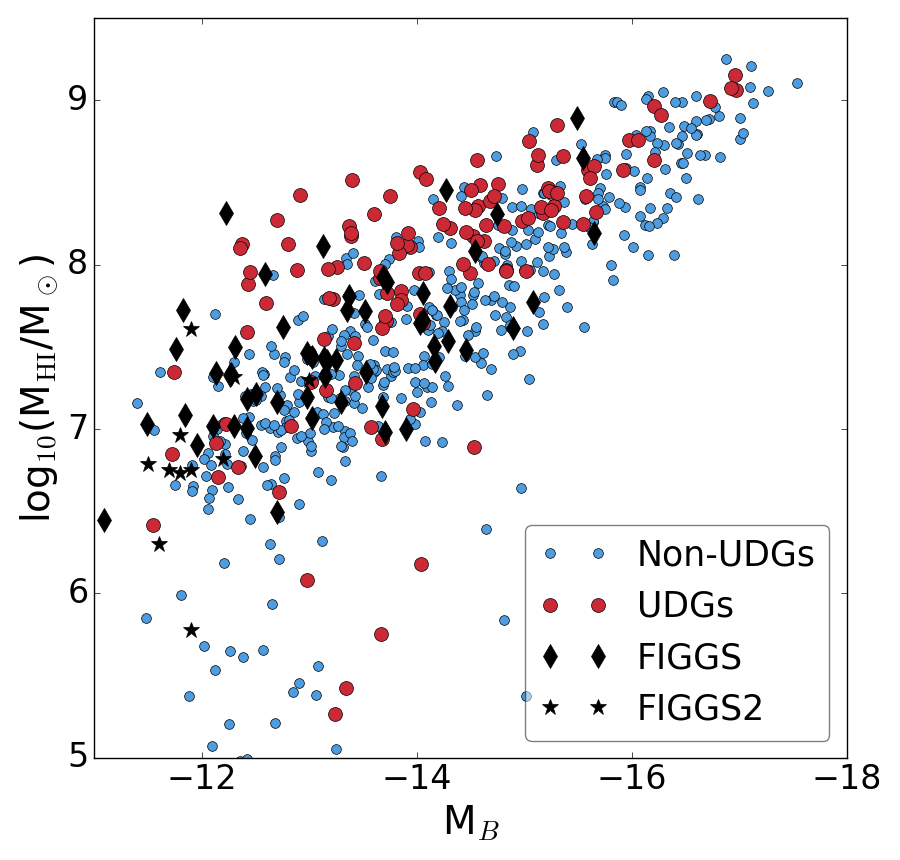}
\caption{B-band magnitudes and HI masses of isolated UDGs (shown in red), our non-UDG comparison sample (shown in blue), and the FIGGS \citep[][black diamonds]{begum2008} and FIGGS2 \citep[][black stars]{patra2016} samples. Although consistent with the FIGGS and FIGGS2 samples, UDGs are modestly HI-rich compared to non-UDGs at the same luminosity, with the disparity increasing towards lower luminosities.}
\label{fig:MHI}
\end{figure}

\begin{figure}
\includegraphics[trim= 6mm 5mm 4mm 5mm, clip, width=0.47\textwidth]{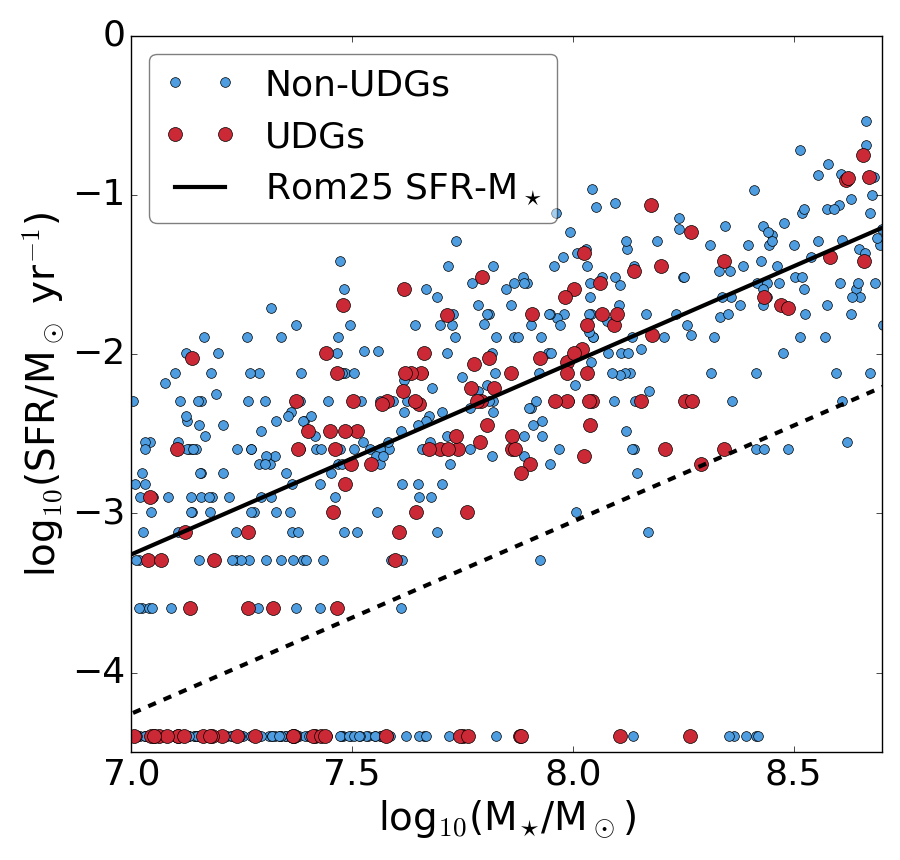}
\caption{Star formation rates for isolated UDGs (shown in red) and our non-UDG comparison sample (shown in blue). Galaxies that have not formed a star particle within the last 250 Myr are arbitrarily placed at log$_{10}$(SFR/M$_\odot$yr$^{-1}$)~$=-4.4$. We also show the SFR-M$_{\star}$ relation for \textsc{Romulus25}, as calculated in \protect\citet[][solid black line]{tremmel2019introducing}. The dashed black line is 1 dex below the relation and indicates the boundary between star-forming and quenched galaxies. The global SFRs of isolated UDGs are very similar to those of more typical isolated dwarfs. Both lie along the \textsc{Romulus25} SFR-M$_{\star}$ relation.}
\label{fig:SFR}
\end{figure}

All of the isolated UDGs identified in \textsc{Romulus25} have M$_{\star}$<10$^{9}$ M$_\odot$. This is consistent with observations, which have shown UDGs to have stellar masses typical of dwarf galaxies, regardless of environment \citep[e.g.,][]{trujillo2017,leisman2017,lee2017,sifon2018}. In Table \ref{tab:massfrac}, we show the fraction of isolated galaxies that are UDGs as a function of stellar mass. Broadly, UDGs make up 20\% of all isolated galaxies with 7<log$_{10}$(M$_{\star}$/M$_\odot$)<9 in \textsc{Romulus25}. However, they are most common at stellar masses between 10$^{7.5}$ and 10$^{8.5}$ M$_\odot$, where they constitute 25-29\% of all isolated galaxies. \\
\begin{table}
\centering
\caption{Number of total isolated dwarf galaxies and UDGs in different mass bins in {\sc Romulus25}. The errors in UDG fraction are Poisson errors.}
\begin{tabular}{cccc}
\hline
log(M$_{\star}$/M$_{\odot}$) & N$_{\mathrm{total}}$ & N$_{\mathrm{UDG}}$ & UDG Fraction\\
\hline
$7-7.5$ & 267 & 46 & $0.17\pm0.03$\\
$7.5-8$ & 178 & 51 & $0.29\pm0.04$\\
$8-8.5$ & 124 & 31 & $0.25\pm0.05$\\
$8.5-9$ & 102 & 6 & $0.06\pm0.02$\\
\hline
\end{tabular}
\label{tab:massfrac}
\end{table}
\indent At all stellar masses where they are present, UDGs have above-average sizes. In the bottom panel of Figure \ref{fig:smreffmu0}, we plot the effective radii of all of the isolated galaxies in \textsc{Romulus25} with M$_{\star}$<10$^{9.5}$ M$_\odot$ as a function of stellar mass alongside an observed relation from \citet{lange2016}. Because this relation is based on r-band data from the GAMA survey, the effective radii that we show here are derived from fits to r-band surface brightness profiles. This is why some UDGs, which are classified using g-band data, appear to have r$_{\mathrm{eff}}$<1.5 kpc. \\
\indent Our simulated galaxies follow the \citet{lange2016} relation reasonably well down to M$_\star\sim$10$^{7.5}$ M$_\odot$. However, our lower mass galaxies are consistently biased high with respect to the (now extrapolated) relation. This is due to the fact that, while we impose no surface brightness limits on our sample, the GAMA survey is insensitive to r-band surface brightnesses fainter than 24.5 mag/arcsec$^2$ \citep{lange2015}. In the bottom panel of Figure \ref{fig:smreffmu0}, we have marked those galaxies that are likely to be observable by GAMA (i.e., those galaxies with $\mu_\mathrm{eff,r}$<24.5 mag/arcsec$^2$ and M$_\star$>10$^8$ M$_\odot$) with a thick black outline. As we might expect, these galaxies adhere more closely to the observed relation than does the \textsc{Romulus25} sample as a whole. However, the majority of our sample - including all of the UDGs - would not be detected by GAMA. For M$_\star$<10$^9$ M$_\odot$, those galaxies that are most likely to be observed tend to be unusually compact, suggesting that the relation may be too steep at lower masses.\\
\indent However, compounding this is the genuine absence in our simulation of the population of very compact low-mass dwarf galaxies that has been observed in the field \cite[e.g.,][]{sung2002,zitrin2009}. While we form galaxies at a variety of sizes for M$_\star$>10$^{7.5}$ M$_\odot$, there is considerably less scatter in r$_\mathrm{eff}$ at lower masses. As discussed in \citet{tremmel2020}, the limited resolution of \textsc{Romulus25} likely plays an important role in determining the sizes of galaxies with M$_{\star}$<10$^{7.5}$ M$_\odot$. The force softening length of the simulation is 350 pc and does not converge to a Newtonian force until twice this. We therefore do not expect to resolve structures smaller than 700 pc, which is marked as our resolution limit in the bottom panel of Figure \ref{fig:smreffmu0}. We begin to see galaxies approaching this limit at M$_\star\approx$10$^{7.5}$ M$_\odot$. Consequently, it is likely that the resolution of our simulation biases these galaxies' sizes high and thereby contributes to the formation of UDGs at M$_\star$<10$^{7.5}$ M$_\odot$. However, it should be noted that this lack of diversity in the sizes of low-mass dwarf galaxies is not unique to the \textsc{Romulus} simulations, but is, rather, a relatively common problem within cosmological simulations, including many that are of considerably higher resolution \citep[e.g.,][]{santossantos2018,garrisonkimmel2019}. \\
\indent In the top panel of Figure \ref{fig:smreffmu0}, we show the g-band central surface brightnesses as a function of stellar mass for all of the isolated galaxies in \textsc{Romulus25} with M$_\star$<10$^{9.5}$ M$_\odot$. Above M$_\star\sim$10$^{7.5}$ M$_\odot$, UDGs, which are shown in red, represent the low surface brightness end of the galaxy distribution. On the other hand, most galaxies with M$_\star$<10$^{7.5}$ M$_\odot$ have $\mu_{0,g}\gtrsim$24 mag/arcsec$^2$. The UDG classification for galaxies in this mass range is thus mainly due to effective radius rather than central surface brightness. There is no distinct separation between UDGs and non-UDGs in effective radius or central surface brightness. In agreement with a number of other authors \citep[e.g.,][]{vandokkum2015forty,vanderburg2016,wittmann2017,conselice2018,mancerapina2019weave}, we find that UDGs are part of a continuous distribution of galaxies. \\
\indent We see further evidence of this in Figures \ref{fig:SI} and \ref{fig:smhm}. In the former, we compare the distribution of S\'ersic indices for UDGs to that of non-UDGs. We restrict our comparison to galaxies with M$_\star$<10$^{8.7}$ M$_\odot$ because this corresponds to the stellar mass of our most massive UDG, allowing for a direct comparison between the two samples. The distributions are very similar, although UDGs do tend to have slightly lower S\'ersic indices than non-UDGs. The distribution peaks at $n=1$, suggesting that field UDGs tend to have exponential profiles. This is consistent with observations of group and cluster UDGs, the S\'ersic indices of which have been found to lie primarily in the range 0.6<$n$<1.2 \citep[e.g.,][]{vandokkum2015forty,koda2015,mihos2015,yagi2016,roman2017,mancerapina2019weave}. The few field UDGs with published S\'ersic indices also lie in this range \citep{greco2018study}. \\
\indent In Figure \ref{fig:smhm}, we plot the stellar mass - halo mass relation for isolated galaxies in \textsc{Romulus25} with M$_\star$<10$^{9.5}$ M$_\odot$. In order to make an apples-to-apples comparison to the abundance matching relations of \citet{Moster2013} and \citet{kravtsov2018}, we correct our halo masses following \citet{munshi2013}, who find that baryon mass loss due to feedback causes halos in baryonic simulations to be up to 30\% less massive than their dark-matter-only counterparts. Our galaxies are consistent with these relations. All of our UDGs are genuine dwarfs galaxies, inhabiting dark matter halos with M$_{\mathrm{200}}$<10$^{11}$ M$_\odot$. They are neither more nor less dark-matter-dominated than non-UDGs of similar stellar mass.\\
\indent One caveat to this assertion is that our halo finder may not identify a completely dark-matter-devoid object and, even if it did, the galaxy would not be considered resolved. The upper end of estimated dark matter masses for the dark-matter-deficient galaxies in the NGC 1052 group (M$_\mathrm{dm}\lesssim$10$^8$ M$_\odot$; \citeauthor{vandokkum2018} \citeyear{vandokkum2018}), would contain too few dark matter particles to meet our resolution criteria. By definition, then, any extremely dark-matter-deficient objects would be excluded from our sample.\\
\indent At the other end of the spectrum, we cannot fully rule out the existence of more massive UDGs. \textsc{Romulus25} is a relatively small volume containing only 39 galaxies with M$_{\mathrm{200}}$>10$^{12}$ M$_\odot$. However, the fact that we do not see any UDGs with M$_{\mathrm{200}}$>10$^{11}$ M$_\odot$ and that this is consistent with the results of a number of other simulations \citep[e.g.,][]{dicintio2017,liao2019} suggests that, if they do exist, massive UDGs must either be rare or require physics not implemented within any of these simulations.\\
\indent Having established that our simulated UDGs are a subset of the dwarf galaxy population, we can select an appropriate sample of galaxies to which to compare them. Because all of our UDGs have stellar masses in the range 10$^7$-10$^{8.7}$ M$_\odot$, we adopt the 484 other isolated galaxies in this mass range as our `non-UDG dwarf' comparison sample. To mitigate the effects of mass trends, we separate our UDG and non-UDG dwarfs into three mass bins: 10$^7$<M$_\star$/M$_\odot$<10$^{7.5}$, 10$^{7.5}$<M$_\star$/M$_\odot$<10$^{8}$, and 10$^8$<M$_\star$/M$_\odot$<10$^{8.7}$. Note that the highest mass bin is slightly broader than the other two. This is because, as may be seen in Table \ref{tab:massfrac}, there are only 6 UDGs with M$_\star$>10$^{8.5}$ M$_\odot$. \\
\indent In Figure \ref{fig:MHI}, we plot the HI masses of the galaxies in our samples against their B-band magnitudes and compare them to galaxies from the FIGGS and FIGGS2 samples \citep{begum2008,patra2016}. Although we find that UDGs are slightly more HI-rich than non-UDGs at a given luminosity, the effect is relatively subtle over most of the luminosity range. At luminosities brighter than M$_\mathrm{B}$=-14, the median UDG has 1.7 times more HI than the median non-UDG. At fainter luminosities, however, this factor jumps to 4. We also have a population of HI-poor UDGs and non-UDGs. This is consistent with both observations \citep[e.g.,][]{papastergis2017} and previous simulations of field UDGs from the NIHAO group, who find that UDGs tend to be modestly HI-rich, but can also be gas-poor and quiescent \citep{dicintio2017,jiang2019}. In our simulations, this gas-poor population is dominated by galaxies that have had a significant encounter with a more massive halo and/or periods of AGN activity \citep[e.g.,][]{dickey2019,sharma2020}. We note that UDGs and non-UDGs are equally likely to have central SMBHs, as well as to have experienced AGN activity at some point over their lifetimes. \\
\indent Our isolated UDGs have average $z=0$ star formation rates (SFRs) for their stellar masses. As shown in Figure \ref{fig:SFR}, they, like our non-UDG isolated dwarf sample, follow the SFR-M$_\star$ relation for \textsc{Romulus25}, as calculated in \citet{tremmel2019introducing}. There is no discernible difference between the SFRs of the UDGs and those of the non-UDGs, except, perhaps, in the low mass bin. For M$_\star$>10$^{7.5}$ M$_\odot$, the fraction of isolated galaxies that are quenched is low ($<$0.1; $<$0.05 for M$_\star$>10$^{8}$ M$_\odot$) and corresponds to the gas-poor population in Figure \ref{fig:MHI}. This is as we might expect, given the extremely low quenched fraction that has been observed among field dwarf galaxies \citep[e.g.,][]{geha2012}. The fraction of galaxies that are quenched also varies very little between UDGs ($\sim$0.09) and non-UDGs ($\sim$0.1). \\
\indent Below M$_\star$=10$^{7.5}$ M$_\odot$, however, the quenched fraction jumps to just under 0.45, with UDGs 10\% more likely than non-UDGs to have ceased forming stars. This is evidence that we are beginning to see the effects of our limited resolution within this mass range. Because stars can form from relatively low density gas in the \textsc{Romulus} simulations, stellar feedback is more effective, leading to more efficient gas removal - particularly within the lower gravitational potentials of these low-mass dwarfs. As quenched galaxies tend to be fainter than star-forming galaxies as a result of their ageing stellar population, it is likely that artificial over-quenching in galaxies with M$_\star$<10$^{7.5}$ M$_\odot$ has led to an inflation of the UDG population in our low mass bin. However, it is also worth noting that, because these quenched low-mass dwarf galaxies are extremely faint (the median $\mu_\mathrm{eff,g}$ for quenched low-mass UDGs is 28.3 mag/arcsec$^2$ vs 27.7 mag/arcsec$^2$ for unquenched low-mass UDGs), it is possible that the observed quenched fraction is underestimated in this mass range. \\
\indent Like most field dwarfs, isolated UDGs are relatively blue. Their median g-r color is 0.22, with the bluest galaxies in the low mass bin (median g-r = 0.19) and the reddest galaxies in the high mass bin (median g-r = 0.25). As we might expect, given their similar HI masses and SFRs, the color distributions of our non-UDG dwarfs are indistinguishable from those of our isolated UDG sample. The colors of the sample as a whole are consistent with those of the galaxies from FIGGS \citep{begum2008}.
\subsection{Evolution of UDGs}
\label{evo}
While the global $z=0$ properties of UDGs and non-UDGs within the same mass range are broadly very similar, we do see significant differences in their evolution.
\subsubsection{Redistribution of Star Formation}
\label{sfmigration}
\begin{figure*}
\includegraphics[trim= 18mm 0mm 18mm 7mm, clip, width=0.97\textwidth]{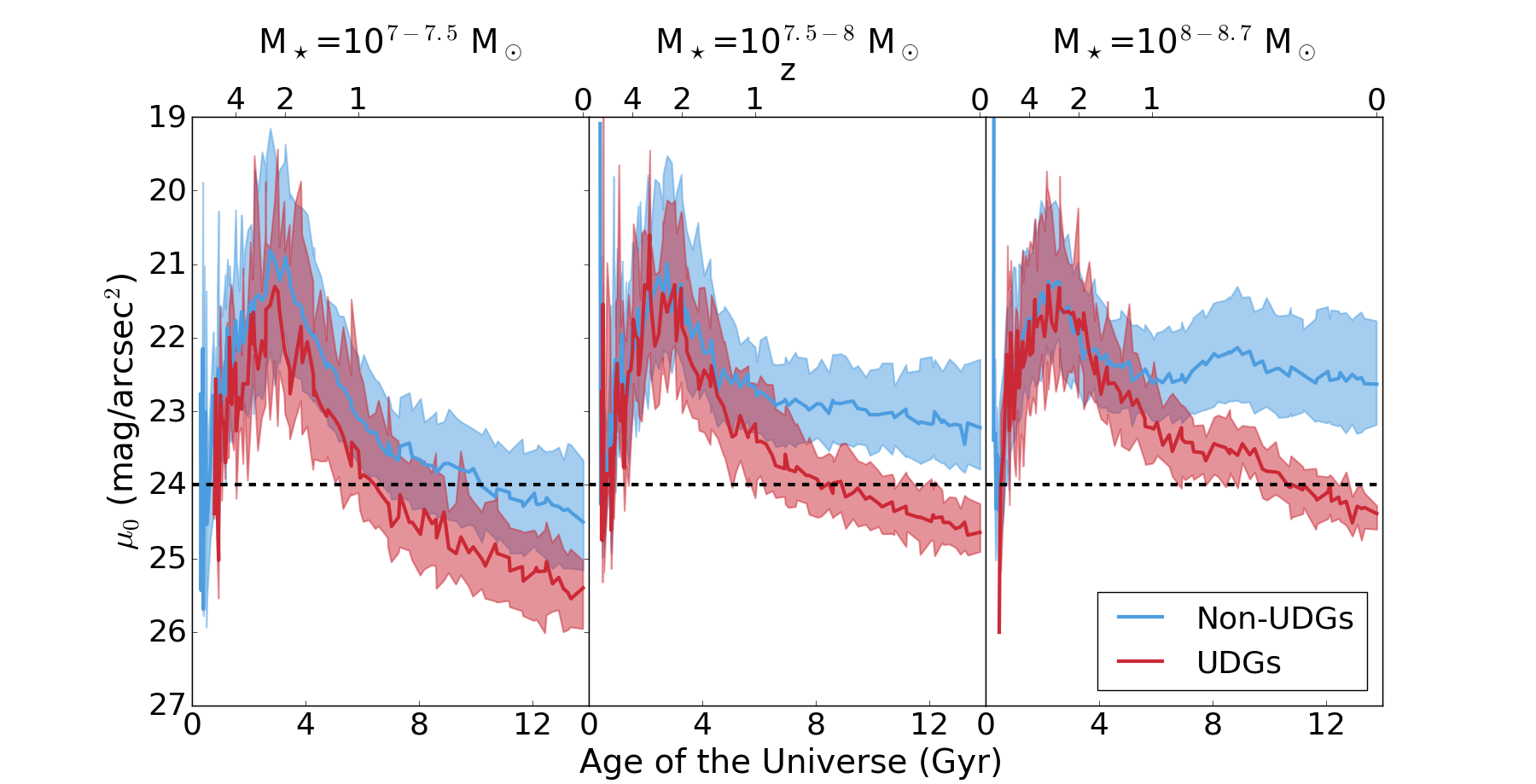}
\caption{The evolution of central g-band surface brightness in our isolated UDGs (red) and our non-UDG comparison sample (blue) for our low mass (left), intermediate mass (center), and high mass (right) bins. The thick solid lines track the evolution of the median while the shading indicates the interquartile range at each step. Anything below the dashed line at $\mu_0$ = 24 mag/arcsec$^2$ is low surface brightness enough to be classified as a UDG. In the high and intermediate mass bins, the evolution of $\mu_0$ is very similar in the UDG and non-UDG progenitors until $\sim$4 Gyr into the simulation. Past this point, the central surface brightnesses of the non-UDG progenitors stay roughly constant, while those of the UDG progenitors continue to dim, leading to considerable differences in final $\mu_0$. In the low mass bin, the evolution of central surface brightness is similar in both samples and nearly all galaxies are low surface brightness enough to be classified as UDGs at $z=0$.}
\label{fig:mu0evo}
\end{figure*}

\begin{figure*}
\includegraphics[trim= 18mm 0mm 18mm 7mm, clip, width=0.97\textwidth]{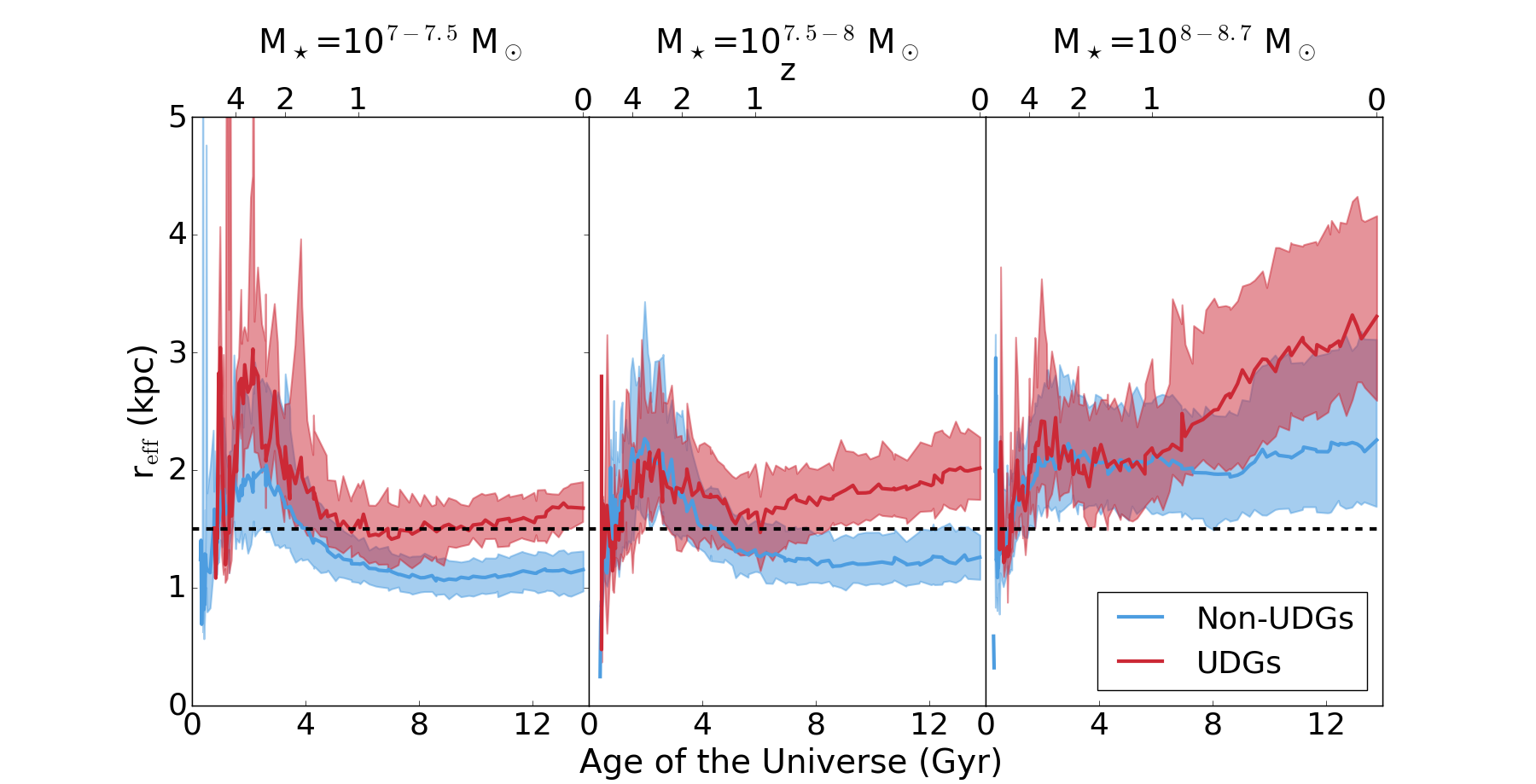}
\caption{The evolution of effective radius in our isolated UDGs (red) and our non-UDG comparison sample (blue) for our low mass (left), intermediate mass (center), and high mass (right) bins. The thick solid lines track the evolution of the median while the shading indicates the interquartile range at each step. Anything above the dashed line at r$_{\mathrm{eff}}$ = 1.5 kpc is large enough to be classified as a UDG. In the high and intermediate mass bins, the initial evolution of effective radius is very similar in both samples. However, in all of the mass groups, the effective radii of the non-UDG progenitors plateau several Gyr before the simulation terminates, while those of the UDG progenitors continue to increase through $z=0$.}
\label{fig:reffevo}
\end{figure*}

\begin{figure*}
\includegraphics[trim= 18mm 0mm 18mm 7mm, clip, width=0.97\textwidth]{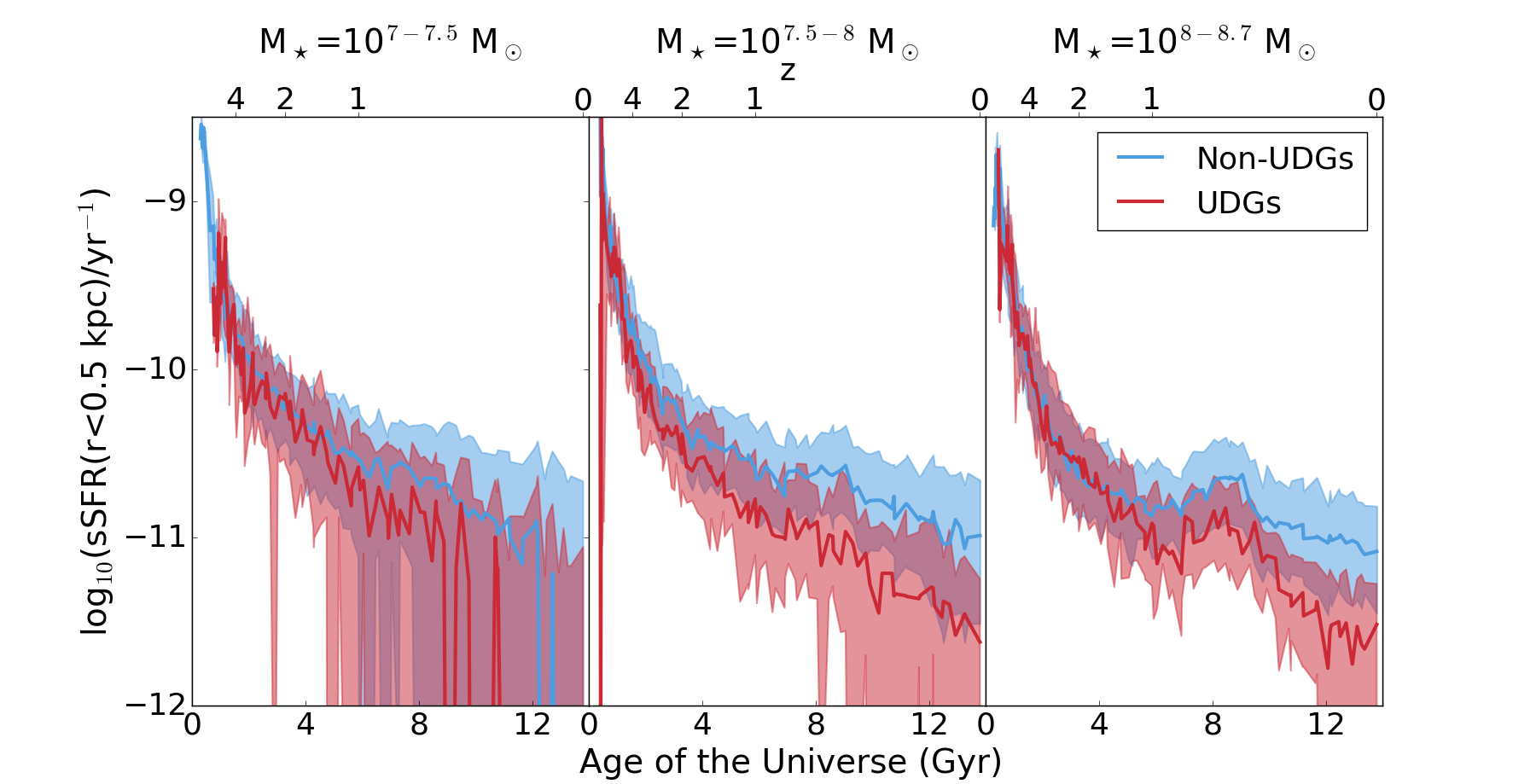}
\caption{The evolution of central (r<0.5 kpc) specific star formation rate in our isolated UDGs (red) and our non-UDG comparison sample (blue) for our low mass (left), intermediate mass (center), and high mass (right) bins. The thick solid lines track the evolution of the median while the shading indicates the interquartile range at each step. Particularly in the high and intermediate mass bins, those galaxies that will be UDGs by $z=0$ tend to evolve to lower central star formation rates than non-UDGs, leading to lower central surface brightnesses due to the passive evolution of an ageing stellar population in their centers. While UDGs have less central star formation compared to non-UDGs, their total star formation rates are similar (Figure~\ref{fig:SFR}), indicating that star formation in UDGs is more spread out compared to non-UDGs.}
\label{fig:cssfrevo}
\end{figure*}

\begin{figure*}
\includegraphics[trim= 26mm 2mm 26mm 5mm, clip, width=0.97\textwidth]{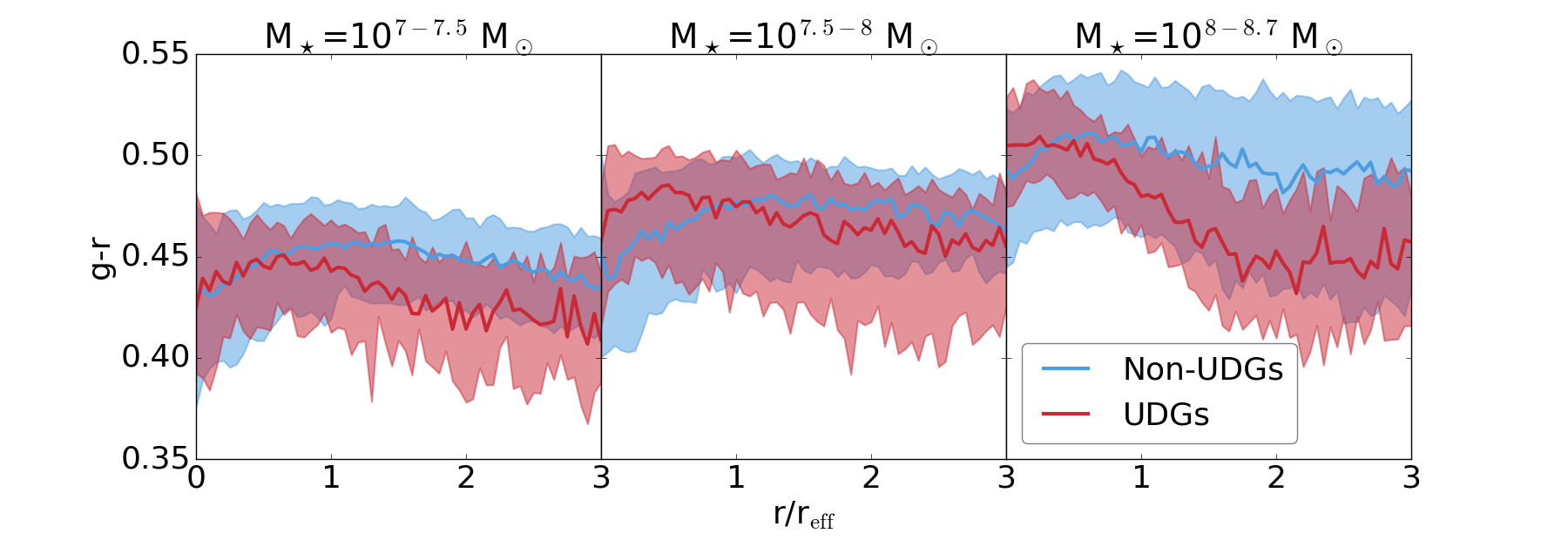}
\caption{g-r color profiles for our isolated UDGs (red) and our non-UDG comparison sample (blue) for our low mass (left), intermediate mass (center), and high mass (right) bins. The thick solid lines track the median profile while the shading indicates the interquartile range in each radial bin. Redistribution of star formation from the centers of isolated UDGs to larger radii leads to redder central regions and bluer outskirts in these galaxies. Particularly in the high and intermediate mass bins, this results in steeper negative color gradients.}
\label{fig:grgrad}
\end{figure*}

In Figure \ref{fig:mu0evo}, we show the evolution of central surface brightness within our UDG and non-UDG comparison samples. At each timestep, we identify the main progenitor of each galaxy in our $z=0$ sample and calculate its central surface brightness using the procedure described in Section \ref{class}. We then calculate the median values (solid lines) and interquartile ranges (shading) of $\mu_0$ at each timestep and for each sample using the central surface brightnesses of the main progenitors of the galaxies in that sample. The evolution of the median values and interquartile ranges over time is hereafter referred to as the `evolutionary track'.\\
\indent In the high mass and intermediate mass bins, the evolution of central surface brightness is initially very similar for progenitors of both UDGs and non-UDGs. However, for non-UDG progenitors, very little evolution in $\mu_0$ occurs after $\sim$4 Gyr into the simulation. By contrast, the centers of UDG progenitors continue to fade over the course of the simulation, resulting in significantly fainter central surface brightnesses at $z=0$. In the low mass bin, the shapes of the UDG and non-UDG progenitor evolutionary tracks are very similar: both groups experience an initial brightening period, but fade starting $\sim$3 Gyr into the simulation. Although the UDG progenitors are, on average, fainter than the non-UDG progenitors, the vast majority of the galaxies in this mass group are low surface brightness enough to be classified as UDGs at $z=0$. \\
\indent Very similar patterns appear in the evolutionary tracks of the effective radii of our two samples, which are shown in Figure \ref{fig:reffevo}. The evolution of UDG and non-UDG progenitors is nearly indistinguishable until $\sim$4-5 Gyr into the simulation. Past this point, non-UDG progenitors experience very little evolution in effective radius, while those galaxies that will be UDGs by $z=0$ continue to grow in size. Although this difference is most dramatic in the high mass bin, this is also the mass range in which it is least relevant to UDG classification: nearly every galaxy with M$_\star$>10$^8$ M$_\odot$ is physically large enough to be a UDG and has been over most of its evolution. In the low and intermediate mass bins, however, a large effective radius is truly a defining characteristic of the UDG sample. \\
\indent Although UDGs at all masses are, on average, larger and lower surface brightness than typical isolated dwarfs, the reasons why a given galaxy might be classified as a UDG are mass-dependent. High mass dwarfs almost always have r$_\mathrm{eff}\geq$1.5 kpc; those that are UDGs are classified as such entirely as a result of their low central surface brightnesses. The opposite is true of low mass dwarfs, which are predominantly faint enough to be UDGs, but require an effective radius $\sim$0.5 kpc larger than the median value to be part of the UDG sample. It is only in the intermediate mass group that both central surface brightness and effective radius are relevant to UDG classification. We see this same trend in the UDGs identified in our cluster simulation, \textsc{RomulusC} \citep{tremmel2020}.
\\
\indent We find that the reason isolated UDGs are fainter than typical galaxies of the same stellar masses is that they have low central SFRs. As may be seen in Figure \ref{fig:cssfrevo}, those galaxies that will be UDGs at $z=0$ evolve to lower central specific SFRs (that is, SFR within the inner 0.5 kpc of each galaxy divided by the stellar mass of the galaxy) than do the galaxies in our non-UDG comparison sample. This is particularly noticeable in the high and intermediate mass bins, where the difference in final central surface brightness is also most significant. As with the effective radius and central surface brightness evolutionary tracks, we begin to see a difference between UDG and non-UDG progenitors in these mass groups at $\sim$4 Gyr into the simulation, which suggests a common root cause.\\
\indent This decrease in central SFR leads to comparatively lower central stellar densities, as well as older - and therefore fainter - central stellar populations. While stellar mass continues to build up in the centers of non-UDGs, it quickly levels off in the centers of UDGs. However, it does not decrease, which suggests that the central surface brightness evolution that we see is primarily driven by the ageing stellar populations in the centers of UDGs. UDGs within the high and intermediate mass bins have central stellar populations that are, on average, $\sim$2 Gyr older than those of non-UDGs of similar mass. This is fundamentally the same process that we see in the cluster UDGs identified in \textsc{RomulusC}. In both simulations, we observe a dimming of surface brightness as stellar populations passively evolve, leading to a clear correlation between the age of a galaxy's central stellar population and its central surface brightness \citep[see Figure 17 in][]{tremmel2020}. However, while this passive evolution is a consequence of quenching via ram pressure stripping in the cluster environment, it must have a different origin in the field.  \\
\indent We have already established that isolated UDGs and non-UDGs have similar \textit{global} HI masses and SFRs. The fact that UDGs have lower \textit{central} SFRs therefore indicates that star-forming gas and star formation have moved outward. We would expect this to be most noticeable in the high and intermediate mass groups, where the differences in central SFR are most significant.\\
\indent We see evidence of this in Figure \ref{fig:grgrad}, where we show the g-r color profiles for our UDG and non-UDG comparison samples at $z=0$. Broadly speaking, our simulated dwarf galaxies tend to have relatively shallow color gradients, consistent with those of observed star-forming dwarfs \cite[e.g.,][]{hunter2006,tortora2010}. However, while non-UDGs tend to have flat or slightly positive radial gradients, UDGs typically have steeper negative gradients due to their redder centers and bluer outskirts. This difference is most apparent when color gradients are measured out to $\sim$1.5 r$_\mathrm{eff}$, where the difference in the median color gradient between UDGs and non-UDGs is 0.03 mag/r$_\mathrm{eff}$ within the high and intermediate mass bins and 0.02 mag/r$_\mathrm{eff}$ in the low mass bin. These steeper gradients persist, albeit to a slightly lesser degree, in the high and intermediate mass bins when color profiles are scaled by R$_\mathrm{vir}$, rather than r$_\mathrm{eff}$. Determining whether or not this distinction exists among real dwarf galaxies may, however, prove difficult. Tracing already low surface brightness galaxies out to $\geq$1.5 r$_\mathrm{eff}$ is a non-trivial task. Additionally, we have not accounted for internal reddening due to dust, which may lessen the already small differences between the color gradients of UDGs and non-UDGs.\\
\indent While we might also expect to find that the metallicity gradients of isolated UDGs differ from those of non-UDGs, metallicity gradients in \textsc{Romulus25} are typically close to zero. Although we examined the cold gas metallicity and stellar metallicity profiles of our UDG and non-UDG comparison samples, we find that there is little difference between the two populations. UDGs do tend to have slightly offset metallicity profiles (lower by $\sim$0.1 dex), which is consistent with observational findings that more extended galaxies are typically less metal-rich than more compact ones \citep[e.g.,][]{ellison2008} and that low surface brightness galaxies are often metal-poor \citep[e.g.,][]{mcgaugh1994}. However, the overall metallicity gradients of both UDGs and non-UDGs are on the order of -0.1 dex/r$_\mathrm{eff}$ - consistent with the largely flat profiles found in observed dwarf irregular galaxies \citep[e.g.,][]{hunter1999}. Although it is possible that we will find steeper metallicity gradients among UDGs in higher resolution zoom-in simulations, their absence in \textsc{Romulus25} may be an indication that any differences between UDGs and non-UDGs will be subtle. \\
\indent More qualitative hallmarks of redistribution of star formation to larger radii may, however, be more easily observable. Mock UVI images of many of our UDGs, such as those shown in the top right and middle panels of Figure \ref{fig:rep}, reveal that, while the centers of the UDGs are nearly quenched, asymmetrical bursts of star formation appear along the edges of the galaxies. We would therefore expect star-forming field UDGs to have relatively irregular appearances, consistent with the images shown in \citet{leisman2017}. Similar off-center bursts of star formation have also been observed in UDGs on the edges of groups and clusters \citep[e.g.,][]{martinezdelgado2016,roman2017}. We investigate possible explanations for this redistribution of star formation in isolated UDGs in the next two sections.
\subsubsection{Spin-up}
\label{sec:spin}
\begin{figure}
\includegraphics[trim = 5mm 4mm 5mm 1mm, clip, width=0.47\textwidth]{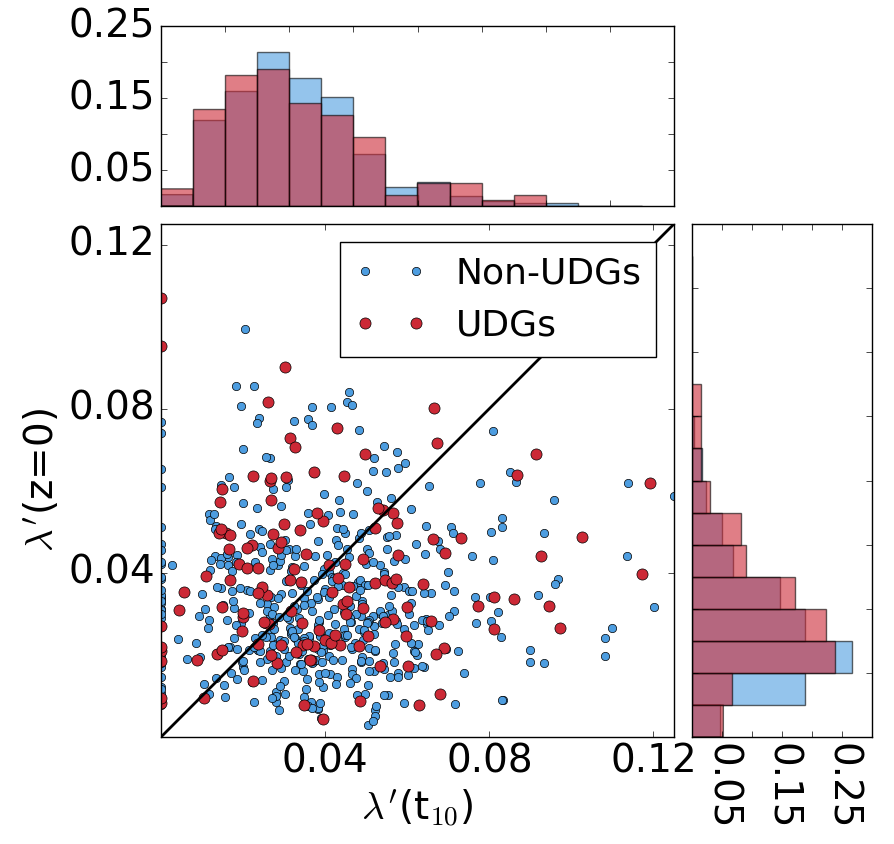}
\caption{The Bullock spin parameters of our isolated UDGs (red) and our non-UDG comparison sample (blue) measured at the time at which each galaxy had formed 10\% of its stars (t$_{10}$) and at $z=0$. The solid black line denotes where both spins are equal. The scatter around this line indicates that the final spin of a given galaxy has little to no dependence on the spin with which the galaxy formed. The projected histograms of the spin distributions at t$_{10}$ (top) and $z=0$ (right) show the fraction of galaxies in each sample within a given range of spins. Galaxies that are not traced back to t$_{10}$ are arbitrarily placed at $\lambda'$(t$_{10}$) = 0, but are not included in the histogram of $\lambda'$(t$_{10}$) values. Although UDGs are more likely to have slightly higher than average angular momentum at $z=0$, they do not form exclusively in high spin halos. Even at $z=0$, the difference between the spins of UDGs and non-UDGs is small.}
\label{fig:spinevo}
\end{figure}

\begin{figure*}
\includegraphics[trim=17mm 6mm 17mm 15mm, clip, width=0.75\textwidth]{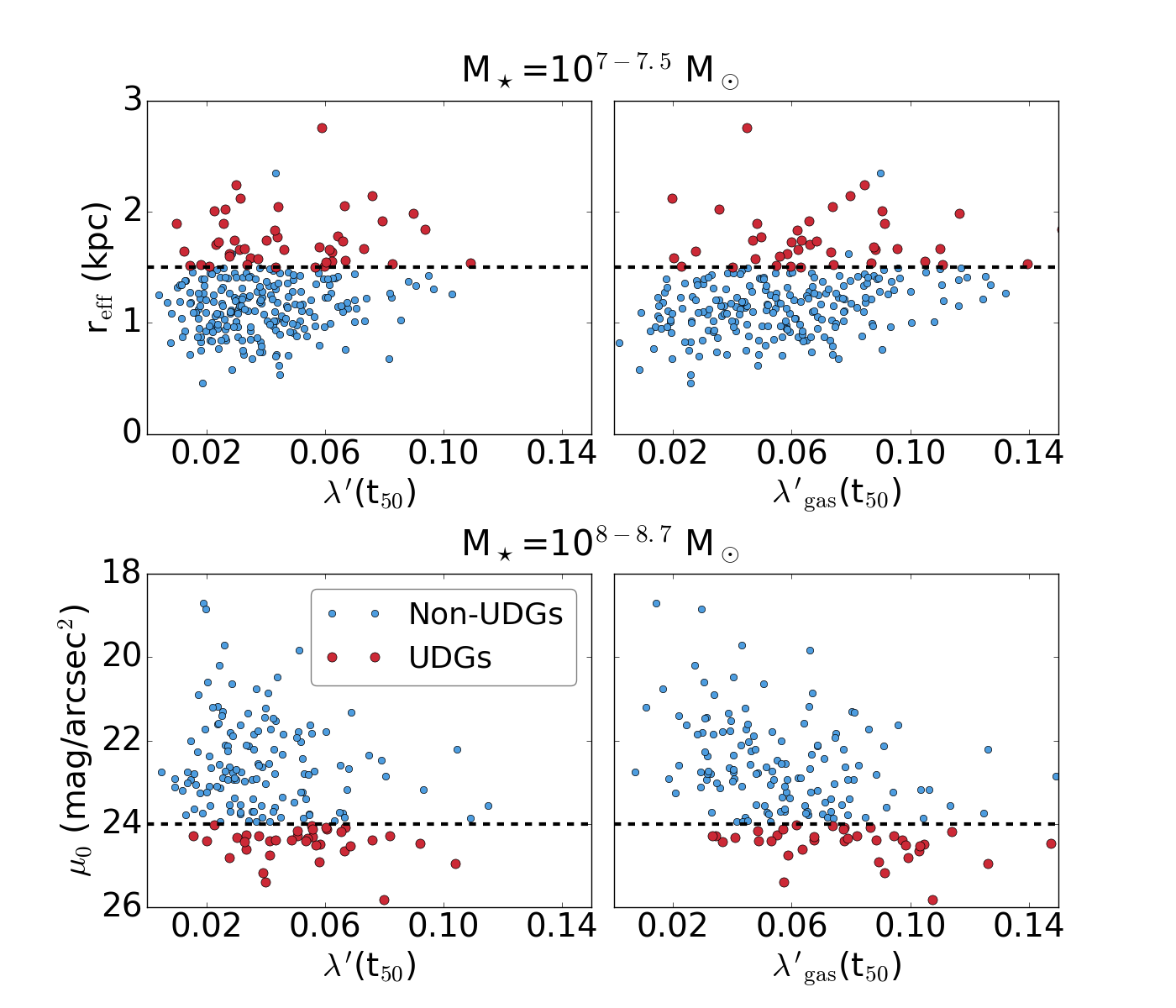}
\caption{\textit{Top:} Effective radius at $z=0$ vs the total spin of the halo (left) and the spin of the gas (right) at the time at which each galaxy had formed 50\% of its stars (t$_{50}$) for galaxies in our low mass bin. \textit{Bottom:} Central surface brightness at $z=0$ vs the total spin of the halo (left) and the spin of the gas (right) at t$_{50}$ for galaxies in our high mass bin. Dashed horizontal lines indicate the UDG classification boundaries: r$_\mathrm{eff}$=1.5 kpc and $\mu_\mathrm{0,g}$=24 mag/arcsec$^2$. Our low mass UDGs are classified as UDGs almost exclusively because of their large effective radii, while our high mass UDGs are defined by their low central surface brightnesses. However, high spin is not the primary cause of either defining characteristic. UDGs exist at all but the lowest spins and there are no strong correlations between spin and either effective radius or central surface brightness in these mass bins.}
\label{fig:spincor}
\end{figure*}

It is thought that the rotation of modern galaxies has its origin in tidal torquing of clumps of dark matter and baryonic material in the early universe \citep[e.g.,][]{hoyle1951,peebles1969,doroshkevich1970,white1984,barnes1987}. The specific angular momentum of galaxies is typically characterized by the dimensionless spin parameter: the original ($\lambda$) defined in \citet{peebles1969} and a revised version \citep[$\lambda'$;][]{bullock2001} lacking the explicit energy - and implicit redshift - dependence \cite[e.g.,][]{hetznecker2006} of the original, which is used throughout this analysis. Both versions of the spin parameter are found to follow a log-normal distribution \cite[e.g,][]{barnes1987}, with the distribution of $\lambda'$ peaking at $\sim$0.035. \\
\indent Numerical models by \citet{dalcanton1997} suggest that low surface brightness galaxies are the natural inhabitants of high spin dark matter halos. Conservation of angular momentum dictates that a more quickly rotating disk will extend to larger radii than a more slowly rotating one, resulting in the same amount of baryonic material being spread over a larger area. A galaxy that formed in a high spin halo would therefore be expected to have lower gas surface densities and accordingly lower star formation rates and stellar densities. \citet{amorisco2016} extended this work to UDGs and showed that their abundance and size distribution could be recovered by assuming that cluster UDGs formed in dwarf-mass halos with higher-than-average angular momentum.\\ 
\indent In Figure \ref{fig:spinevo}, we show the distribution of $\lambda'$ values among our UDG and non-UDG comparison samples at the time at which each galaxy had formed 10\% of its stars (t$_{10}$) and at $z=0$. At both times, UDGs exist at all spins. Although they are marginally more likely to occupy halos with slightly higher-than-average angular momentum at $z=0$, UDGs do not form in exclusively high spin halos. At t$_{10}$, the spin distributions for our two samples are indistinguishable. In fact, we find very little correlation between the spin each galaxy forms with and the spin it has at $z=0$, despite the fact that these galaxies are all relatively isolated. That UDGs do not form in halos with abnormal properties is consistent with our earlier findings about their evolution. As shown in Figures \ref{fig:mu0evo}-\ref{fig:cssfrevo}, the effective radii, central surface brightnesses, and central SFRs of UDGs and non-UDGs are extremely similar for at least the first few billion years of the simulation. If UDGs had formed in high spin halos, we would expect to see an immediate difference, particularly in effective radius. \\
\indent The conclusion that UDGs did not form exclusively in high spin dark matter halos does not necessarily mean that spin is irrelevant, particularly as we see evidence of slightly above-average angular momentum at $z=0$. If the baryonic components of UDGs were significantly spun up at some point during their histories, the resulting galaxies likely wouldn't be terribly different from those that had formed with high spin. We might, therefore, expect present-day UDGs to bear some imprint of the era during which they are maximally spun-up, which we find to be $\sim$t$_{50}$, the time at which each galaxy has formed 50\% of its stars. 
\\
\indent In Figure \ref{fig:spincor}, we show the total spin of the halo and the spin of the gas at t$_{50}$ for the low and high mass galaxy bins. For both mass bins, we plot spin against the characteristic that determines whether or not the galaxies in the group are UDGs: r$_\mathrm{eff}$ for the low mass bin and $\mu_0$ for the high mass bin. If spin were the primary factor determining whether or not these galaxies became UDGs, we would expect to see a strong correlation between it and these characteristics. Looking at the top two panels, we can see that there is very little correlation between spin and effective radius for the galaxies in the low mass bin. This is not to say that size is completely independent of spin for these galaxies: relatively compact galaxies (r$_\mathrm{eff}$<1 kpc) do not exist for $\lambda'_{\mathrm{(gas)}}$>0.1 and extremely diffuse galaxies (r$_\mathrm{eff}$>2 kpc) do not appear at $\lambda'_{\mathrm{(gas)}}$<0.02. However, beyond these extremes, spin seems to have little effect on the sizes of galaxies in the low mass bin. \\
\indent Moving to the lower two panels, we do see a weak correlation between spin and central surface brightness for the galaxies in the high mass bin. As before, it is primarily driven by the extremes: bright galaxies ($\mu_0$<21.5 mag/arcsec$^2$) do not exist at $\lambda'$>0.07 ($\lambda'_{\mathrm{gas}}$>0.085) and faint galaxies ($\mu_0$>25 mag/arcsec$^2$) are not found at $\lambda'$<0.04 ($\lambda'_{\mathrm{gas}}$<0.055). The correlation is slightly stronger for the spin of the gas than for the spin of the total halo. This is likely because surface brightness inherently measures the distribution of the baryonic component of the galaxy; if the galaxy's gas is spun up (and therefore more widely dispersed) we would expect the stars that form from it to be correspondingly diffuse. Additionally, while the total halo spin may be influenced by substructure as far out as the virial radius, the gas component of the spin is more likely to reflect the spin of the inner ``disk'' of the galaxy where star formation is actively occurring (e.g., \citeauthor{zjupa2017} \citeyear{zjupa2017}; \citeauthor{jiang2019size} \citeyear{jiang2019size}, although see \citeauthor{roskar2010} \citeyear{roskar2010}). The weak correlation that we observe between the spin of the galaxy's gas and its central surface brightness may therefore indicate that increased spin is contributing to the decrease in gas density in the centers of UDG progenitors and the subsequent redistribution of star formation to larger radii. However, it is unlikely to be the only factor at play.
\subsubsection{Early Mergers}
\begin{figure*}
\includegraphics[trim=12mm 5mm 12mm 0mm, clip,width=0.86\textwidth]{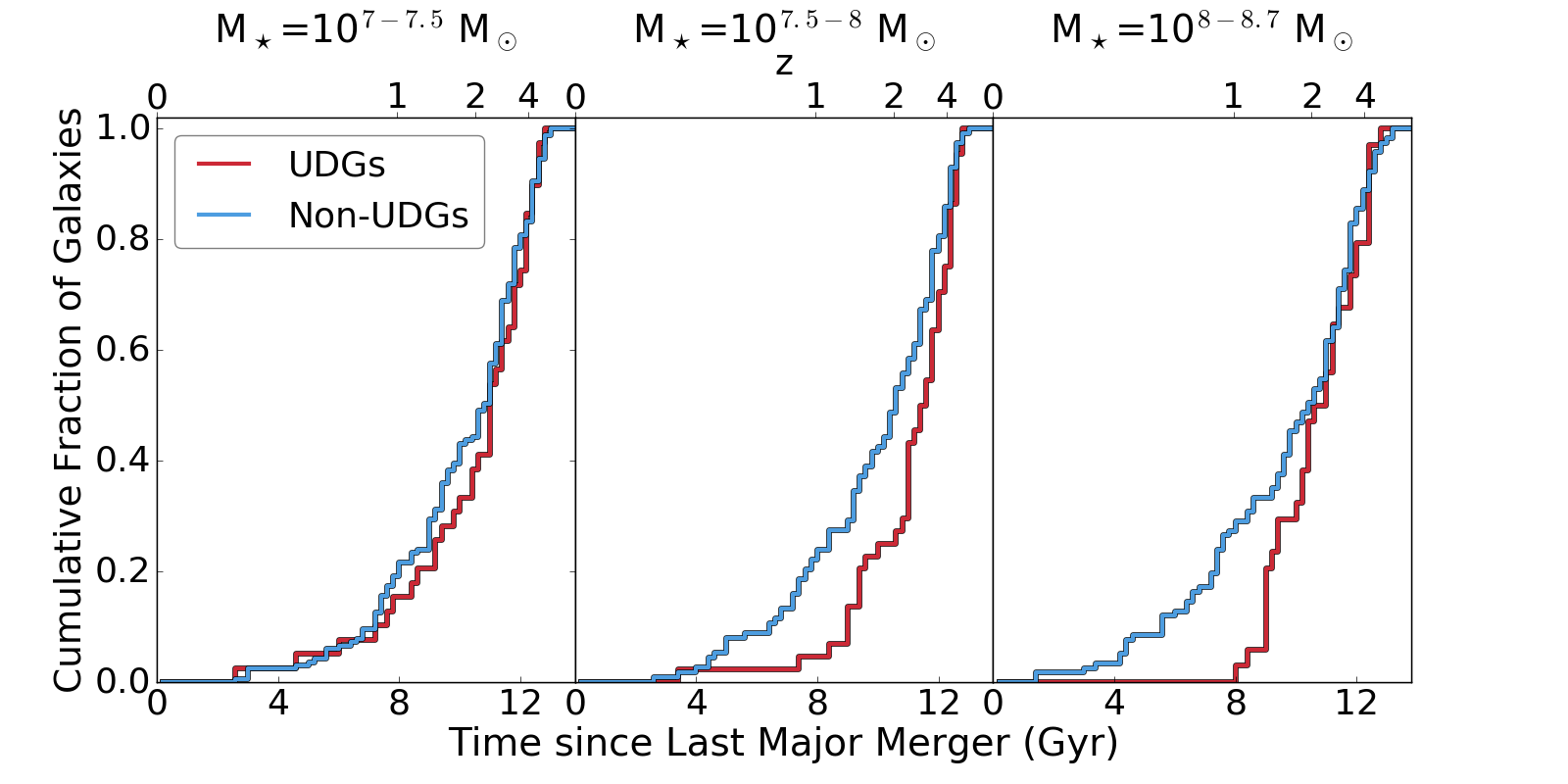}
\caption{Cumulative distribution of the time that has elapsed since each galaxy in our UDG and non-UDG comparison samples experienced its most recent major merger. Here, a major merger is defined as any merger in which the total mass of the secondary galaxy is at least 20\% of the mass of the primary. Although UDGs and non-UDGs have very similar global merger histories (i.e., they have similar numbers of minor and major mergers over their lifetimes and have primarily interacted with gas-rich companions), UDGs in the high and intermediate mass bins do not tend to have experienced major mergers within the last 8 Gyr. Rather, the majority of UDGs last underwent a major merger 8-11 Gyr ago. This is also when we see a departure between the evolutionary tracks of UDGs and non-UDGs in Figures \ref{fig:mu0evo}-\ref{fig:cssfrevo}.}
\label{fig:tslmm}
\end{figure*}

\begin{figure*}
\includegraphics[width=0.91\textwidth]{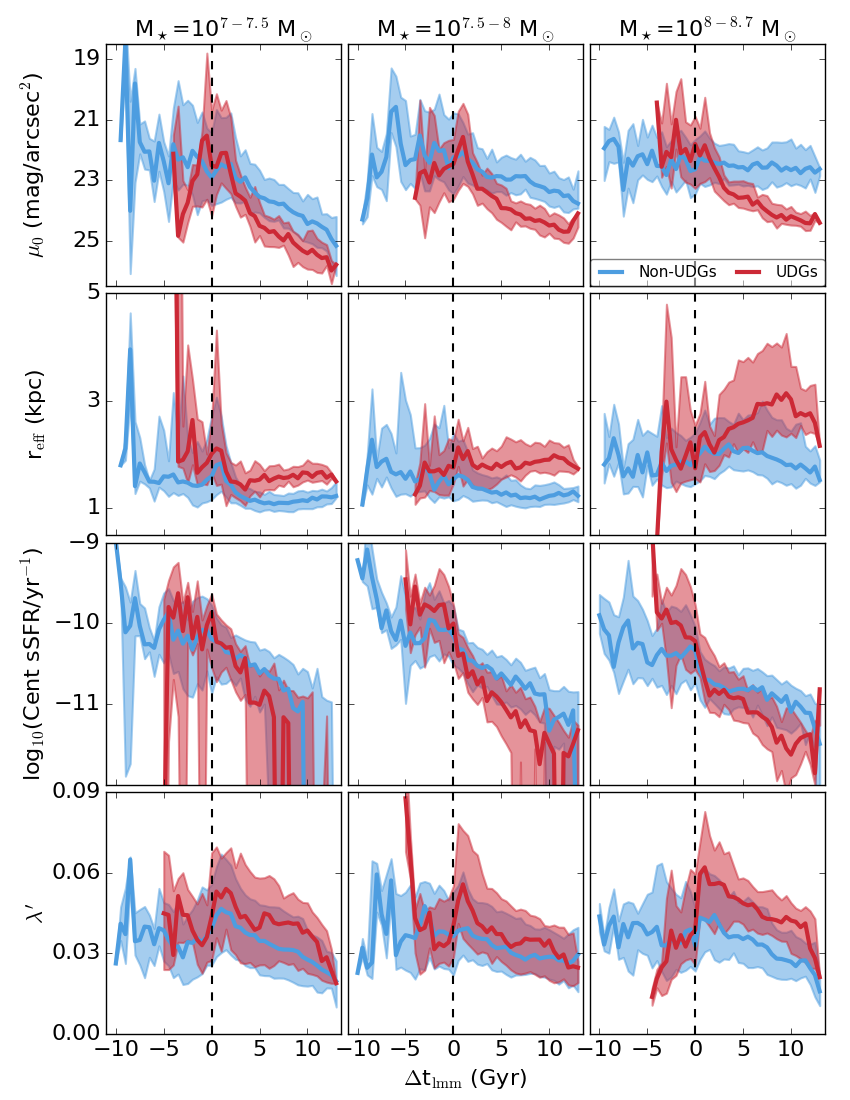}
\caption{Central g-band surface brightness, effective radius, central specific SFR (cf. Figures \ref{fig:mu0evo}-\ref{fig:cssfrevo}), and total spin as a function of time relative to that of the last major merger for dwarfs in our three mass bins. Negative time values indicate the time until the merger takes place, while positive values indicate that time that has passed since it occurred. The dashed black vertical lines at $\Delta$t$_{\mathrm{lmm}}=0$ Gyr show the time at which the merger begins, i.e., the time at which the virial radii of the primary and secondary galaxies first overlap. The thick colored lines denote the median evolutionary track for each mass bin, while the shading indicates the interquartile range. Only time bins containing at least 5 galaxies in each sample and mass bin are plotted. Major mergers strongly impact the evolution of UDG progenitors - particularly those within the high and intermediate mass bins, spinning them up, increasing their effective radii, and reducing their central specific SFRs and surface brightnesses. Similar effects appear in the non-UDG progenitors, but to a much lesser extent.}
\label{fig:evorelmer}
\end{figure*}

UDGs and non-UDGs experience similar numbers of both minor and major mergers over their lifetimes and predominantly interact with gas-rich companions. However, the UDGs in our high and intermediate mass bins tend to have had considerably quieter merger histories since z$\sim$1 compared to non-UDGs. In Figure \ref{fig:tslmm}, we show the cumulative distribution of the time that has elapsed since each galaxy in our sample last experienced a major merger. Here, a major merger is defined as any merger in which the total mass of the secondary galaxy is at least 20\% the mass of the primary galaxy. While roughly one-fourth of the non-UDGs in the high and intermediate mass bins has experienced a major merger within the last 8 Gyr, only 2\% of UDGs of similar mass have a major merger this recent. Instead, the majority of UDGs within the high and intermediate mass bins last underwent a major merger between 8 and 11 Gyr ago. Notably, this is also around the time that we start to see a departure between the UDG and non-UDG evolutionary tracks in these groups in Figs \ref{fig:mu0evo}-\ref{fig:cssfrevo}. \\
\indent We can more directly assess the impact of major mergers on the evolution of our galaxies by examining their effects on key quantities. In the top 3 rows of Figure \ref{fig:evorelmer}, we reproduce Figures \ref{fig:mu0evo}-\ref{fig:cssfrevo}, re-scaling the evolutionary tracks for central surface brightness, effective radius, and central specific SFR such that they are relative to the time at which each galaxy last experienced a major merger ($\Delta$t$_{\mathrm{lmm}}$). Negative time values therefore indicate the time until the merger takes place, while positive time values indicate the time that has taken place since the merger occurred. We take $\Delta$t$_{\mathrm{lmm}}=0$ to be the time at which the virial radii of the primary and secondary galaxies first overlap, following \citet[][]{hetznecker2006}. We also include evolutionary tracks for total spin ($\lambda'$) in the bottom row of Fig \ref{fig:evorelmer}. \\
\indent For each panel of Figure \ref{fig:evorelmer}, we only plot time bins that include at least 5 galaxies. Because so few UDGs have late mergers, we are only able to trace the median evolutionary tracks back to $\Delta$t$_{\mathrm{lmm}}=-5$ Gyr.  By contrast, non-UDGs have a much wider distribution of merger times and their evolutionary tracks therefore cover a considerably broader range of times. Those non-UDGs that contribute to the earliest time bin ($\sim$-10.5 Gyr) last underwent a major merger $\sim$3.3 Gyr ago, so their individual tracks run from $\Delta$t$_{\mathrm{lmm}}=-10.5-3.3$ Gyr. Accordingly, while all UDGs that contribute to the $\Delta$t$_{\mathrm{lmm}}=-5$ Gyr time bin are at high redshift, the same time bin contains a mixture of redshifts for the non-UDGs: high redshifts for those galaxies with early mergers and lower ones for those with later mergers. The behavior of the UDG tracks in these earliest time bins is therefore most appropriately compared to the behavior of the non-UDG tracks at $\Delta$t$_{\mathrm{lmm}}\sim-10.5$ Gyr, as these exclusively contain high-redshift galaxies. This is particularly evident for quantities that are inherently highly redshift-dependent, like specific SFR. The later time bins are simpler to interpret, as both samples include galaxies that have early mergers. The evolutionary tracks therefore have a more similar mix of redshifts at $\Delta$t$_{\mathrm{lmm}}>-2$~Gyr.\\
\indent In our discussion of Figures \ref{fig:mu0evo}-\ref{fig:cssfrevo}, we noted that the evolutionary tracks of UDG and non-UDG progenitors were indistinguishable for the first several billion years of the simulation. We see very similar trends in Figure \ref{fig:evorelmer}. Prior to the mergers, the behaviors of the UDG and non-UDG evolutionary tracks are extremely similar. Following the mergers, however, there is an abrupt alteration in the evolution of the UDG populations. \\
\indent Looking at the top row, we can see that the global evolution of central surface brightness in the non-UDG population is largely undisturbed by the mergers. Those galaxies in the high mass bin maintain a steady central surface brightness throughout the entire simulation, while those in the lower mass bins gradually dim over time. By contrast, the UDGs experience a significant drop in central surface brightness following their mergers. The contrast between the responses of the two populations is particularly stark in the high mass bin, where we know that central surface brightness is the determining factor in UDG classification. This difference in behavior also explains why we don't see UDGs that have experienced recent mergers in the high and intermediate mass bins. For these UDGs, there is a stronger correlation between the time at which a galaxy last underwent a major merger and its central surface brightness. Those galaxies that have recently experienced a merger are likely still too bright to be classified as UDGs, but might become UDGs in the future if they dim following the merger. \\
\indent In the second row, we plot the evolutionary tracks for the effective radii of our UDG and non-UDG samples. As with central surface brightness, we see that the effective radii of UDGs are much more strongly affected by mergers than those of non-UDGs. As we noted in our earlier discussion, the effective radii of non-UDGs tend to remain relatively steady over the course of the simulation. We do see a slight temporary increase in their effective radii immediately following the mergers, but the jump is considerably more pronounced - and far more permanent - in the UDG population.\\
\indent As we might expect, given our finding that the faintness of UDGs is connected to their low central SFRs, the patterns that we see in the third row, where we plot evolutionary tracks for central specific SFR, are nearly identical to those that we see in central surface brightness. The central specific SFRs of UDGs are similar to those of non-UDGs until the mergers, when they experience a significant drop. This tells us that mergers are responsible for moving star formation out of the centers of UDGs. Movies\footnote{https://www.youtube.com/watch?v=lp8VG3OYgL0} of these interactions reveal that the mergers lower the density of gas at the centers of UDGs and compress gas along their outer regions, resulting in asymmetric bursts of star formation along the edges of the galaxies that can persist for billions of years. This behavior is most apparent in the highest mass UDGs, where we can see that the drop in central specific SFR in response to the mergers is most extreme. \\
\indent This is not what we would initially expect. Most observations and simulations - particularly of higher mass galaxies - show that mergers tend to funnel gas to the centers of galaxies, causing them to become more centrally concentrated and even igniting AGN activity \citep[e.g.,][]{toomre1972,barnes1988,noguchi1988,hernquist1989,mihos1996,springel2005,stierwalt2013}. Our findings may, however, be consistent with observations by \citet{privon2017}, which suggest that interactions between field dwarf galaxies can result in more widely distributed bursts of star formation than interactions involving more massive galaxies.\\
\indent In the bottom row of Figure \ref{fig:evorelmer}, we plot the evolutionary tracks for the spin parameters of our galaxies. We find that both UDGs and non-UDGs are spun-up by major mergers. This behavior is consistent with previous findings regarding more massive galaxies. Large-scale N-body simulations have been used to show that the conversion from the orbital angular momentum of the secondary galaxy to the internal spin of the remnant galaxy that takes place during major mergers typically increases the spin of the primary galaxy by 25-30\% \citep[e.g.,][]{vitvitska2002,hetznecker2006}. We find that the mergers that create UDGs spin the primary galaxy up more than those that result in non-UDGs. In the high mass bin, for instance, the median post-merger spin-up experienced by non-UDG progenitors is 25\%, while the median spin-up experienced by UDG progenitors is 75\%. However, as discussed in Section \ref{sec:spin}, the spin-up of UDGs is largely temporary: by $z=0$, UDGs and non-UDGs have similar spin distributions.\\
\indent Figure \ref{fig:evorelmer} shows that mergers induce the changes that ultimately lead to the distinctions between our UDG and non-UDG samples. However, these alterations are most pronounced in the high and intermediate mass bins. Although those galaxies that ultimately end up in our low mass UDG bin are affected by mergers in the same way as those in the higher mass bins, the magnitude of the change is lower, leading to less disparate evolutionary paths for low mass UDGs and non-UDGs. To some extent, this may be physical: there is considerably less spread in effective radius (the defining feature of these galaxies) at low mass than at high mass. \\
\indent However, this may also be influenced by our resolution. In order to ensure that we are considering only well-resolved mergers, our merger trees only trace objects down to 1000 particles ($\sim$10$^8$ M$_\odot$). Although this is an order of magnitude below our fiducial resolution limit, it is conservative compared to similar simulations, which often go down to 50-100 particles. This means that, while we should be capturing the vast majority of major mergers in the high and intermediate mass bins, it is likely that we are missing significant interactions in the low mass bin - particularly at high redshift, where galaxies typically have fewer particles. We are able to identify major mergers for only $\sim$76\% of low mass isolated galaxies, compared to $\sim$87\% of high and intermediate mass galaxies. \\
\indent Another factor that is not reflected in Figure \ref{fig:evorelmer} is the prevalence of unbound interactions. Many of our galaxies - including a number of those that have never had a major merger - execute flybys that alter their evolution in much the same way that a merger might \citep[e.g.,][]{sinha2015,martin2020}. We will explore the role that flybys play in the creation of UDGs in an upcoming paper on the broader low surface brightness galaxy population in \textsc{Romulus25} (Wright et al., in prep). 
\subsubsection{Co-planar Mergers?}
\begin{figure*}
\includegraphics[trim=22mm 0mm 22mm 10mm, clip, width=0.86\textwidth]{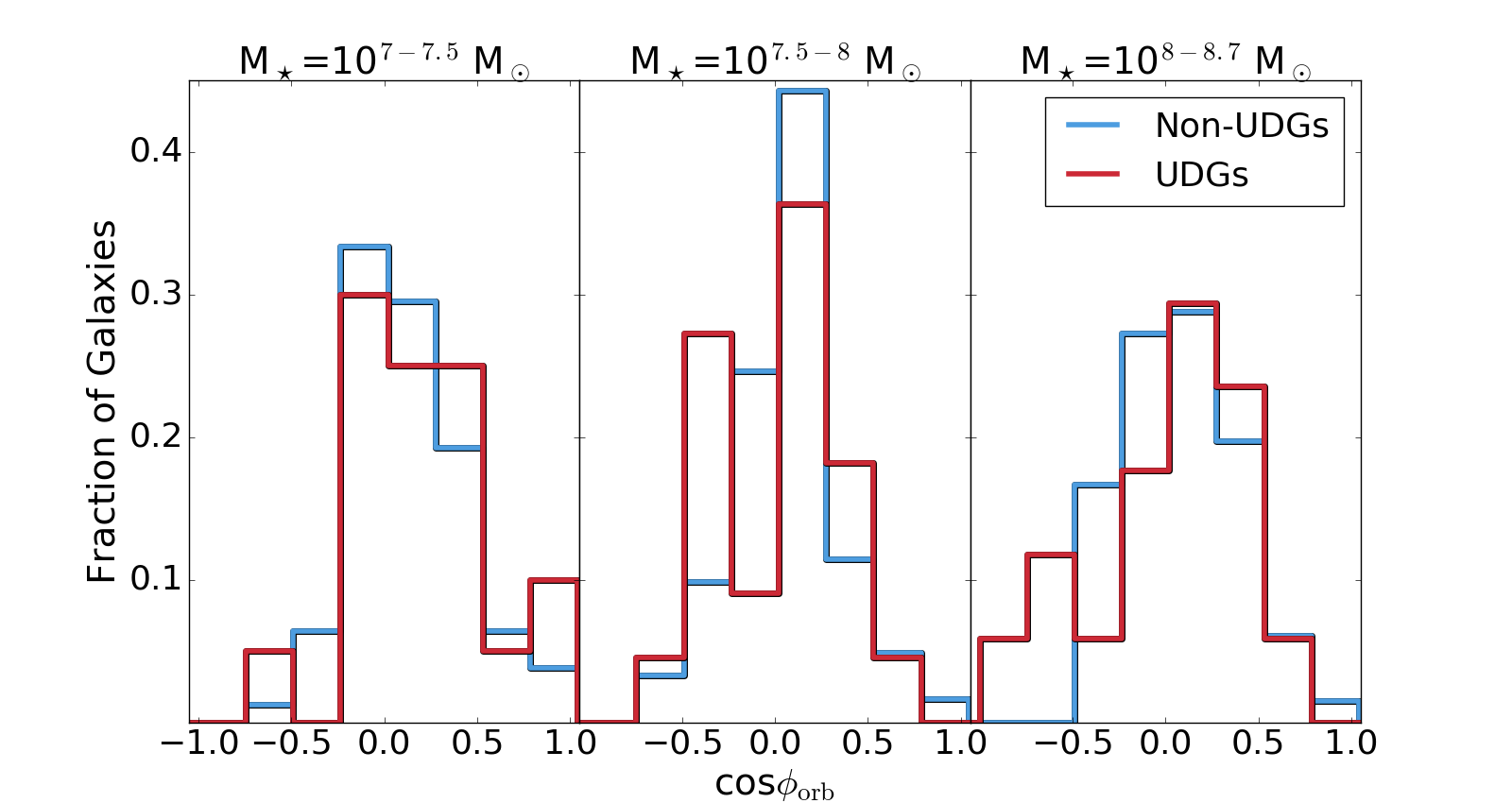}
\caption{Distribution of merger configurations for our UDG and non-UDG comparison samples. Following \protect\citet{dicintio2019}, $\phi_\mathrm{orb}$ is the angle between the orbital angular momentum vector of the secondary galaxy and the specific angular momentum vector of the gas of the primary galaxy. Accordingly, cos$\phi_\mathrm{orb} =$ 1 indicates a co-rotating, co-planar merger, cos$\phi_\mathrm{orb} =$ -1 indicates a counter-rotating, co-planar merger, and cos$\phi_\mathrm{orb} =$ 0 indicates a perpendicular merger. Again following \protect\citet{dicintio2019}, only galaxies that have experienced a >3:1 merger since $z=2.5$ are shown. The distributions of merger configurations that produce UDGs and non-UDGs are very similar, although UDG progenitors are slightly less likely to have experienced an extreme perpendicular merger.}
\label{fig:phidist}
\end{figure*}

The notion that mergers might contribute to the formation of low surface brightness galaxies is not entirely new. However, in the past, mergers have largely been invoked as a means by which the extended low surface brightness disks that characterize the rare class of objects known as giant low surface brightness galaxies (M$_\star\geq$10$^{11}$) might have formed \citep[e.g.,][]{bruevich2010,reshetnikov2010,saburova2018}. In these scenarios, a large low density disk forms around a more traditional high surface brightness galaxy through stimulated accretion of gas from the circumgalactic medium of the primary galaxy \citep{zhu2018} and/or accretion of ex-situ stars from the secondary galaxy \citep{penarrubia2006,hagen2016,kulier2020}. Isolated N-body simulations by \citet{mapelli2008} have even suggested that P-type ring galaxies - thought to be the products of head-on galactic collisions - might eventually evolve into giant low surface brightness galaxies.\\
\indent Our results are more directly comparable to those of \citet{dicintio2019}, who find that classical low surface brightness galaxies (M$_\star$>10$^{9.5}$ M$_\odot$) in the NIHAO simulations are the products of co-planar, co-rotating, major mergers. Such configurations allow for a more efficient conversion of orbital angular momentum to internal angular momentum, increasing the spin (and therefore size) of the remnant while decreasing its central surface brightness. As these effects are precisely what we observe in Figure \ref{fig:evorelmer}, we might expect to find that orientation is the key difference between mergers that produce UDGs and those that produce non-UDGs. However, we do not see a correlation between orbital alignment and central surface brightness or effective radius. Rather, we find that there is very little difference between the merger configurations of UDG and non-UDG progenitors.\\
\indent In Figure \ref{fig:phidist}, we show the distribution of merger orientations for our UDG and non-UDG comparison samples. Following \citet{dicintio2019}, we characterize merger configurations using $\phi_\mathrm{orb}$, the angle between the orbital angular momentum vector of the secondary galaxy ($\vec{J}_\mathrm{orb} = m\vec{r}\times\vec{v}$) and the angular momentum vector of the gas of the primary galaxy ($\vec{J}_\mathrm{gas,primary}$):
\begin{equation}
    \phi_\mathrm{orb} = \mathrm{acos}(\vec{J}_\mathrm{orb} \cdot \vec{J}_\mathrm{gas,primary}). 
\end{equation}
Accordingly, cos$\phi_\mathrm{orb} =$ 1 indicates a co-rotating, co-planar merger, cos$\phi_\mathrm{orb} =$ -1 indicates a counter-rotating, co-planar merger, and cos$\phi_\mathrm{orb} =$ 0 indicates a perpendicular merger. We find that dwarf-dwarf interactions are more prone to perpendicular configurations than planar ones. The distribution of merger configurations is very similar across all three mass bins. The only significant difference that exists between the UDG and non-UDG samples is within the intermediate mass bin, where UDGs are 10\% more likely to have experienced a counter-rotating merger, while non-UDGs are 80\% more likely to have experienced a perpendicular merger. Again following \citet{dicintio2019}, we only show the most recent $\leq$3:1 merger in Figure \ref{fig:phidist} and exclude any such mergers that occur prior to $z=2.5$. However, using cuts consistent with those applied throughout the rest of this discussion (i.e., the most recent $\leq$5:1 merger, regardless of redshift) does not significantly alter these findings. It is not orbital alignment that determines whether a merger produces a UDG or a non-UDG.\\
\indent Ultimately, though, this may be unsurprising. \citet{dicintio2019} posit that M$_\star\sim$10$^9$ M$_\odot$ represents a transition mass above which galaxy evolution is angular momentum dominated and below which galaxy evolution is feedback dominated. In fact, \citet{dekel2020} show that galaxies below this threshold in the VELA simulations are likely to experience a spin flip and subsequent destruction of their gaseous disk following a merger. We also find that mergers often lead to substantial changes in the direction of a dwarf galaxy's angular momentum vector. Although dwarf-dwarf mergers more commonly result in disk creation than destruction in \textsc{Romulus25}, this may explain why an initial alignment (or lack thereof) between the spin of the primary galaxy and the orbit of the secondary galaxy is largely irrelevant in this mass range.\\
\indent The UDGs that form in the NIHAO simulations have prolonged, bursty star formation histories that lead to the creation of dark matter cores and the expansion of their stellar components \citep{dicintio2017}. Because \textsc{Romulus25} does not have the resolution to form stars at the densities required to produce dark matter cores, we cannot directly compare our results to those of \citet{dicintio2017}. However, the merger-driven scenario we have presented is compatible with their findings. Although the UDGs in \textsc{Romulus25} do not have atypical global star formation histories, it's possible that at higher resolution we would see an initial increase in central star formation following the merger that would produce a dark matter core and help drive the redistribution of star formation discussed in Section \ref{sfmigration}. We have, however, shown that core formation is not required for UDG formation. The question of what types of mergers lead to the formation of low surface brightness galaxies in the dwarf regime will be addressed in more detail in our upcoming paper on the low surface brightness galaxy population in \textsc{Romulus25} (Wright et al., in prep). \\
\indent On the observational side, there is evidence to suggest that some UDGs may be the products of gravitational interactions, but this is usually taken to be tidal interactions rather than direct mergers. While the majority of the UDGs that have been discovered are exceptionally round in appearance and lack any clear tidal features \citep[e.g.,][]{mowla2017}, there are a number whose elongated shapes and/or associated tidal streams indicate that a significant disturbance has taken place \citep[e.g.,][]{collins2013,mihos2015,toloba2016,crnojevic2016,merritt2016,greco2018sumo,toloba2018,bennet2018}. However, these galaxies are all members of groups or clusters and are thought to be tidally interacting with their parent halos, rather than their fellow dwarfs. While many of the UDGs that populate our sample are similarly irregular in appearance, they have no obvious source of perturbation at $z=0$, having long ago consumed their companions. 
\section{Comparison to HUDS}
\label{disc}
\begin{figure}
\includegraphics[trim = 5mm 4mm 5mm 4mm, clip, width=0.47\textwidth]{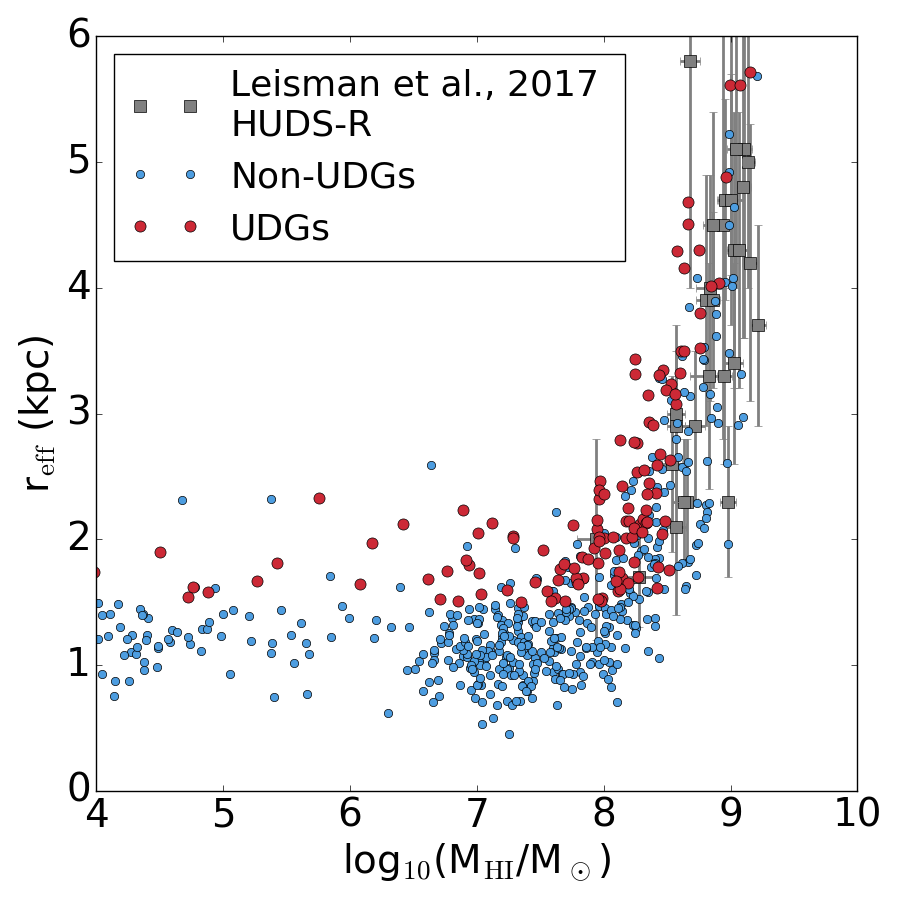}
\caption{HI masses and effective radii of isolated UDGs and non-UDGs from \textsc{Romulus25} compared to the restricted HI-bearing UDG (HUDS-R) sample from \protect{\citet{leisman2017}}. Both the observed galaxies and the simulated galaxies follow the same trend and occupy roughly the same space. However, the majority of our UDG sample lies below \protect{\citet{leisman2017}}'s detection threshold, indicating that the population of field UDGs may be considerably larger than HI-selected surveys might suggest.}
\label{fig:lcomp}
\end{figure}

\indent Our findings are largely consistent with the few observations of isolated UDGs that exist. \citet{leisman2017}, who used ALFALFA data to identify a sample of 115 HI-bearing UDGs (HUDS) in the field, find that isolated UDGs are dwarf galaxies with blue colors and irregular morphologies. Like the UDGs in our sample, their galaxies have average SFRs and tend to be slightly HI-rich for their stellar masses. In Figure \ref{fig:lcomp}, we compare the HI masses and effective radii of our isolated UDG and non-UDG samples to \citet{leisman2017}'s sample. Note that we include only the 30 galaxies in the restricted HUDS sample (HUDS-R), which were selected via central surface brightness and effective radius criteria identical to those used in this work. In both the simulated and the observed galaxies, there is a general trend for physically larger galaxies to be more HI-rich, although there is considerable scatter at M$_{\mathrm{HI}}$>10$^{8.5}$~M$_\odot$. This trend is also noted in \citet{dicintio2017}. \\
\indent The high mass \textsc{Romulus25} UDGs broadly occupy the same space as the \citet{leisman2017} sample. However, more than half of our UDG sample actually lies below the detection threshold of the \citet{leisman2017} sample (M$_{\mathrm{HI}}\sim$10$^{8.2}$ M$_\odot$). We therefore predict the existence of a large number of UDGs that would not have been observed by ALFALFA. Although our UDGs, like the HUDS, are slightly HI-rich for their stellar masses, they are actually HI-poor for their effective radii when compared to the non-UDGs in our sample. This is likely because UDGs tend to have high effective radii for their stellar masses, so, at a given effective radius, we are comparing the HI masses of slightly less massive UDGs to those of slightly more massive non-UDGs. \\
\indent However, the UDGs in our sample appear slightly HI-poor for their effective radii even when compared to the HUDS-R sample. This is most likely a consequence of the differences in how r$_\mathrm{eff}$ is calculated in the simulations vs the observations. Because we rotate our galaxies such that they are face-on before fitting S\'ersic profiles to their surface brightness profiles, we have assigned each galaxy the maximum r$_\mathrm{eff}$ that could possibly be measured for it. Although the galaxies in the HUDS-R sample are fitted as if they are face-on, their inclinations are poorly constrained and it is unlikely that this assumption is valid for the whole sample. The reported r$_\mathrm{eff}$ values are therefore likely to be an underestimate of the true values. Shifting the HUDS-R galaxies to higher r$_\mathrm{eff}$ values would bring our samples into better agreement. \\
\indent Another potential source of bias is the relatively small size of \textsc{Romulus25}. Our survey is of a (25 Mpc)$^3$ volume, while \citet{leisman2017}'s is of a ($\sim$93 Mpc)$^3$ volume - roughly 50 times larger\footnote{Although note that ALFALFA is HI-flux-limited, so the lowest mass galaxies within the HUDS-R sample would not be detected across this entire volume \citep{jones2018}}. HI-selected surveys tend to be biased high compared to optically-selected surveys \citep[e.g.,][]{catinella2010}, so they are most likely to identify the high tail of the M$_{\mathrm{HI}}$ distribution. Our box simply may not be big enough to have a reasonable chance of containing such galaxies. \\
\indent Follow-up work by \citet{jones2018} uses \citet{leisman2017}'s sample to estimate the number density of HI-bearing field UDGs. If we restrict our sample to the parameter space where the HUDS are considered to be complete (10$^{8.5}$<M$_\mathrm{HI}$/M$_\odot$<10$^{9.5}$), the prevalence of UDGs within \textsc{Romulus25} matches the values reported in \citet{jones2018} remarkably well. They calculate a cosmic number density of (1.5$\pm$0.6$)\times$10$^{-3}$ Mpc$^{-3}$, suggesting that, in a (25 Mpc)$^3$ volume, we should expect to see 23$\pm$9 isolated UDGs with HI masses in the aforementioned range. In \textsc{Romulus25}, we find 22 such UDGs, yielding a cosmic number density of 1.4$\times$10$^{-3}$ Mpc$^{-3}$. Our simulated field UDGs make up 9\% of galaxies with M$_\mathrm{HI}$=10$^{8.5-9.5}$ M$_\odot$, compared to the 6\% reported by \citet{jones2018}.\\
\indent However, there are two significant caveats to these findings. The first is that the values reported in \citet{jones2018} are based on nearly the full HUDS sample (only slightly pared down for completeness), rather than the HUDS-R sample, which we compare to in Figure \ref{fig:lcomp}. Their galaxies are therefore classified as UDGs if the average surface brightness within the effective radius (<$\mu_\mathrm{eff}$>) is fainter than 24 mag/arcsec$^2$. This is a significantly less stringent criterion than that used to select our sample ($\mu_\mathrm{0,g}$>24 mag/arcsec$^2$) and the HUDS-R sample. We might, therefore, expect to find far fewer UDGs than \citet{jones2018}, given that we are excluding from our sample many galaxies that they would include.\\
\indent The second caveat, however, acts in the opposite direction. As previously mentioned, we rotate all galaxies such that they are face-on prior to fitting a S\'ersic profile to their surface brightness profiles. Because this orientation maximizes the effective radii and minimizes the central surface brightnesses of galaxies, we are effectively maximizing the number of UDGs that we find. In the field, it is likely that there are a large number of galaxies that would be identified as UDGs were they at lower inclinations. However, because galaxies tend to appear brighter when viewed edge-on, they would not be included in the HUDS sample \citep[although see][]{he2019}. This suggests that we should expect to find considerably \textit{more} HI-bearing UDGs in \textsc{Romulus25} than are observed by surveys like ALFALFA. We will explore the effects of inclination and alternative classification criteria on our sample in more detail in Van Nest et al., in prep. 
\section{Summary}
\label{wrapup}
We have identified a sample of 134 isolated UDGs in the \textsc{Romulus25} cosmological simulation. We find that these galaxies have stellar masses between 10$^7$ and 10$^9$ M$_\odot$ and account for 20\% of all isolated galaxies in this mass range. They are true dwarf galaxies, occupying dark matter halos with M$_{200}$ <10$^{11}$ M$_\odot$, and are consistent with the stellar mass-halo mass relation \citep[e.g.,][]{Moster2013}. Despite their low surface brightnesses and large effective radii, the isolated UDGs in \textsc{Romulus25} have star formation rates and colors that are typical for their stellar masses and environment. Like \citet{dicintio2017}, we also find that they are moderately HI-rich for their luminosities, having 70\% more HI than non-UDGs at luminosities brighter than M$_\mathrm{B}$=-14 and 300\% more at luminosities fainter than M$_\mathrm{B}$=-14. These findings are consistent with existent observations of field UDGs. \\
\indent Although UDGs are broadly characterized by low surface brightnesses and large effective radii relative to other galaxies of similar stellar mass, the specific criteria typically used for UDG selection do not necessarily limit UDG samples to the tails of both of these distributions at all masses. At high mass (M$_\star$>10$^{8}$ M$_\odot$), most galaxies have r$_\mathrm{eff}$>1.5 kpc. UDGs are therefore those galaxies that are unusually low surface brightness. At low mass (M$_\star$<10$^{7.5}$ M$_\odot$), most galaxies have $\mu_{\mathrm{0,g}}$>24 mag arcsec$^{-2}$. UDGs are therefore those galaxies that have unusually large effective radii. It is only really at intermediate masses that both the central surface brightness and the effective radius criteria are relevant. \\
\indent We find that UDGs are primarily the products of major mergers at relatively early times (>8 Gyr ago). These mergers increase the effective radii and total spin of UDG progenitors, while decreasing their central SFRs and surface brightnesses. The mergers cause star formation to redistributed to the outskirts of galaxies, resulting in lower central SFRs and therefore older and fainter central stellar populations. Particularly in the high mass UDGs, these mergers often produce asymmetric bursts of star formation along the edges of the galaxies that persist for billions of years. Accordingly, even at $z=0$, the centers of UDGs are redder than those of non-UDGs, while their outskirts are bluer, resulting in steeper negative g-r color gradients among UDGs.\\
\indent Although both UDGs and non-UDGs are spun up by major mergers, UDGs experience as much as 200\% more spin-up. This likely contributes to the change in the distribution of star-forming gas within UDGs and therefore to their low central surface brightnesses and large effective radii. However, UDGs do not form in exclusively high spin halos and we do not find a strong correlation between the maximum spin of UDG progenitors and the characteristics that make them UDGs at $z=0$ (i.e., effective radius for low mass UDGs and central surface brightness for high mass UDGs). We also find that the spin-up produced by the mergers is largely temporary. By $z=0$, UDGs and non-UDGs have a similar distribution of spin parameters.\\
\indent Prior to the mergers, the evolution of the effective radii, central surface brightnesses, central star formation rates, and spins of UDG progenitors are indistinguishable from those of non-UDG progenitors. Both populations of galaxies also have similar numbers of major and minor mergers and primarily merge with gas-rich companions. This suggests that it is the specific dynamical properties of a given merger that determine whether or not the resultant galaxy eventually becomes a UDG. We explore the role that mergers and flybys play in this process in more detail in an upcoming paper on the broader LSB galaxy population of \textsc{Romulus25} (Wright et al., in prep).\\
\indent Perhaps the most significant finding of this paper is that UDGs are not a small population, even in the field. We predict a cosmic number density of 8.6$\times$10$^{-3}$ isolated UDGs/Mpc$^3$ (1.4$\times$10$^{-3}$ Mpc$^{-3}$ for those detectable by ALFALFA). Their extreme diffuseness means that UDGs are unlikely to have been included in previous surveys of field dwarf galaxies. While this may have few implications for relations like the SFR-M$_*$ and M$_*$-M$_\mathrm{HI}$ relations, where UDGs are quite average, we expect that this deficit significantly biases the low-mass end of many other relations (e.g., M$_*$-r$_\mathrm{eff}$). However, future surveys like LSST and the Dragonfly Wide Field Survey \citep{danieli2020} will have an unprecedented view of the low surface brightness universe. Their discoveries will shed new light on the formation and evolution of diverse dwarf galaxies.
\section*{Acknowledgments}
 The authors thank the anonymous referee for their careful read of this paper and helpful suggestions that improved its clarity. We also thank Johnny Greco, Luke Leisman, Arianna Di Cintio, Marla Geha, Michael Jones, Amanda Moffett, Nushkia Chamba, Frank van den Bosch, and Jordan Van Nest for useful discussions related to this work. ACW is supported by an ACM SIGHPC/Intel Computational \& Data Science fellowship. AMB, DN, and FM acknowledge the hospitality at the Aspen Center for Physics, which is supported by National Science Foundation grant PHY-1607611. The python packages {\sc matplotlib} \citep{hunter2007}, {\sc numpy} \citep{walt2011numpy}, {\sc tangos} \citep{pontzen2018}, {\sc pynbody} \citep{pynbody}, \textsc{scipy} \citep{scipy2020}, and {\sc Astropy} \citep{astropy2013,astropy2018} were all used in parts of this analysis. This research is part of the Blue Waters sustained-petascale computing project, which is supported by the National Science Foundation (awards OCI-0725070 and ACI-1238993) and the state of Illinois. Blue Waters is a joint effort of the University of Illinois at Urbana-Champaign and its National Center for Supercomputing Applications. This work is also part of a PRAC allocation support by the National Science Foundation (award number OAC-1613674).
 \section*{Data Availability}
The data for this work was generated from a proprietary branch of the {\sc ChaNGa} N-Body+SPH code \citep{menon2015}. The public repository for {\sc ChaNGa} is available on github https://github.com/N-BodyShop/changa). Analysis was conducted using the publicly available software pynbody \citep[][https://github.com/pynbody/pynbody]{pynbody} and TANGOS \citep[][https://github.com/pynbody/tangos]{pontzen2018}. These results were generated from the RomulusC and Romulus25 cosmological simulations. The raw output from these simulations can be accessed upon request from Michael Tremmel (michael.tremmel@yale.edu), along with the TANGOS database files that were generated from these outputs and directly used for this analysis.
\bibliographystyle{mnras}
\bibliography{UDGs}

\begin{thebibliography}{}
\makeatletter
\relax
\def\mn@urlcharsother{\let\do\@makeother \do\$\do\&\do\#\do\^\do\_\do\%\do\~}
\def\mn@doi{\begingroup\mn@urlcharsother \@ifnextchar [ {\mn@doi@}
  {\mn@doi@[]}}
\def\mn@doi@[#1]#2{\def\@tempa{#1}\ifx\@tempa\@empty \href
  {http://dx.doi.org/#2} {doi:#2}\else \href {http://dx.doi.org/#2} {#1}\fi
  \endgroup}
\def\mn@eprint#1#2{\mn@eprint@#1:#2::\@nil}
\def\mn@eprint@arXiv#1{\href {http://arxiv.org/abs/#1} {{\tt arXiv:#1}}}
\def\mn@eprint@dblp#1{\href {http://dblp.uni-trier.de/rec/bibtex/#1.xml}
  {dblp:#1}}
\def\mn@eprint@#1:#2:#3:#4\@nil{\def\@tempa {#1}\def\@tempb {#2}\def\@tempc
  {#3}\ifx \@tempc \@empty \let \@tempc \@tempb \let \@tempb \@tempa \fi \ifx
  \@tempb \@empty \def\@tempb {arXiv}\fi \@ifundefined
  {mn@eprint@\@tempb}{\@tempb:\@tempc}{\expandafter \expandafter \csname
  mn@eprint@\@tempb\endcsname \expandafter{\@tempc}}}

\bibitem[\protect\citeauthoryear{{Abel}, {Anninos}, {Zhang}  \&
  {Norman}}{{Abel} et~al.}{1997}]{abel1997}
{Abel} T.,  {Anninos} P.,  {Zhang} Y.,   {Norman} M.~L.,  1997, \mn@doi [\na]
  {10.1016/S1384-1076(97)00010-9}, \href
  {https://ui.adsabs.harvard.edu/abs/1997NewA....2..181A} {2, 181}

\bibitem[\protect\citeauthoryear{{Abraham} \& {van Dokkum}}{{Abraham} \& {van
  Dokkum}}{2014}]{abraham2014}
{Abraham} R.~G.,  {van Dokkum} P.~G.,  2014, \mn@doi [\pasp] {10.1086/674875},
  \href {https://ui.adsabs.harvard.edu/abs/2014PASP..126...55A} {126, 55}

\bibitem[\protect\citeauthoryear{{Amorisco} \& {Loeb}}{{Amorisco} \&
  {Loeb}}{2016}]{amorisco2016}
{Amorisco} N.~C.,  {Loeb} A.,  2016, \mn@doi [\mnras] {10.1093/mnrasl/slw055},
  \href {https://ui.adsabs.harvard.edu/abs/2016MNRAS.459L..51A} {459, L51}

\bibitem[\protect\citeauthoryear{{Amorisco}, {Monachesi}, {Agnello}  \&
  {White}}{{Amorisco} et~al.}{2018}]{amorisco2018}
{Amorisco} N.~C.,  {Monachesi} A.,  {Agnello} A.,   {White} S.~D.~M.,  2018,
  \mn@doi [\mnras] {10.1093/mnras/sty116}, \href
  {https://ui.adsabs.harvard.edu/abs/2018MNRAS.475.4235A} {475, 4235}

\bibitem[\protect\citeauthoryear{{Astropy Collaboration} et~al.,}{{Astropy
  Collaboration} et~al.}{2013}]{astropy2013}
{Astropy Collaboration} et~al., 2013, \mn@doi [\aap]
  {10.1051/0004-6361/201322068}, \href
  {https://ui.adsabs.harvard.edu/abs/2013A&A...558A..33A} {558, A33}

\bibitem[\protect\citeauthoryear{{Astropy Collaboration} et~al.,}{{Astropy
  Collaboration} et~al.}{2018}]{astropy2018}
{Astropy Collaboration} et~al., 2018, \mn@doi [\aj] {10.3847/1538-3881/aabc4f},
  \href {https://ui.adsabs.harvard.edu/abs/2018AJ....156..123A} {156, 123}

\bibitem[\protect\citeauthoryear{{Barbosa} et~al.,}{{Barbosa}
  et~al.}{2020}]{barbosa2020}
{Barbosa} C.~E.,  et~al., 2020, \mn@doi [\apjs] {10.3847/1538-4365/ab7660},
  \href {https://ui.adsabs.harvard.edu/abs/2020ApJS..247...46B} {247, 46}

\bibitem[\protect\citeauthoryear{{Barnes}}{{Barnes}}{1988}]{barnes1988}
{Barnes} J.~E.,  1988, \mn@doi [\apj] {10.1086/166593}, \href
  {https://ui.adsabs.harvard.edu/abs/1988ApJ...331..699B} {331, 699}

\bibitem[\protect\citeauthoryear{{Barnes} \& {Efstathiou}}{{Barnes} \&
  {Efstathiou}}{1987}]{barnes1987}
{Barnes} J.,  {Efstathiou} G.,  1987, \mn@doi [\apj] {10.1086/165480}, \href
  {https://ui.adsabs.harvard.edu/abs/1987ApJ...319..575B} {319, 575}

\bibitem[\protect\citeauthoryear{{Beasley} \& {Trujillo}}{{Beasley} \&
  {Trujillo}}{2016}]{beasleytrujillo2016}
{Beasley} M.~A.,  {Trujillo} I.,  2016, \mn@doi [\apj]
  {10.3847/0004-637X/830/1/23}, \href
  {https://ui.adsabs.harvard.edu/abs/2016ApJ...830...23B} {830, 23}

\bibitem[\protect\citeauthoryear{{Beasley}, {Romanowsky}, {Pota}, {Navarro},
  {Martinez Delgado}, {Neyer}  \& {Deich}}{{Beasley}
  et~al.}{2016}]{beasley2016}
{Beasley} M.~A.,  {Romanowsky} A.~J.,  {Pota} V.,  {Navarro} I.~M.,  {Martinez
  Delgado} D.,  {Neyer} F.,   {Deich} A.~L.,  2016, \mn@doi [\apjl]
  {10.3847/2041-8205/819/2/L20}, \href
  {https://ui.adsabs.harvard.edu/abs/2016ApJ...819L..20B} {819, L20}

\bibitem[\protect\citeauthoryear{{Begum}, {Chengalur}, {Karachentsev},
  {Sharina}  \& {Kaisin}}{{Begum} et~al.}{2008}]{begum2008}
{Begum} A.,  {Chengalur} J.~N.,  {Karachentsev} I.~D.,  {Sharina} M.~E.,
  {Kaisin} S.~S.,  2008, \mn@doi [\mnras] {10.1111/j.1365-2966.2008.13150.x},
  \href {https://ui.adsabs.harvard.edu/abs/2008MNRAS.386.1667B} {386, 1667}

\bibitem[\protect\citeauthoryear{{Bennet}, {Sand}, {Zaritsky}, {Crnojevi{\'c}},
  {Spekkens}  \& {Karunakaran}}{{Bennet} et~al.}{2018}]{bennet2018}
{Bennet} P.,  {Sand} D.~J.,  {Zaritsky} D.,  {Crnojevi{\'c}} D.,  {Spekkens}
  K.,   {Karunakaran} A.,  2018, \mn@doi [\apjl] {10.3847/2041-8213/aadedf},
  \href {https://ui.adsabs.harvard.edu/abs/2018ApJ...866L..11B} {866, L11}

\bibitem[\protect\citeauthoryear{{Binggeli}, {Sandage}  \&
  {Tammann}}{{Binggeli} et~al.}{1985}]{binggeli1985}
{Binggeli} B.,  {Sandage} A.,   {Tammann} G.~A.,  1985, \mn@doi [\aj]
  {10.1086/113874}, \href
  {https://ui.adsabs.harvard.edu/abs/1985AJ.....90.1681B} {90, 1681}

\bibitem[\protect\citeauthoryear{{Black}}{{Black}}{1981}]{black1981}
{Black} J.~H.,  1981, \mn@doi [\mnras] {10.1093/mnras/197.3.553}, \href
  {https://ui.adsabs.harvard.edu/abs/1981MNRAS.197..553B} {197, 553}

\bibitem[\protect\citeauthoryear{{Borlaff} et~al.,}{{Borlaff}
  et~al.}{2019}]{borlaff2019}
{Borlaff} A.,  et~al., 2019, \mn@doi [\aap] {10.1051/0004-6361/201834312},
  \href {https://ui.adsabs.harvard.edu/abs/2019A&A...621A.133B} {621, A133}

\bibitem[\protect\citeauthoryear{{Bromm}, {Ferrara}, {Coppi}  \&
  {Larson}}{{Bromm} et~al.}{2001}]{bromm2001}
{Bromm} V.,  {Ferrara} A.,  {Coppi} P.~S.,   {Larson} R.~B.,  2001, \mn@doi
  [\mnras] {10.1046/j.1365-8711.2001.04915.x}, \href
  {https://ui.adsabs.harvard.edu/abs/2001MNRAS.328..969B} {328, 969}

\bibitem[\protect\citeauthoryear{{Bruevich}, {Gusev}  \&
  {Guslyakova}}{{Bruevich} et~al.}{2010}]{bruevich2010}
{Bruevich} V.~V.,  {Gusev} A.~S.,   {Guslyakova} S.~A.,  2010, \mn@doi
  [Astronomy Reports] {10.1134/S106377291005001X}, \href
  {https://ui.adsabs.harvard.edu/abs/2010ARep...54..375B} {54, 375}

\bibitem[\protect\citeauthoryear{{Bryan} \& {Norman}}{{Bryan} \&
  {Norman}}{1998}]{bryan1998}
{Bryan} G.~L.,  {Norman} M.~L.,  1998, \mn@doi [\apj] {10.1086/305262}, \href
  {https://ui.adsabs.harvard.edu/abs/1998ApJ...495...80B} {495, 80}

\bibitem[\protect\citeauthoryear{{Bullock}, {Dekel}, {Kolatt}, {Kravtsov},
  {Klypin}, {Porciani}  \& {Primack}}{{Bullock} et~al.}{2001}]{bullock2001}
{Bullock} J.~S.,  {Dekel} A.,  {Kolatt} T.~S.,  {Kravtsov} A.~V.,  {Klypin}
  A.~A.,  {Porciani} C.,   {Primack} J.~R.,  2001, \mn@doi [\apj]
  {10.1086/321477}, \href
  {https://ui.adsabs.harvard.edu/abs/2001ApJ...555..240B} {555, 240}

\bibitem[\protect\citeauthoryear{{Caldwell} \& {Bothun}}{{Caldwell} \&
  {Bothun}}{1987}]{caldwell1987}
{Caldwell} N.,  {Bothun} G.~D.,  1987, \mn@doi [\aj] {10.1086/114550}, \href
  {https://ui.adsabs.harvard.edu/abs/1987AJ.....94.1126C} {94, 1126}

\bibitem[\protect\citeauthoryear{{Capaccioli}}{{Capaccioli}}{1989}]{capaccioli1989}
{Capaccioli} M.,  1989, in {Corwin} Harold~G. J.,  {Bottinelli} L.,  eds, World
  of Galaxies (Le Monde des Galaxies). pp 208--227

\bibitem[\protect\citeauthoryear{{Cardona-Barrero}, {Di Cintio}, {Brook},
  {Ruiz-Lara}, {Beasley}, {Falc{\'o}n-Barroso}  \&
  {Macci{\`o}}}{{Cardona-Barrero} et~al.}{2020}]{cardonabarrero2020}
{Cardona-Barrero} S.,  {Di Cintio} A.,  {Brook} C. B.~A.,  {Ruiz-Lara} T.,
  {Beasley} M.~A.,  {Falc{\'o}n-Barroso} J.,   {Macci{\`o}} A.~V.,  2020,
  \mn@doi [\mnras] {10.1093/mnras/staa2094}, \href
  {https://ui.adsabs.harvard.edu/abs/2020MNRAS.497.4282C} {497, 4282}

\bibitem[\protect\citeauthoryear{{Carleton}, {Errani}, {Cooper}, {Kaplinghat},
  {Pe{\~n}arrubia}  \& {Guo}}{{Carleton} et~al.}{2019}]{carleton2019}
{Carleton} T.,  {Errani} R.,  {Cooper} M.,  {Kaplinghat} M.,  {Pe{\~n}arrubia}
  J.,   {Guo} Y.,  2019, \mn@doi [\mnras] {10.1093/mnras/stz383}, \href
  {https://ui.adsabs.harvard.edu/abs/2019MNRAS.485..382C} {485, 382}

\bibitem[\protect\citeauthoryear{{Catinella} et~al.,}{{Catinella}
  et~al.}{2010}]{catinella2010}
{Catinella} B.,  et~al., 2010, \mn@doi [\mnras]
  {10.1111/j.1365-2966.2009.16180.x}, \href
  {https://ui.adsabs.harvard.edu/abs/2010MNRAS.403..683C} {403, 683}

\bibitem[\protect\citeauthoryear{{Cen}}{{Cen}}{1992}]{cen1992}
{Cen} R.,  1992, \mn@doi [\apjs] {10.1086/191630}, \href
  {https://ui.adsabs.harvard.edu/abs/1992ApJS...78..341C} {78, 341}

\bibitem[\protect\citeauthoryear{{Chamba}, {Trujillo}  \& {Knapen}}{{Chamba}
  et~al.}{2020}]{chamba2020}
{Chamba} N.,  {Trujillo} I.,   {Knapen} J.~H.,  2020, \mn@doi [\aap]
  {10.1051/0004-6361/201936821}, \href
  {https://ui.adsabs.harvard.edu/abs/2020A&A...633L...3C} {633, L3}

\bibitem[\protect\citeauthoryear{{Chan}, {Kere{\v{s}}}, {O{\~n}orbe},
  {Hopkins}, {Muratov}, {Faucher-Gigu{\`e}re}  \& {Quataert}}{{Chan}
  et~al.}{2015}]{chan2015}
{Chan} T.~K.,  {Kere{\v{s}}} D.,  {O{\~n}orbe} J.,  {Hopkins} P.~F.,  {Muratov}
  A.~L.,  {Faucher-Gigu{\`e}re} C.~A.,   {Quataert} E.,  2015, \mn@doi [\mnras]
  {10.1093/mnras/stv2165}, \href
  {https://ui.adsabs.harvard.edu/abs/2015MNRAS.454.2981C} {454, 2981}

\bibitem[\protect\citeauthoryear{{Chan}, {Kere{\v{s}}}, {Wetzel}, {Hopkins},
  {Faucher-Gigu{\`e}re}, {El-Badry}, {Garrison-Kimmel}  \&
  {Boylan-Kolchin}}{{Chan} et~al.}{2018}]{chan2018}
{Chan} T.~K.,  {Kere{\v{s}}} D.,  {Wetzel} A.,  {Hopkins} P.~F.,
  {Faucher-Gigu{\`e}re} C.~A.,  {El-Badry} K.,  {Garrison-Kimmel} S.,
  {Boylan-Kolchin} M.,  2018, \mn@doi [\mnras] {10.1093/mnras/sty1153}, \href
  {https://ui.adsabs.harvard.edu/abs/2018MNRAS.478..906C} {478, 906}

\bibitem[\protect\citeauthoryear{{Christensen}, {Governato}, {Quinn}, {Brooks},
  {Shen}, {McCleary}, {Fisher}  \& {Wadsley}}{{Christensen}
  et~al.}{2014}]{christensen2014}
{Christensen} C.~R.,  {Governato} F.,  {Quinn} T.,  {Brooks} A.~M.,  {Shen} S.,
   {McCleary} J.,  {Fisher} D.~B.,   {Wadsley} J.,  2014, \mn@doi [\mnras]
  {10.1093/mnras/stu399}, \href
  {https://ui.adsabs.harvard.edu/abs/2014MNRAS.440.2843C} {440, 2843}

\bibitem[\protect\citeauthoryear{{Collins} et~al.,}{{Collins}
  et~al.}{2013}]{collins2013}
{Collins} M. L.~M.,  et~al., 2013, \mn@doi [\apj]
  {10.1088/0004-637X/768/2/172}, \href
  {https://ui.adsabs.harvard.edu/abs/2013ApJ...768..172C} {768, 172}

\bibitem[\protect\citeauthoryear{{Conselice}}{{Conselice}}{2018}]{conselice2018}
{Conselice} C.~J.,  2018, \mn@doi [Research Notes of the American Astronomical
  Society] {10.3847/2515-5172/aab7f6}, \href
  {https://ui.adsabs.harvard.edu/abs/2018RNAAS...2...43C} {2, 43}

\bibitem[\protect\citeauthoryear{{Conselice}, {O'Neil}, {Gallagher}  \&
  {Wyse}}{{Conselice} et~al.}{2003}]{conselice2003}
{Conselice} C.~J.,  {O'Neil} K.,  {Gallagher} J.~S.,   {Wyse} R. F.~G.,  2003,
  \mn@doi [\apj] {10.1086/375216}, \href
  {https://ui.adsabs.harvard.edu/abs/2003ApJ...591..167C} {591, 167}

\bibitem[\protect\citeauthoryear{{Crnojevi{\'c}} et~al.,}{{Crnojevi{\'c}}
  et~al.}{2016}]{crnojevic2016}
{Crnojevi{\'c}} D.,  et~al., 2016, \mn@doi [\apj] {10.3847/0004-637X/823/1/19},
  \href {https://ui.adsabs.harvard.edu/abs/2016ApJ...823...19C} {823, 19}

\bibitem[\protect\citeauthoryear{{Dalcanton}, {Spergel}  \&
  {Summers}}{{Dalcanton} et~al.}{1997}]{dalcanton1997}
{Dalcanton} J.~J.,  {Spergel} D.~N.,   {Summers} F.~J.,  1997, \mn@doi [\apj]
  {10.1086/304182}, \href
  {https://ui.adsabs.harvard.edu/abs/1997ApJ...482..659D} {482, 659}

\bibitem[\protect\citeauthoryear{{Danieli}, {van Dokkum}  \&
  {Conroy}}{{Danieli} et~al.}{2018}]{danieli2018}
{Danieli} S.,  {van Dokkum} P.,   {Conroy} C.,  2018, \mn@doi [\apj]
  {10.3847/1538-4357/aaadfb}, \href
  {https://ui.adsabs.harvard.edu/abs/2018ApJ...856...69D} {856, 69}

\bibitem[\protect\citeauthoryear{{Danieli}, {van Dokkum}, {Conroy}, {Abraham}
  \& {Romanowsky}}{{Danieli} et~al.}{2019}]{danieli2019}
{Danieli} S.,  {van Dokkum} P.,  {Conroy} C.,  {Abraham} R.,   {Romanowsky}
  A.~J.,  2019, \mn@doi [\apjl] {10.3847/2041-8213/ab0e8c}, \href
  {https://ui.adsabs.harvard.edu/abs/2019ApJ...874L..12D} {874, L12}

\bibitem[\protect\citeauthoryear{{Danieli} et~al.,}{{Danieli}
  et~al.}{2020}]{danieli2020}
{Danieli} S.,  et~al., 2020, \mn@doi [\apj] {10.3847/1538-4357/ab88a8}, \href
  {https://ui.adsabs.harvard.edu/abs/2020ApJ...894..119D} {894, 119}

\bibitem[\protect\citeauthoryear{{Dekel}, {Ginzburg}, {Jiang}, {Freundlich},
  {Lapiner}, {Ceverino}  \& {Primack}}{{Dekel} et~al.}{2020}]{dekel2020}
{Dekel} A.,  {Ginzburg} O.,  {Jiang} F.,  {Freundlich} J.,  {Lapiner} S.,
  {Ceverino} D.,   {Primack} J.,  2020, \mn@doi [\mnras]
  {10.1093/mnras/staa470}, \href
  {https://ui.adsabs.harvard.edu/abs/2020MNRAS.493.4126D} {493, 4126}

\bibitem[\protect\citeauthoryear{{Di Cintio}, {Brook}, {Macci{\`o}}, {Stinson},
  {Knebe}, {Dutton}  \& {Wadsley}}{{Di Cintio} et~al.}{2014}]{dicintio2014}
{Di Cintio} A.,  {Brook} C.~B.,  {Macci{\`o}} A.~V.,  {Stinson} G.~S.,  {Knebe}
  A.,  {Dutton} A.~A.,   {Wadsley} J.,  2014, \mn@doi [\mnras]
  {10.1093/mnras/stt1891}, \href
  {https://ui.adsabs.harvard.edu/abs/2014MNRAS.437..415D} {437, 415}

\bibitem[\protect\citeauthoryear{{Di Cintio}, {Brook}, {Dutton}, {Macci{\`o}},
  {Obreja}  \& {Dekel}}{{Di Cintio} et~al.}{2017}]{dicintio2017}
{Di Cintio} A.,  {Brook} C.~B.,  {Dutton} A.~A.,  {Macci{\`o}} A.~V.,  {Obreja}
  A.,   {Dekel} A.,  2017, \mn@doi [\mnras] {10.1093/mnrasl/slw210}, \href
  {https://ui.adsabs.harvard.edu/abs/2017MNRAS.466L...1D} {466, L1}

\bibitem[\protect\citeauthoryear{{Di Cintio}, {Brook}, {Macci{\`o}}, {Dutton}
  \& {Cardona-Barrero}}{{Di Cintio} et~al.}{2019}]{dicintio2019}
{Di Cintio} A.,  {Brook} C.~B.,  {Macci{\`o}} A.~V.,  {Dutton} A.~A.,
  {Cardona-Barrero} S.,  2019, \mn@doi [\mnras] {10.1093/mnras/stz985}, \href
  {https://ui.adsabs.harvard.edu/abs/2019MNRAS.486.2535D} {486, 2535}

\bibitem[\protect\citeauthoryear{{Dickey}, {Geha}, {Wetzel}  \&
  {El-Badry}}{{Dickey} et~al.}{2019}]{dickey2019}
{Dickey} C.~M.,  {Geha} M.,  {Wetzel} A.,   {El-Badry} K.,  2019, \mn@doi
  [\apj] {10.3847/1538-4357/ab3220}, \href
  {https://ui.adsabs.harvard.edu/abs/2019ApJ...884..180D} {884, 180}

\bibitem[\protect\citeauthoryear{{Disney}}{{Disney}}{1976}]{disney1976}
{Disney} M.~J.,  1976, \mn@doi [\nat] {10.1038/263573a0}, \href
  {https://ui.adsabs.harvard.edu/abs/1976Natur.263..573D} {263, 573}

\bibitem[\protect\citeauthoryear{{Doroshkevich}}{{Doroshkevich}}{1970}]{doroshkevich1970}
{Doroshkevich} A.~G.,  1970, Astrofizika, \href
  {https://ui.adsabs.harvard.edu/abs/1970Afz.....6..581D} {6, 581}

\bibitem[\protect\citeauthoryear{{Ellison}, {Patton}, {Simard}  \&
  {McConnachie}}{{Ellison} et~al.}{2008}]{ellison2008}
{Ellison} S.~L.,  {Patton} D.~R.,  {Simard} L.,   {McConnachie} A.~W.,  2008,
  \mn@doi [\apjl] {10.1086/527296}, \href
  {https://ui.adsabs.harvard.edu/abs/2008ApJ...672L.107E} {672, L107}

\bibitem[\protect\citeauthoryear{{Forbes}, {Alabi}, {Romanowsky}, {Brodie}  \&
  {Arimoto}}{{Forbes} et~al.}{2020}]{forbes2020}
{Forbes} D.~A.,  {Alabi} A.,  {Romanowsky} A.~J.,  {Brodie} J.~P.,   {Arimoto}
  N.,  2020, \mn@doi [\mnras] {10.1093/mnras/staa180}, \href
  {https://ui.adsabs.harvard.edu/abs/2020MNRAS.tmp..175F} {p.~175}

\bibitem[\protect\citeauthoryear{{Freundlich}, {Dekel}, {Jiang}, {Ishai},
  {Cornuault}, {Lapiner}, {Dutton}  \& {Macci{\`o}}}{{Freundlich}
  et~al.}{2020}]{freundlich2020}
{Freundlich} J.,  {Dekel} A.,  {Jiang} F.,  {Ishai} G.,  {Cornuault} N.,
  {Lapiner} S.,  {Dutton} A.~A.,   {Macci{\`o}} A.~V.,  2020, \mn@doi [\mnras]
  {10.1093/mnras/stz3306}, \href
  {https://ui.adsabs.harvard.edu/abs/2020MNRAS.491.4523F} {491, 4523}

\bibitem[\protect\citeauthoryear{{Garrison-Kimmel} et~al.,}{{Garrison-Kimmel}
  et~al.}{2019}]{garrisonkimmel2019}
{Garrison-Kimmel} S.,  et~al., 2019, \mn@doi [\mnras] {10.1093/mnras/stz1317},
  \href {https://ui.adsabs.harvard.edu/abs/2019MNRAS.487.1380G} {487, 1380}

\bibitem[\protect\citeauthoryear{{Geha}, {Blanton}, {Yan}  \& {Tinker}}{{Geha}
  et~al.}{2012}]{geha2012}
{Geha} M.,  {Blanton} M.~R.,  {Yan} R.,   {Tinker} J.~L.,  2012, \mn@doi [\apj]
  {10.1088/0004-637X/757/1/85}, \href
  {https://ui.adsabs.harvard.edu/abs/2012ApJ...757...85G} {757, 85}

\bibitem[\protect\citeauthoryear{{Gill}, {Knebe}  \& {Gibson}}{{Gill}
  et~al.}{2004}]{gill2004}
{Gill} S. P.~D.,  {Knebe} A.,   {Gibson} B.~K.,  2004, \mn@doi [\mnras]
  {10.1111/j.1365-2966.2004.07786.x}, \href
  {https://ui.adsabs.harvard.edu/abs/2004MNRAS.351..399G} {351, 399}

\bibitem[\protect\citeauthoryear{{Girardi} et~al.,}{{Girardi}
  et~al.}{2010}]{Girardi2010}
{Girardi} L.,  et~al., 2010, \mn@doi [\apj] {10.1088/0004-637X/724/2/1030},
  \href {https://ui.adsabs.harvard.edu/abs/2010ApJ...724.1030G} {724, 1030}

\bibitem[\protect\citeauthoryear{{Gnedin}}{{Gnedin}}{2003}]{gnedin2003}
{Gnedin} O.~Y.,  2003, \mn@doi [\apj] {10.1086/374774}, \href
  {https://ui.adsabs.harvard.edu/abs/2003ApJ...589..752G} {589, 752}

\bibitem[\protect\citeauthoryear{{Governato} et~al.,}{{Governato}
  et~al.}{2010}]{governato2010}
{Governato} F.,  et~al., 2010, \mn@doi [\nat] {10.1038/nature08640}, \href
  {https://ui.adsabs.harvard.edu/abs/2010Natur.463..203G} {463, 203}

\bibitem[\protect\citeauthoryear{{Governato} et~al.,}{{Governato}
  et~al.}{2015}]{governato2015}
{Governato} F.,  et~al., 2015, \mn@doi [\mnras] {10.1093/mnras/stu2720}, \href
  {https://ui.adsabs.harvard.edu/abs/2015MNRAS.448..792G} {448, 792}

\bibitem[\protect\citeauthoryear{{Greco} et~al.,}{{Greco}
  et~al.}{2018a}]{greco2018sumo}
{Greco} J.~P.,  et~al., 2018a, \mn@doi [\pasj] {10.1093/pasj/psx051}, \href
  {https://ui.adsabs.harvard.edu/abs/2018PASJ...70S..19G} {70, S19}

\bibitem[\protect\citeauthoryear{{Greco} et~al.,}{{Greco}
  et~al.}{2018b}]{greco2018illuminating}
{Greco} J.~P.,  et~al., 2018b, \mn@doi [\apj] {10.3847/1538-4357/aab842}, \href
  {https://ui.adsabs.harvard.edu/abs/2018ApJ...857..104G} {857, 104}

\bibitem[\protect\citeauthoryear{{Greco}, {Goulding}, {Greene}, {Strauss},
  {Huang}, {Kim}  \& {Komiyama}}{{Greco} et~al.}{2018c}]{greco2018study}
{Greco} J.~P.,  {Goulding} A.~D.,  {Greene} J.~E.,  {Strauss} M.~A.,  {Huang}
  S.,  {Kim} J.~H.,   {Komiyama} Y.,  2018c, \mn@doi [\apj]
  {10.3847/1538-4357/aae0f4}, \href
  {https://ui.adsabs.harvard.edu/abs/2018ApJ...866..112G} {866, 112}

\bibitem[\protect\citeauthoryear{{Greco}, {van Dokkum}, {Danieli}, {Carlsten}
  \& {Conroy}}{{Greco} et~al.}{2020}]{greco2020}
{Greco} J.~P.,  {van Dokkum} P.,  {Danieli} S.,  {Carlsten} S.~G.,   {Conroy}
  C.,  2020, arXiv e-prints, \href
  {https://ui.adsabs.harvard.edu/abs/2020arXiv200407273G} {p. arXiv:2004.07273}

\bibitem[\protect\citeauthoryear{{Haardt} \& {Madau}}{{Haardt} \&
  {Madau}}{2012}]{haardt2012}
{Haardt} F.,  {Madau} P.,  2012, \mn@doi [\apj] {10.1088/0004-637X/746/2/125},
  \href {https://ui.adsabs.harvard.edu/abs/2012ApJ...746..125H} {746, 125}

\bibitem[\protect\citeauthoryear{{Hagen} et~al.,}{{Hagen}
  et~al.}{2016}]{hagen2016}
{Hagen} L. M.~Z.,  et~al., 2016, \mn@doi [\apj] {10.3847/0004-637X/826/2/210},
  \href {https://ui.adsabs.harvard.edu/abs/2016ApJ...826..210H} {826, 210}

\bibitem[\protect\citeauthoryear{{He}, {Wu}, {Du}, {Wicker}, {Zhao}, {Lei}  \&
  {Liu}}{{He} et~al.}{2019}]{he2019}
{He} M.,  {Wu} H.,  {Du} W.,  {Wicker} J.,  {Zhao} P.,  {Lei} F.,   {Liu} J.,
  2019, \mn@doi [\apj] {10.3847/1538-4357/ab2710}, \href
  {https://ui.adsabs.harvard.edu/abs/2019ApJ...880...30H} {880, 30}

\bibitem[\protect\citeauthoryear{{Hernquist}}{{Hernquist}}{1989}]{hernquist1989}
{Hernquist} L.,  1989, \mn@doi [\nat] {10.1038/340687a0}, \href
  {https://ui.adsabs.harvard.edu/abs/1989Natur.340..687H} {340, 687}

\bibitem[\protect\citeauthoryear{{Hetznecker} \& {Burkert}}{{Hetznecker} \&
  {Burkert}}{2006}]{hetznecker2006}
{Hetznecker} H.,  {Burkert} A.,  2006, \mn@doi [\mnras]
  {10.1111/j.1365-2966.2006.10616.x}, \href
  {https://ui.adsabs.harvard.edu/abs/2006MNRAS.370.1905H} {370, 1905}

\bibitem[\protect\citeauthoryear{{Hoyle}}{{Hoyle}}{1951}]{hoyle1951}
{Hoyle} F.,  1951, in Problems of Cosmical Aerodynamics. p.~195

\bibitem[\protect\citeauthoryear{{Hunter}}{{Hunter}}{2007}]{hunter2007}
{Hunter} J.~D.,  2007, \mn@doi [Computing in Science and Engineering]
  {10.1109/MCSE.2007.55}, \href
  {https://ui.adsabs.harvard.edu/abs/2007CSE.....9...90H} {9, 90}

\bibitem[\protect\citeauthoryear{{Hunter} \& {Elmegreen}}{{Hunter} \&
  {Elmegreen}}{2006}]{hunter2006}
{Hunter} D.~A.,  {Elmegreen} B.~G.,  2006, \mn@doi [\apjs] {10.1086/498096},
  \href {https://ui.adsabs.harvard.edu/abs/2006ApJS..162...49H} {162, 49}

\bibitem[\protect\citeauthoryear{{Hunter} \& {Hoffman}}{{Hunter} \&
  {Hoffman}}{1999}]{hunter1999}
{Hunter} D.~A.,  {Hoffman} L.,  1999, \mn@doi [\aj] {10.1086/300885}, \href
  {https://ui.adsabs.harvard.edu/abs/1999AJ....117.2789H} {117, 2789}

\bibitem[\protect\citeauthoryear{{Impey}, {Bothun}  \& {Malin}}{{Impey}
  et~al.}{1988}]{impey1988}
{Impey} C.,  {Bothun} G.,   {Malin} D.,  1988, \mn@doi [\apj] {10.1086/166500},
  \href {https://ui.adsabs.harvard.edu/abs/1988ApJ...330..634I} {330, 634}

\bibitem[\protect\citeauthoryear{{Jackson} et~al.,}{{Jackson}
  et~al.}{2020}]{jackson2020}
{Jackson} R.~A.,  et~al., 2020, arXiv e-prints, \href
  {https://ui.adsabs.harvard.edu/abs/2020arXiv200706581J} {p. arXiv:2007.06581}

\bibitem[\protect\citeauthoryear{{Janowiecki}, {Jones}, {Leisman}  \&
  {Webb}}{{Janowiecki} et~al.}{2019}]{janowiecki2019}
{Janowiecki} S.,  {Jones} M.~G.,  {Leisman} L.,   {Webb} A.,  2019, \mn@doi
  [\mnras] {10.1093/mnras/stz1868}, \href
  {https://ui.adsabs.harvard.edu/abs/2019MNRAS.490..566J} {490, 566}

\bibitem[\protect\citeauthoryear{{Jiang}, {Dekel}, {Freundlich}, {Romanowsky},
  {Dutton}, {Macci{\`o}}  \& {Di Cintio}}{{Jiang} et~al.}{2019a}]{jiang2019}
{Jiang} F.,  {Dekel} A.,  {Freundlich} J.,  {Romanowsky} A.~J.,  {Dutton}
  A.~A.,  {Macci{\`o}} A.~V.,   {Di Cintio} A.,  2019a, \mn@doi [\mnras]
  {10.1093/mnras/stz1499}, \href
  {https://ui.adsabs.harvard.edu/abs/2019MNRAS.487.5272J} {487, 5272}

\bibitem[\protect\citeauthoryear{{Jiang} et~al.,}{{Jiang}
  et~al.}{2019b}]{jiang2019size}
{Jiang} F.,  et~al., 2019b, \mn@doi [\mnras] {10.1093/mnras/stz1952}, \href
  {https://ui.adsabs.harvard.edu/abs/2019MNRAS.488.4801J} {488, 4801}

\bibitem[\protect\citeauthoryear{{Jones}, {Papastergis}, {Pandya}, {Leisman},
  {Romanowsky}, {Yung}, {Somerville}  \& {Adams}}{{Jones}
  et~al.}{2018}]{jones2018}
{Jones} M.~G.,  {Papastergis} E.,  {Pandya} V.,  {Leisman} L.,  {Romanowsky}
  A.~J.,  {Yung} L.~Y.~A.,  {Somerville} R.~S.,   {Adams} E.~A.~K.,  2018,
  \mn@doi [\aap] {10.1051/0004-6361/201732409}, \href
  {https://ui.adsabs.harvard.edu/abs/2018A&A...614A..21J} {614, A21}

\bibitem[\protect\citeauthoryear{{Kadowaki}, {Zaritsky}  \&
  {Donnerstein}}{{Kadowaki} et~al.}{2017}]{kadowaki2017}
{Kadowaki} J.,  {Zaritsky} D.,   {Donnerstein} R.~L.,  2017, \mn@doi [\apjl]
  {10.3847/2041-8213/aa653d}, \href
  {https://ui.adsabs.harvard.edu/abs/2017ApJ...838L..21K} {838, L21}

\bibitem[\protect\citeauthoryear{{Kennicutt}, {Tamblyn}  \&
  {Congdon}}{{Kennicutt} et~al.}{1994}]{kennicutt1994}
{Kennicutt} Robert~C. J.,  {Tamblyn} P.,   {Congdon} C.~E.,  1994, \mn@doi
  [\apj] {10.1086/174790}, \href
  {https://ui.adsabs.harvard.edu/abs/1994ApJ...435...22K} {435, 22}

\bibitem[\protect\citeauthoryear{{Knebe}, {Green}  \& {Binney}}{{Knebe}
  et~al.}{2001}]{knebe2001}
{Knebe} A.,  {Green} A.,   {Binney} J.,  2001, \mn@doi [\mnras]
  {10.1046/j.1365-8711.2001.04532.x}, \href
  {https://ui.adsabs.harvard.edu/abs/2001MNRAS.325..845K} {325, 845}

\bibitem[\protect\citeauthoryear{{Koda}, {Yagi}, {Yamanoi}  \&
  {Komiyama}}{{Koda} et~al.}{2015}]{koda2015}
{Koda} J.,  {Yagi} M.,  {Yamanoi} H.,   {Komiyama} Y.,  2015, \mn@doi [\apjl]
  {10.1088/2041-8205/807/1/L2}, \href
  {https://ui.adsabs.harvard.edu/abs/2015ApJ...807L...2K} {807, L2}

\bibitem[\protect\citeauthoryear{{Kormendy} \& {Bahcall}}{{Kormendy} \&
  {Bahcall}}{1974}]{kormendy1974}
{Kormendy} J.,  {Bahcall} J.~N.,  1974, \mn@doi [\aj] {10.1086/111595}, \href
  {https://ui.adsabs.harvard.edu/abs/1974AJ.....79..671K} {79, 671}

\bibitem[\protect\citeauthoryear{{Kov{\'a}cs}, {Bogd{\'a}n}  \&
  {Canning}}{{Kov{\'a}cs} et~al.}{2019}]{kovacs2019}
{Kov{\'a}cs} O.~E.,  {Bogd{\'a}n} {\'A}.,   {Canning} R. E.~A.,  2019, \mn@doi
  [\apjl] {10.3847/2041-8213/ab2916}, \href
  {https://ui.adsabs.harvard.edu/abs/2019ApJ...879L..12K} {879, L12}

\bibitem[\protect\citeauthoryear{{Kravtsov}, {Vikhlinin}  \&
  {Meshcheryakov}}{{Kravtsov} et~al.}{2018}]{kravtsov2018}
{Kravtsov} A.~V.,  {Vikhlinin} A.~A.,   {Meshcheryakov} A.~V.,  2018, \mn@doi
  [Astronomy Letters] {10.1134/S1063773717120015}, \href
  {https://ui.adsabs.harvard.edu/abs/2018AstL...44....8K} {44, 8}

\bibitem[\protect\citeauthoryear{{Kroupa}}{{Kroupa}}{2001}]{kroupa2001}
{Kroupa} P.,  2001, \mn@doi [\mnras] {10.1046/j.1365-8711.2001.04022.x}, \href
  {https://ui.adsabs.harvard.edu/abs/2001MNRAS.322..231K} {322, 231}

\bibitem[\protect\citeauthoryear{{Kulier}, {Galaz}, {Padilla}  \&
  {Trayford}}{{Kulier} et~al.}{2020}]{kulier2020}
{Kulier} A.,  {Galaz} G.,  {Padilla} N.~D.,   {Trayford} J.~W.,  2020, \mn@doi
  [\mnras] {10.1093/mnras/staa1798}, \href
  {https://ui.adsabs.harvard.edu/abs/2020MNRAS.496.3996K} {496, 3996}

\bibitem[\protect\citeauthoryear{{Lange} et~al.,}{{Lange}
  et~al.}{2015}]{lange2015}
{Lange} R.,  et~al., 2015, \mn@doi [\mnras] {10.1093/mnras/stu2467}, \href
  {https://ui.adsabs.harvard.edu/abs/2015MNRAS.447.2603L} {447, 2603}

\bibitem[\protect\citeauthoryear{{Lange} et~al.,}{{Lange}
  et~al.}{2016}]{lange2016}
{Lange} R.,  et~al., 2016, \mn@doi [\mnras] {10.1093/mnras/stw1495}, \href
  {https://ui.adsabs.harvard.edu/abs/2016MNRAS.462.1470L} {462, 1470}

\bibitem[\protect\citeauthoryear{{Lee}, {Kang}, {Lee}  \& {Jang}}{{Lee}
  et~al.}{2017}]{lee2017}
{Lee} M.~G.,  {Kang} J.,  {Lee} J.~H.,   {Jang} I.~S.,  2017, \mn@doi [\apj]
  {10.3847/1538-4357/aa78fb}, \href
  {https://ui.adsabs.harvard.edu/abs/2017ApJ...844..157L} {844, 157}

\bibitem[\protect\citeauthoryear{{Leisman} et~al.,}{{Leisman}
  et~al.}{2017}]{leisman2017}
{Leisman} L.,  et~al., 2017, \mn@doi [\apj] {10.3847/1538-4357/aa7575}, \href
  {https://ui.adsabs.harvard.edu/abs/2017ApJ...842..133L} {842, 133}

\bibitem[\protect\citeauthoryear{{Liao} et~al.,}{{Liao}
  et~al.}{2019}]{liao2019}
{Liao} S.,  et~al., 2019, \mn@doi [\mnras] {10.1093/mnras/stz2969}, \href
  {https://ui.adsabs.harvard.edu/abs/2019MNRAS.490.5182L} {490, 5182}

\bibitem[\protect\citeauthoryear{{Lim}, {Peng}, {C{\^o}t{\'e}}, {Sales}, {den
  Brok}, {Blakeslee}  \& {Guhathakurta}}{{Lim} et~al.}{2018}]{lim2018}
{Lim} S.,  {Peng} E.~W.,  {C{\^o}t{\'e}} P.,  {Sales} L.~V.,  {den Brok} M.,
  {Blakeslee} J.~P.,   {Guhathakurta} P.,  2018, \mn@doi [\apj]
  {10.3847/1538-4357/aacb81}, \href
  {https://ui.adsabs.harvard.edu/abs/2018ApJ...862...82L} {862, 82}

\bibitem[\protect\citeauthoryear{{Ludlow}, {Schaye}, {Schaller}  \&
  {Richings}}{{Ludlow} et~al.}{2019}]{ludlow2019}
{Ludlow} A.~D.,  {Schaye} J.,  {Schaller} M.,   {Richings} J.,  2019, \mn@doi
  [\mnras] {10.1093/mnrasl/slz110}, \href
  {https://ui.adsabs.harvard.edu/abs/2019MNRAS.488L.123L} {488, L123}

\bibitem[\protect\citeauthoryear{{Malin}}{{Malin}}{1978}]{malin1978}
{Malin} D.~F.,  1978, \mn@doi [\nat] {10.1038/276591a0}, \href
  {https://ui.adsabs.harvard.edu/abs/1978Natur.276..591M} {276, 591}

\bibitem[\protect\citeauthoryear{{Mancera Pi{\~n}a}, {Aguerri}, {Peletier},
  {Venhola}, {Trager}  \& {Choque Challapa}}{{Mancera Pi{\~n}a}
  et~al.}{2019a}]{mancerapina2019weave}
{Mancera Pi{\~n}a} P.~E.,  {Aguerri} J.~A.~L.,  {Peletier} R.~F.,  {Venhola}
  A.,  {Trager} S.,   {Choque Challapa} N.,  2019a, \mn@doi [\mnras]
  {10.1093/mnras/stz238}, \href
  {https://ui.adsabs.harvard.edu/abs/2019MNRAS.485.1036M} {485, 1036}

\bibitem[\protect\citeauthoryear{{Mancera Pi{\~n}a} et~al.,}{{Mancera Pi{\~n}a}
  et~al.}{2019b}]{mancerapina2019btfr}
{Mancera Pi{\~n}a} P.~E.,  et~al., 2019b, \mn@doi [\apjl]
  {10.3847/2041-8213/ab40c7}, \href
  {https://ui.adsabs.harvard.edu/abs/2019ApJ...883L..33M} {883, L33}

\bibitem[\protect\citeauthoryear{{Mancera Pi{\~n}a} et~al.,}{{Mancera Pi{\~n}a}
  et~al.}{2020}]{mancerapina2020}
{Mancera Pi{\~n}a} P.~E.,  et~al., 2020, \mn@doi [\mnras]
  {10.1093/mnras/staa1256}, \href
  {https://ui.adsabs.harvard.edu/abs/2020MNRAS.495.3636M} {495, 3636}

\bibitem[\protect\citeauthoryear{{Mapelli}, {Moore}, {Ripamonti}, {Mayer},
  {Colpi}  \& {Giordano}}{{Mapelli} et~al.}{2008}]{mapelli2008}
{Mapelli} M.,  {Moore} B.,  {Ripamonti} E.,  {Mayer} L.,  {Colpi} M.,
  {Giordano} L.,  2008, \mn@doi [\mnras] {10.1111/j.1365-2966.2007.12650.x},
  \href {https://ui.adsabs.harvard.edu/abs/2008MNRAS.383.1223M} {383, 1223}

\bibitem[\protect\citeauthoryear{{Marigo}, {Girardi}, {Bressan}, {Groenewegen},
  {Silva}  \& {Granato}}{{Marigo} et~al.}{2008}]{Marigo2008}
{Marigo} P.,  {Girardi} L.,  {Bressan} A.,  {Groenewegen} M.~A.~T.,  {Silva}
  L.,   {Granato} G.~L.,  2008, \mn@doi [\aap] {10.1051/0004-6361:20078467},
  \href {https://ui.adsabs.harvard.edu/abs/2008A&A...482..883M} {482, 883}

\bibitem[\protect\citeauthoryear{{Mart{\'\i}n-Navarro}
  et~al.,}{{Mart{\'\i}n-Navarro} et~al.}{2019}]{martinnavarro2019}
{Mart{\'\i}n-Navarro} I.,  et~al., 2019, \mn@doi [\mnras]
  {10.1093/mnras/stz252}, \href
  {https://ui.adsabs.harvard.edu/abs/2019MNRAS.484.3425M} {484, 3425}

\bibitem[\protect\citeauthoryear{{Martin} et~al.,}{{Martin}
  et~al.}{2019}]{martin2019}
{Martin} G.,  et~al., 2019, \mn@doi [\mnras] {10.1093/mnras/stz356}, \href
  {https://ui.adsabs.harvard.edu/abs/2019MNRAS.485..796M} {485, 796}

\bibitem[\protect\citeauthoryear{{Martin} et~al.,}{{Martin}
  et~al.}{2020}]{martin2020}
{Martin} G.,  et~al., 2020, \mn@doi [\mnras] {10.1093/mnras/staa3443}, \href
  {https://ui.adsabs.harvard.edu/abs/2020MNRAS.tmp.3262M} {}

\bibitem[\protect\citeauthoryear{{Mart{\'\i}nez-Delgado}
  et~al.,}{{Mart{\'\i}nez-Delgado} et~al.}{2016}]{martinezdelgado2016}
{Mart{\'\i}nez-Delgado} D.,  et~al., 2016, \mn@doi [\aj]
  {10.3847/0004-6256/151/4/96}, \href
  {https://ui.adsabs.harvard.edu/abs/2016AJ....151...96M} {151, 96}

\bibitem[\protect\citeauthoryear{{McGaugh}}{{McGaugh}}{1994}]{mcgaugh1994}
{McGaugh} S.~S.,  1994, \mn@doi [\apj] {10.1086/174049}, \href
  {https://ui.adsabs.harvard.edu/abs/1994ApJ...426..135M} {426, 135}

\bibitem[\protect\citeauthoryear{{Menon}, {Wesolowski}, {Zheng}, {Jetley},
  {Kale}, {Quinn}  \& {Governato}}{{Menon} et~al.}{2015}]{menon2015}
{Menon} H.,  {Wesolowski} L.,  {Zheng} G.,  {Jetley} P.,  {Kale} L.,  {Quinn}
  T.,   {Governato} F.,  2015, \mn@doi [Computational Astrophysics and
  Cosmology] {10.1186/s40668-015-0007-9}, \href
  {https://ui.adsabs.harvard.edu/abs/2015ComAC...2....1M} {2, 1}

\bibitem[\protect\citeauthoryear{{Merritt}, {van Dokkum}, {Danieli}, {Abraham},
  {Zhang}, {Karachentsev}  \& {Makarova}}{{Merritt} et~al.}{2016}]{merritt2016}
{Merritt} A.,  {van Dokkum} P.,  {Danieli} S.,  {Abraham} R.,  {Zhang} J.,
  {Karachentsev} I.~D.,   {Makarova} L.~N.,  2016, \mn@doi [\apj]
  {10.3847/1538-4357/833/2/168}, \href
  {https://ui.adsabs.harvard.edu/abs/2016ApJ...833..168M} {833, 168}

\bibitem[\protect\citeauthoryear{{Mihos} \& {Hernquist}}{{Mihos} \&
  {Hernquist}}{1996}]{mihos1996}
{Mihos} J.~C.,  {Hernquist} L.,  1996, \mn@doi [\apj] {10.1086/177353}, \href
  {https://ui.adsabs.harvard.edu/abs/1996ApJ...464..641M} {464, 641}

\bibitem[\protect\citeauthoryear{{Mihos} et~al.,}{{Mihos}
  et~al.}{2015}]{mihos2015}
{Mihos} J.~C.,  et~al., 2015, \mn@doi [\apjl] {10.1088/2041-8205/809/2/L21},
  \href {https://ui.adsabs.harvard.edu/abs/2015ApJ...809L..21M} {809, L21}

\bibitem[\protect\citeauthoryear{{Mihos}, {Harding}, {Feldmeier}, {Rudick},
  {Janowiecki}, {Morrison}, {Slater}  \& {Watkins}}{{Mihos}
  et~al.}{2017}]{mihos2017}
{Mihos} J.~C.,  {Harding} P.,  {Feldmeier} J.~J.,  {Rudick} C.,  {Janowiecki}
  S.,  {Morrison} H.,  {Slater} C.,   {Watkins} A.,  2017, \mn@doi [\apj]
  {10.3847/1538-4357/834/1/16}, \href
  {https://ui.adsabs.harvard.edu/abs/2017ApJ...834...16M} {834, 16}

\bibitem[\protect\citeauthoryear{{Moore}, {Katz}, {Lake}, {Dressler}  \&
  {Oemler}}{{Moore} et~al.}{1996}]{moore1996}
{Moore} B.,  {Katz} N.,  {Lake} G.,  {Dressler} A.,   {Oemler} A.,  1996,
  \mn@doi [\nat] {10.1038/379613a0}, \href
  {https://ui.adsabs.harvard.edu/abs/1996Natur.379..613M} {379, 613}

\bibitem[\protect\citeauthoryear{{Moster}, {Naab}  \& {White}}{{Moster}
  et~al.}{2013}]{Moster2013}
{Moster} B.~P.,  {Naab} T.,   {White} S. D.~M.,  2013, \mn@doi [\mnras]
  {10.1093/mnras/sts261}, \href
  {https://ui.adsabs.harvard.edu/abs/2013MNRAS.428.3121M} {428, 3121}

\bibitem[\protect\citeauthoryear{{Mowla}, {van Dokkum}, {Merritt}, {Abraham},
  {Yagi}  \& {Koda}}{{Mowla} et~al.}{2017}]{mowla2017}
{Mowla} L.,  {van Dokkum} P.,  {Merritt} A.,  {Abraham} R.,  {Yagi} M.,
  {Koda} J.,  2017, \mn@doi [\apj] {10.3847/1538-4357/aa961b}, \href
  {https://ui.adsabs.harvard.edu/abs/2017ApJ...851...27M} {851, 27}

\bibitem[\protect\citeauthoryear{{Munshi} et~al.,}{{Munshi}
  et~al.}{2013}]{munshi2013}
{Munshi} F.,  et~al., 2013, \mn@doi [\apj] {10.1088/0004-637X/766/1/56}, \href
  {https://ui.adsabs.harvard.edu/abs/2013ApJ...766...56M} {766, 56}

\bibitem[\protect\citeauthoryear{{Noguchi}}{{Noguchi}}{1988}]{noguchi1988}
{Noguchi} M.,  1988, \aap, \href
  {https://ui.adsabs.harvard.edu/abs/1988A&A...203..259N} {203, 259}

\bibitem[\protect\citeauthoryear{{Ogiya}}{{Ogiya}}{2018}]{ogiya2018}
{Ogiya} G.,  2018, \mn@doi [\mnras] {10.1093/mnrasl/sly138}, \href
  {https://ui.adsabs.harvard.edu/abs/2018MNRAS.480L.106O} {480, L106}

\bibitem[\protect\citeauthoryear{{Papastergis}, {Adams}  \&
  {Romanowsky}}{{Papastergis} et~al.}{2017}]{papastergis2017}
{Papastergis} E.,  {Adams} E.~A.~K.,   {Romanowsky} A.~J.,  2017, \mn@doi
  [\aap] {10.1051/0004-6361/201730795}, \href
  {https://ui.adsabs.harvard.edu/abs/2017A&A...601L..10P} {601, L10}

\bibitem[\protect\citeauthoryear{{Patra}, {Chengalur}, {Karachentsev}  \&
  {Sharina}}{{Patra} et~al.}{2016}]{patra2016}
{Patra} N.~N.,  {Chengalur} J.~N.,  {Karachentsev} I.~D.,   {Sharina} M.~E.,
  2016, \mn@doi [Astrophysical Bulletin] {10.1134/S1990341316040040}, \href
  {https://ui.adsabs.harvard.edu/abs/2016AstBu..71..408P} {71, 408}

\bibitem[\protect\citeauthoryear{{Pe{\~n}arrubia}, {McConnachie}  \&
  {Babul}}{{Pe{\~n}arrubia} et~al.}{2006}]{penarrubia2006}
{Pe{\~n}arrubia} J.,  {McConnachie} A.,   {Babul} A.,  2006, \mn@doi [\apjl]
  {10.1086/508656}, \href
  {https://ui.adsabs.harvard.edu/abs/2006ApJ...650L..33P} {650, L33}

\bibitem[\protect\citeauthoryear{{Peebles}}{{Peebles}}{1969}]{peebles1969}
{Peebles} P.~J.~E.,  1969, \mn@doi [\apj] {10.1086/149876}, \href
  {https://ui.adsabs.harvard.edu/abs/1969ApJ...155..393P} {155, 393}

\bibitem[\protect\citeauthoryear{{Peng} \& {Lim}}{{Peng} \&
  {Lim}}{2016}]{peng2016}
{Peng} E.~W.,  {Lim} S.,  2016, \mn@doi [\apjl] {10.3847/2041-8205/822/2/L31},
  \href {https://ui.adsabs.harvard.edu/abs/2016ApJ...822L..31P} {822, L31}

\bibitem[\protect\citeauthoryear{{Penny} et~al.,}{{Penny}
  et~al.}{2016}]{penny2016}
{Penny} S.~J.,  et~al., 2016, \mn@doi [\mnras] {10.1093/mnras/stw1913}, \href
  {https://ui.adsabs.harvard.edu/abs/2016MNRAS.462.3955P} {462, 3955}

\bibitem[\protect\citeauthoryear{{Pillepich} et~al.,}{{Pillepich}
  et~al.}{2019}]{pillepich2019}
{Pillepich} A.,  et~al., 2019, \mn@doi [\mnras] {10.1093/mnras/stz2338}, \href
  {https://ui.adsabs.harvard.edu/abs/2019MNRAS.490.3196P} {490, 3196}

\bibitem[\protect\citeauthoryear{{Planck Collaboration} et~al.,}{{Planck
  Collaboration} et~al.}{2014}]{planck2014}
{Planck Collaboration} et~al., 2014, \mn@doi [\aap]
  {10.1051/0004-6361/201321591}, \href
  {https://ui.adsabs.harvard.edu/abs/2014A&A...571A..16P} {571, A16}

\bibitem[\protect\citeauthoryear{{Pontzen} \& {Governato}}{{Pontzen} \&
  {Governato}}{2012}]{pontzen2012}
{Pontzen} A.,  {Governato} F.,  2012, \mn@doi [\mnras]
  {10.1111/j.1365-2966.2012.20571.x}, \href
  {https://ui.adsabs.harvard.edu/abs/2012MNRAS.421.3464P} {421, 3464}

\bibitem[\protect\citeauthoryear{{Pontzen} \& {Tremmel}}{{Pontzen} \&
  {Tremmel}}{2018}]{pontzen2018}
{Pontzen} A.,  {Tremmel} M.,  2018, \mn@doi [\apjs] {10.3847/1538-4365/aac832},
  \href {https://ui.adsabs.harvard.edu/abs/2018ApJS..237...23P} {237, 23}

\bibitem[\protect\citeauthoryear{{Pontzen} et~al.,}{{Pontzen}
  et~al.}{2008}]{pontzen2008}
{Pontzen} A.,  et~al., 2008, \mn@doi [\mnras]
  {10.1111/j.1365-2966.2008.13782.x}, \href
  {https://ui.adsabs.harvard.edu/abs/2008MNRAS.390.1349P} {390, 1349}

\bibitem[\protect\citeauthoryear{{Pontzen}, {Ro{\v s}kar}, {Stinson}, {Woods},
  {Reed}, {Coles}  \& {Quinn}}{{Pontzen} et~al.}{2013}]{pynbody}
{Pontzen} A.,  {Ro{\v s}kar} R.,  {Stinson} G.~S.,  {Woods} R.,  {Reed} D.~M.,
  {Coles} J.,   {Quinn} T.~R.,  2013, {pynbody: Astrophysics Simulation
  Analysis for Python}

\bibitem[\protect\citeauthoryear{{Privon} et~al.,}{{Privon}
  et~al.}{2017}]{privon2017}
{Privon} G.~C.,  et~al., 2017, \mn@doi [\apj] {10.3847/1538-4357/aa8560}, \href
  {https://ui.adsabs.harvard.edu/abs/2017ApJ...846...74P} {846, 74}

\bibitem[\protect\citeauthoryear{{Prole}, {van der Burg}, {Hilker}  \&
  {Davies}}{{Prole} et~al.}{2019}]{prole2019}
{Prole} D.~J.,  {van der Burg} R.~F.~J.,  {Hilker} M.,   {Davies} J.~I.,  2019,
  \mn@doi [\mnras] {10.1093/mnras/stz1843}, \href
  {https://ui.adsabs.harvard.edu/abs/2019MNRAS.488.2143P} {488, 2143}

\bibitem[\protect\citeauthoryear{{Rasmussen}, {Mulchaey}, {Bai}, {Ponman},
  {Raychaudhury}  \& {Dariush}}{{Rasmussen} et~al.}{2012}]{rasmussen2012}
{Rasmussen} J.,  {Mulchaey} J.~S.,  {Bai} L.,  {Ponman} T.~J.,  {Raychaudhury}
  S.,   {Dariush} A.,  2012, \mn@doi [\apj] {10.1088/0004-637X/757/2/122},
  \href {https://ui.adsabs.harvard.edu/abs/2012ApJ...757..122R} {757, 122}

\bibitem[\protect\citeauthoryear{{Reshetnikov}, {Moiseev}  \&
  {Sotnikova}}{{Reshetnikov} et~al.}{2010}]{reshetnikov2010}
{Reshetnikov} V.~P.,  {Moiseev} A.~V.,   {Sotnikova} N.~Y.,  2010, \mn@doi
  [\mnras] {10.1111/j.1745-3933.2010.00888.x}, \href
  {https://ui.adsabs.harvard.edu/abs/2010MNRAS.406L..90R} {406, L90}

\bibitem[\protect\citeauthoryear{{Rom{\'a}n} \& {Trujillo}}{{Rom{\'a}n} \&
  {Trujillo}}{2017}]{roman2017}
{Rom{\'a}n} J.,  {Trujillo} I.,  2017, \mn@doi [\mnras] {10.1093/mnras/stx694},
  \href {https://ui.adsabs.harvard.edu/abs/2017MNRAS.468.4039R} {468, 4039}

\bibitem[\protect\citeauthoryear{{Rong}, {Guo}, {Gao}, {Liao}, {Xie}, {Puzia},
  {Sun}  \& {Pan}}{{Rong} et~al.}{2017}]{rong2017}
{Rong} Y.,  {Guo} Q.,  {Gao} L.,  {Liao} S.,  {Xie} L.,  {Puzia} T.~H.,  {Sun}
  S.,   {Pan} J.,  2017, \mn@doi [\mnras] {10.1093/mnras/stx1440}, \href
  {https://ui.adsabs.harvard.edu/abs/2017MNRAS.470.4231R} {470, 4231}

\bibitem[\protect\citeauthoryear{{Ro{\v{s}}kar}, {Debattista}, {Brooks},
  {Quinn}, {Brook}, {Governato}, {Dalcanton}  \& {Wadsley}}{{Ro{\v{s}}kar}
  et~al.}{2010}]{roskar2010}
{Ro{\v{s}}kar} R.,  {Debattista} V.~P.,  {Brooks} A.~M.,  {Quinn} T.~R.,
  {Brook} C.~B.,  {Governato} F.,  {Dalcanton} J.~J.,   {Wadsley} J.,  2010,
  \mn@doi [\mnras] {10.1111/j.1365-2966.2010.17178.x}, \href
  {https://ui.adsabs.harvard.edu/abs/2010MNRAS.408..783R} {408, 783}

\bibitem[\protect\citeauthoryear{{Saburova}, {Chilingarian}, {Katkov},
  {Egorov}, {Kasparova}, {Khoperskov}, {Uklein}  \& {Vozyakova}}{{Saburova}
  et~al.}{2018}]{saburova2018}
{Saburova} A.~S.,  {Chilingarian} I.~V.,  {Katkov} I.~Y.,  {Egorov} O.~V.,
  {Kasparova} A.~V.,  {Khoperskov} S.~A.,  {Uklein} R.~I.,   {Vozyakova} O.~V.,
   2018, \mn@doi [\mnras] {10.1093/mnras/sty2519}, \href
  {https://ui.adsabs.harvard.edu/abs/2018MNRAS.481.3534S} {481, 3534}

\bibitem[\protect\citeauthoryear{{Safarzadeh} \& {Scannapieco}}{{Safarzadeh} \&
  {Scannapieco}}{2017}]{safarzadeh2017}
{Safarzadeh} M.,  {Scannapieco} E.,  2017, \mn@doi [\apj]
  {10.3847/1538-4357/aa94c8}, \href
  {https://ui.adsabs.harvard.edu/abs/2017ApJ...850...99S} {850, 99}

\bibitem[\protect\citeauthoryear{{Sales}, {Navarro}, {Pe{\~n}afiel}, {Peng},
  {Lim}  \& {Hernquist}}{{Sales} et~al.}{2020}]{sales2020}
{Sales} L.~V.,  {Navarro} J.~F.,  {Pe{\~n}afiel} L.,  {Peng} E.~W.,  {Lim} S.,
   {Hernquist} L.,  2020, \mn@doi [\mnras] {10.1093/mnras/staa854}, \href
  {https://ui.adsabs.harvard.edu/abs/2020MNRAS.494.1848S} {494, 1848}

\bibitem[\protect\citeauthoryear{{Sandage} \& {Binggeli}}{{Sandage} \&
  {Binggeli}}{1984}]{sandage1984}
{Sandage} A.,  {Binggeli} B.,  1984, \mn@doi [\aj] {10.1086/113588}, \href
  {https://ui.adsabs.harvard.edu/abs/1984AJ.....89..919S} {89, 919}

\bibitem[\protect\citeauthoryear{{Santos-Santos}, {Di Cintio}, {Brook},
  {Macci{\`o}}, {Dutton}  \& {Dom{\'\i}nguez-Tenreiro}}{{Santos-Santos}
  et~al.}{2018}]{santossantos2018}
{Santos-Santos} I.~M.,  {Di Cintio} A.,  {Brook} C.~B.,  {Macci{\`o}} A.,
  {Dutton} A.,   {Dom{\'\i}nguez-Tenreiro} R.,  2018, \mn@doi [\mnras]
  {10.1093/mnras/stx2660}, \href
  {https://ui.adsabs.harvard.edu/abs/2018MNRAS.473.4392S} {473, 4392}

\bibitem[\protect\citeauthoryear{Schaye et~al.,}{Schaye
  et~al.}{2014}]{schaye2014}
Schaye J.,  et~al., 2014, \mn@doi [Monthly Notices of the Royal Astronomical
  Society] {10.1093/mnras/stu2058}, 446, 521

\bibitem[\protect\citeauthoryear{{Schramm} \& {Silverman}}{{Schramm} \&
  {Silverman}}{2013}]{schramm2013}
{Schramm} M.,  {Silverman} J.~D.,  2013, \mn@doi [\apj]
  {10.1088/0004-637X/767/1/13}, \href
  {https://ui.adsabs.harvard.edu/abs/2013ApJ...767...13S} {767, 13}

\bibitem[\protect\citeauthoryear{{Schwartzenberg}, {Phillipps}  \&
  {Parker}}{{Schwartzenberg} et~al.}{1995}]{schwartzenberg1995}
{Schwartzenberg} J.~M.,  {Phillipps} S.,   {Parker} Q.~A.,  1995, \aap, \href
  {https://ui.adsabs.harvard.edu/abs/1995A&A...293..332S} {293, 332}

\bibitem[\protect\citeauthoryear{{S{\'e}rsic}}{{S{\'e}rsic}}{1963}]{sersic1963}
{S{\'e}rsic} J.~L.,  1963, Boletin de la Asociacion Argentina de Astronomia La
  Plata Argentina, \href
  {https://ui.adsabs.harvard.edu/abs/1963BAAA....6...41S} {6, 41}

\bibitem[\protect\citeauthoryear{{Sharma}, {Brooks}, {Somerville}, {Tremmel},
  {Bellovary}, {Wright}  \& {Quinn}}{{Sharma} et~al.}{2020}]{sharma2020}
{Sharma} R.~S.,  {Brooks} A.~M.,  {Somerville} R.~S.,  {Tremmel} M.,
  {Bellovary} J.,  {Wright} A.~C.,   {Quinn} T.~R.,  2020, \mn@doi [\apj]
  {10.3847/1538-4357/ab960e}, \href
  {https://ui.adsabs.harvard.edu/abs/2020ApJ...897..103S} {897, 103}

\bibitem[\protect\citeauthoryear{{Shen}, {Wadsley}  \& {Stinson}}{{Shen}
  et~al.}{2010}]{shen2010}
{Shen} S.,  {Wadsley} J.,   {Stinson} G.,  2010, \mn@doi [\mnras]
  {10.1111/j.1365-2966.2010.17047.x}, \href
  {https://ui.adsabs.harvard.edu/abs/2010MNRAS.407.1581S} {407, 1581}

\bibitem[\protect\citeauthoryear{{Sif{\'o}n}, {van der Burg}, {Hoekstra},
  {Muzzin}  \& {Herbonnet}}{{Sif{\'o}n} et~al.}{2018}]{sifon2018}
{Sif{\'o}n} C.,  {van der Burg} R. F.~J.,  {Hoekstra} H.,  {Muzzin} A.,
  {Herbonnet} R.,  2018, \mn@doi [\mnras] {10.1093/mnras/stx2648}, \href
  {https://ui.adsabs.harvard.edu/abs/2018MNRAS.473.3747S} {473, 3747}

\bibitem[\protect\citeauthoryear{{Sinha} \& {Holley-Bockelmann}}{{Sinha} \&
  {Holley-Bockelmann}}{2015}]{sinha2015}
{Sinha} M.,  {Holley-Bockelmann} K.,  2015, arXiv e-prints, \href
  {https://ui.adsabs.harvard.edu/abs/2015arXiv150507910S} {p. arXiv:1505.07910}

\bibitem[\protect\citeauthoryear{{Springel}, {Di Matteo}  \&
  {Hernquist}}{{Springel} et~al.}{2005}]{springel2005}
{Springel} V.,  {Di Matteo} T.,   {Hernquist} L.,  2005, \mn@doi [\mnras]
  {10.1111/j.1365-2966.2005.09238.x}, \href
  {https://ui.adsabs.harvard.edu/abs/2005MNRAS.361..776S} {361, 776}

\bibitem[\protect\citeauthoryear{{Stierwalt} et~al.,}{{Stierwalt}
  et~al.}{2013}]{stierwalt2013}
{Stierwalt} S.,  et~al., 2013, \mn@doi [\apjs] {10.1088/0067-0049/206/1/1},
  \href {https://ui.adsabs.harvard.edu/abs/2013ApJS..206....1S} {206, 1}

\bibitem[\protect\citeauthoryear{{Stinson}, {Seth}, {Katz}, {Wadsley},
  {Governato}  \& {Quinn}}{{Stinson} et~al.}{2006}]{stinson2006}
{Stinson} G.,  {Seth} A.,  {Katz} N.,  {Wadsley} J.,  {Governato} F.,   {Quinn}
  T.,  2006, \mn@doi [\mnras] {10.1111/j.1365-2966.2006.11097.x}, \href
  {https://ui.adsabs.harvard.edu/abs/2006MNRAS.373.1074S} {373, 1074}

\bibitem[\protect\citeauthoryear{{Sung}, {Chun}, {Freeman}  \&
  {Chaboyer}}{{Sung} et~al.}{2002}]{sung2002}
{Sung} E.-C.,  {Chun} M.-S.,  {Freeman} K.~C.,   {Chaboyer} B.,  2002, {A New
  Classification Scheme for Blue Compact Dwarf Galaxies}.
{Da Costa}, G.~S. and {Sadler}, E.~M. and {Jerjen}, Helmut, p.~341

\bibitem[\protect\citeauthoryear{{Toloba} et~al.,}{{Toloba}
  et~al.}{2016}]{toloba2016}
{Toloba} E.,  et~al., 2016, \mn@doi [\apjl] {10.3847/2041-8205/816/1/L5}, \href
  {https://ui.adsabs.harvard.edu/abs/2016ApJ...816L...5T} {816, L5}

\bibitem[\protect\citeauthoryear{{Toloba} et~al.,}{{Toloba}
  et~al.}{2018}]{toloba2018}
{Toloba} E.,  et~al., 2018, \mn@doi [\apjl] {10.3847/2041-8213/aab603}, \href
  {https://ui.adsabs.harvard.edu/abs/2018ApJ...856L..31T} {856, L31}

\bibitem[\protect\citeauthoryear{{Toomre} \& {Toomre}}{{Toomre} \&
  {Toomre}}{1972}]{toomre1972}
{Toomre} A.,  {Toomre} J.,  1972, \mn@doi [\apj] {10.1086/151823}, \href
  {https://ui.adsabs.harvard.edu/abs/1972ApJ...178..623T} {178, 623}

\bibitem[\protect\citeauthoryear{{Tortora}, {Napolitano}, {Cardone},
  {Capaccioli}, {Jetzer}  \& {Molinaro}}{{Tortora} et~al.}{2010}]{tortora2010}
{Tortora} C.,  {Napolitano} N.~R.,  {Cardone} V.~F.,  {Capaccioli} M.,
  {Jetzer} P.,   {Molinaro} R.,  2010, \mn@doi [\mnras]
  {10.1111/j.1365-2966.2010.16938.x}, \href
  {https://ui.adsabs.harvard.edu/abs/2010MNRAS.407..144T} {407, 144}

\bibitem[\protect\citeauthoryear{{Tremmel}, {Governato}, {Volonteri}  \&
  {Quinn}}{{Tremmel} et~al.}{2015}]{tremmel2015}
{Tremmel} M.,  {Governato} F.,  {Volonteri} M.,   {Quinn} T.~R.,  2015, \mn@doi
  [\mnras] {10.1093/mnras/stv1060}, \href
  {https://ui.adsabs.harvard.edu/abs/2015MNRAS.451.1868T} {451, 1868}

\bibitem[\protect\citeauthoryear{{Tremmel}, {Karcher}, {Governato},
  {Volonteri}, {Quinn}, {Pontzen}, {Anderson}  \& {Bellovary}}{{Tremmel}
  et~al.}{2017}]{tremmel2017}
{Tremmel} M.,  {Karcher} M.,  {Governato} F.,  {Volonteri} M.,  {Quinn} T.~R.,
  {Pontzen} A.,  {Anderson} L.,   {Bellovary} J.,  2017, \mn@doi [\mnras]
  {10.1093/mnras/stx1160}, \href
  {https://ui.adsabs.harvard.edu/abs/2017MNRAS.470.1121T} {470, 1121}

\bibitem[\protect\citeauthoryear{{Tremmel}, {Governato}, {Volonteri}, {Quinn}
  \& {Pontzen}}{{Tremmel} et~al.}{2018a}]{tremmel2018dancing}
{Tremmel} M.,  {Governato} F.,  {Volonteri} M.,  {Quinn} T.~R.,   {Pontzen} A.,
   2018a, \mn@doi [\mnras] {10.1093/mnras/sty139}, \href
  {https://ui.adsabs.harvard.edu/abs/2018MNRAS.475.4967T} {475, 4967}

\bibitem[\protect\citeauthoryear{{Tremmel}, {Governato}, {Volonteri}, {Pontzen}
   \& {Quinn}}{{Tremmel} et~al.}{2018b}]{tremmel2018wandering}
{Tremmel} M.,  {Governato} F.,  {Volonteri} M.,  {Pontzen} A.,   {Quinn} T.~R.,
   2018b, \mn@doi [\apjl] {10.3847/2041-8213/aabc0a}, \href
  {https://ui.adsabs.harvard.edu/abs/2018ApJ...857L..22T} {857, L22}

\bibitem[\protect\citeauthoryear{{Tremmel} et~al.,}{{Tremmel}
  et~al.}{2019}]{tremmel2019introducing}
{Tremmel} M.,  et~al., 2019, \mn@doi [\mnras] {10.1093/mnras/sty3336}, \href
  {https://ui.adsabs.harvard.edu/abs/2019MNRAS.483.3336T} {483, 3336}

\bibitem[\protect\citeauthoryear{{Tremmel}, {Wright}, {Brooks}, {Munshi},
  {Nagai}  \& {Quinn}}{{Tremmel} et~al.}{2020}]{tremmel2020}
{Tremmel} M.,  {Wright} A.~C.,  {Brooks} A.~M.,  {Munshi} F.,  {Nagai} D.,
  {Quinn} T.~R.,  2020, \mn@doi [\mnras] {10.1093/mnras/staa2015}, \href
  {https://ui.adsabs.harvard.edu/abs/2020MNRAS.497.2786T} {497, 2786}

\bibitem[\protect\citeauthoryear{{Trujillo} \& {Fliri}}{{Trujillo} \&
  {Fliri}}{2016}]{trujillo2016}
{Trujillo} I.,  {Fliri} J.,  2016, \mn@doi [\apj]
  {10.3847/0004-637X/823/2/123}, \href
  {https://ui.adsabs.harvard.edu/abs/2016ApJ...823..123T} {823, 123}

\bibitem[\protect\citeauthoryear{{Trujillo}, {Roman}, {Filho}  \& {S{\'a}nchez
  Almeida}}{{Trujillo} et~al.}{2017}]{trujillo2017}
{Trujillo} I.,  {Roman} J.,  {Filho} M.,   {S{\'a}nchez Almeida} J.,  2017,
  \mn@doi [\apj] {10.3847/1538-4357/aa5cbb}, \href
  {https://ui.adsabs.harvard.edu/abs/2017ApJ...836..191T} {836, 191}

\bibitem[\protect\citeauthoryear{{Trujillo}, {Chamba}  \& {Knapen}}{{Trujillo}
  et~al.}{2020}]{trujillo2020}
{Trujillo} I.,  {Chamba} N.,   {Knapen} J.~H.,  2020, \mn@doi [\mnras]
  {10.1093/mnras/staa236}, \href
  {https://ui.adsabs.harvard.edu/abs/2020MNRAS.493...87T} {493, 87}

\bibitem[\protect\citeauthoryear{{Verner} \& {Ferland}}{{Verner} \&
  {Ferland}}{1996}]{verner1996}
{Verner} D.~A.,  {Ferland} G.~J.,  1996, \mn@doi [\apjs] {10.1086/192284},
  \href {https://ui.adsabs.harvard.edu/abs/1996ApJS..103..467V} {103, 467}

\bibitem[\protect\citeauthoryear{{Virtanen} et~al.,}{{Virtanen}
  et~al.}{2020}]{scipy2020}
{Virtanen} P.,  et~al., 2020, \mn@doi [Nature Methods]
  {10.1038/s41592-019-0686-2}, \href
  {https://ui.adsabs.harvard.edu/abs/2020NatMe..17..261V} {17, 261}

\bibitem[\protect\citeauthoryear{{Vitvitska}, {Klypin}, {Kravtsov}, {Wechsler},
  {Primack}  \& {Bullock}}{{Vitvitska} et~al.}{2002}]{vitvitska2002}
{Vitvitska} M.,  {Klypin} A.~A.,  {Kravtsov} A.~V.,  {Wechsler} R.~H.,
  {Primack} J.~R.,   {Bullock} J.~S.,  2002, \mn@doi [\apj] {10.1086/344361},
  \href {https://ui.adsabs.harvard.edu/abs/2002ApJ...581..799V} {581, 799}

\bibitem[\protect\citeauthoryear{{Wadsley}, {Stadel}  \& {Quinn}}{{Wadsley}
  et~al.}{2004}]{wadsley2004}
{Wadsley} J.~W.,  {Stadel} J.,   {Quinn} T.,  2004, \mn@doi [\na]
  {10.1016/j.newast.2003.08.004}, \href
  {https://ui.adsabs.harvard.edu/abs/2004NewA....9..137W} {9, 137}

\bibitem[\protect\citeauthoryear{{Wadsley}, {Keller}  \& {Quinn}}{{Wadsley}
  et~al.}{2017}]{wadsley2017}
{Wadsley} J.~W.,  {Keller} B.~W.,   {Quinn} T.~R.,  2017, \mn@doi [\mnras]
  {10.1093/mnras/stx1643}, \href
  {https://ui.adsabs.harvard.edu/abs/2017MNRAS.471.2357W} {471, 2357}

\bibitem[\protect\citeauthoryear{Walt, Colbert  \& Varoquaux}{Walt
  et~al.}{2011}]{walt2011numpy}
Walt S. v.~d.,  Colbert S.~C.,   Varoquaux G.,  2011, Computing in Science \&
  Engineering, 13, 22

\bibitem[\protect\citeauthoryear{{White}}{{White}}{1984}]{white1984}
{White} S.~D.~M.,  1984, \mn@doi [\apj] {10.1086/162573}, \href
  {https://ui.adsabs.harvard.edu/abs/1984ApJ...286...38W} {286, 38}

\bibitem[\protect\citeauthoryear{{Wittmann} et~al.,}{{Wittmann}
  et~al.}{2017}]{wittmann2017}
{Wittmann} C.,  et~al., 2017, \mn@doi [\mnras] {10.1093/mnras/stx1229}, \href
  {https://ui.adsabs.harvard.edu/abs/2017MNRAS.470.1512W} {470, 1512}

\bibitem[\protect\citeauthoryear{{Yagi}, {Koda}, {Komiyama}  \&
  {Yamanoi}}{{Yagi} et~al.}{2016}]{yagi2016}
{Yagi} M.,  {Koda} J.,  {Komiyama} Y.,   {Yamanoi} H.,  2016, \mn@doi [\apjs]
  {10.3847/0067-0049/225/1/11}, \href
  {https://ui.adsabs.harvard.edu/abs/2016ApJS..225...11Y} {225, 11}

\bibitem[\protect\citeauthoryear{{Yozin} \& {Bekki}}{{Yozin} \&
  {Bekki}}{2015}]{yozin2015}
{Yozin} C.,  {Bekki} K.,  2015, \mn@doi [\mnras] {10.1093/mnras/stv1073}, \href
  {https://ui.adsabs.harvard.edu/abs/2015MNRAS.452..937Y} {452, 937}

\bibitem[\protect\citeauthoryear{{Zaritsky}}{{Zaritsky}}{2017}]{zaritsky2017}
{Zaritsky} D.,  2017, \mn@doi [\mnras] {10.1093/mnrasl/slw198}, \href
  {https://ui.adsabs.harvard.edu/abs/2017MNRAS.464L.110Z} {464, L110}

\bibitem[\protect\citeauthoryear{{Zhu} et~al.,}{{Zhu} et~al.}{2018}]{zhu2018}
{Zhu} Q.,  et~al., 2018, \mn@doi [\mnras] {10.1093/mnrasl/sly111}, \href
  {https://ui.adsabs.harvard.edu/abs/2018MNRAS.480L..18Z} {480, L18}

\bibitem[\protect\citeauthoryear{{Zitrin}, {Brosch}  \& {Bilenko}}{{Zitrin}
  et~al.}{2009}]{zitrin2009}
{Zitrin} A.,  {Brosch} N.,   {Bilenko} B.,  2009, \mn@doi [\mnras]
  {10.1111/j.1365-2966.2009.15332.x}, \href
  {https://ui.adsabs.harvard.edu/abs/2009MNRAS.399..924Z} {399, 924}

\bibitem[\protect\citeauthoryear{{Zjupa} \& {Springel}}{{Zjupa} \&
  {Springel}}{2017}]{zjupa2017}
{Zjupa} J.,  {Springel} V.,  2017, \mn@doi [\mnras] {10.1093/mnras/stw2945},
  \href {https://ui.adsabs.harvard.edu/abs/2017MNRAS.466.1625Z} {466, 1625}

\bibitem[\protect\citeauthoryear{{Zwicky}}{{Zwicky}}{1957}]{zwicky1957}
{Zwicky} F.,  1957, {Morphological astronomy}

\bibitem[\protect\citeauthoryear{{van Dokkum}, {Abraham}, {Merritt}, {Zhang},
  {Geha}  \& {Conroy}}{{van Dokkum} et~al.}{2015}]{vandokkum2015forty}
{van Dokkum} P.~G.,  {Abraham} R.,  {Merritt} A.,  {Zhang} J.,  {Geha} M.,
  {Conroy} C.,  2015, \mn@doi [\apjl] {10.1088/2041-8205/798/2/L45}, \href
  {https://ui.adsabs.harvard.edu/abs/2015ApJ...798L..45V} {798, L45}

\bibitem[\protect\citeauthoryear{{van Dokkum} et~al.,}{{van Dokkum}
  et~al.}{2016}]{vandokkum2016}
{van Dokkum} P.,  et~al., 2016, \mn@doi [\apjl] {10.3847/2041-8205/828/1/L6},
  \href {https://ui.adsabs.harvard.edu/abs/2016ApJ...828L...6V} {828, L6}

\bibitem[\protect\citeauthoryear{{van Dokkum} et~al.,}{{van Dokkum}
  et~al.}{2017}]{vandokkum2017}
{van Dokkum} P.,  et~al., 2017, \mn@doi [\apjl] {10.3847/2041-8213/aa7ca2},
  \href {https://ui.adsabs.harvard.edu/abs/2017ApJ...844L..11V} {844, L11}

\bibitem[\protect\citeauthoryear{{van Dokkum} et~al.,}{{van Dokkum}
  et~al.}{2018}]{vandokkum2018}
{van Dokkum} P.,  et~al., 2018, \mn@doi [\nat] {10.1038/nature25767}, \href
  {https://ui.adsabs.harvard.edu/abs/2018Natur.555..629V} {555, 629}

\bibitem[\protect\citeauthoryear{{van Dokkum}, {Danieli}, {Abraham}, {Conroy}
  \& {Romanowsky}}{{van Dokkum} et~al.}{2019a}]{vandokkum2019DF4}
{van Dokkum} P.,  {Danieli} S.,  {Abraham} R.,  {Conroy} C.,   {Romanowsky}
  A.~J.,  2019a, \mn@doi [\apjl] {10.3847/2041-8213/ab0d92}, \href
  {https://ui.adsabs.harvard.edu/abs/2019ApJ...874L...5V} {874, L5}

\bibitem[\protect\citeauthoryear{{van Dokkum} et~al.,}{{van Dokkum}
  et~al.}{2019b}]{vandokkum2019}
{van Dokkum} P.,  et~al., 2019b, \mn@doi [\apj] {10.3847/1538-4357/ab2914},
  \href {https://ui.adsabs.harvard.edu/abs/2019ApJ...880...91V} {880, 91}

\bibitem[\protect\citeauthoryear{{van der Burg}, {Muzzin}  \& {Hoekstra}}{{van
  der Burg} et~al.}{2016}]{vanderburg2016}
{van der Burg} R. F.~J.,  {Muzzin} A.,   {Hoekstra} H.,  2016, \mn@doi [\aap]
  {10.1051/0004-6361/201628222}, \href
  {https://ui.adsabs.harvard.edu/abs/2016A&A...590A..20V} {590, A20}

\makeatother
\end{thebibliography}
\label{lastpage}
\end{document}